\documentclass[prd,nofootinbib,11pt,tightenlines,floatfix,eqsecnum,superscriptaddress,showkeys,notitlepage]{revtex4-2}

\usepackage[utf8]{inputenc}          
\usepackage{graphicx,xcolor}         
\usepackage{array}
    \newcolumntype{P}[1]{>{\centering\arraybackslash}p{#1}}
    \newcolumntype{M}[1]{>{\centering\arraybackslash}m{#1}}
\usepackage{dcolumn,longtable}
\usepackage{amsmath,amssymb,slashed} 
\usepackage{bm,xfrac}
\usepackage{multirow}
\usepackage{hyperref}

\providecommand\lfstyle{}                   
\renewcommand\textsc{\MakeUppercase}

\usepackage{siunitx}                                  
\hypersetup{    
    colorlinks=true,       
    linkcolor=black,          
    citecolor=black,        
    filecolor=black,         
    urlcolor=black        
}

\declareslashed{}{\divslash}{0.08}{0}{\mathcal{D}}
\renewcommand\vec[1]{\ensuremath{\bm{#1}}}

\DeclareSIUnit{\fm}{\femto\metre}

\newcommand\snn[1][]{\sqrt{s_\text{NN}}\ifx\\#1\\\else=\SI{#1}{\TeV}\fi}
\newcommand\snnGeV[1][]{\sqrt{s_\text{NN}}\ifx\\#1\\\else=\SI{#1}{\GeV}\fi}

\newcolumntype{C}[1]{>{\vspace{1.5pt}\centering\let\newline\\\arraybackslash\hspace{0pt}}m{#1}}

\begin{document}
\title{One-dimensional complex potentials for quarkonia in a quark-gluon plasma.}
\date{\today}

\author{Roland Katz}
\affiliation{SUBATECH, Universit\'e de Nantes, IMT Atlantique, IN2P3/CNRS, 44307 Nantes, France}
\author{St\'ephane Delorme}
\affiliation{SUBATECH, Universit\'e de Nantes, IMT Atlantique, IN2P3/CNRS, 44307 Nantes, France}
\author{Pol-Bernard Gossiaux}
\affiliation{SUBATECH, Universit\'e de Nantes, IMT Atlantique, IN2P3/CNRS, 44307 Nantes, France}

\begin{abstract}
Master equations of the Lindblad type have recently been derived to describe the dynamics of quarkonium states in the quark-gluon plasma. Because their full resolution in three dimensions is very challenging, the equations are often reduced to one dimension. The main ingredient of these equations is the complex potential that describes the binding of the heavy quark-antiquark pairs and their interactions with the medium. In this work, we propose a one-dimensional complex potential parameterized to reproduce at best two key properties -- the temperature-dependent masses of the eigenstates and their decay widths -- of a three-dimensional lattice QCD inspired potential. Their spectral decompositions are calculated to check their compatibility with the positivity of the master equations.
\end{abstract}

\maketitle

\section*{Introduction} \label{Intro}

The quark-gluon plasma (QGP) produced in high energy heavy ion collisions is considered to be the most extreme medium ever created: mainly composed of deconfined strongly interacting particles, extremely short lived, smallest and most perfect fluid, highest temperature... Among the particles forming this medium, the heavy quarks are known to be excellent probes of its peculiar properties. Being mostly produced at the very beginning of the collision via well known hard processes, the amount of heavy quarks remains constant throughout the collision. Thus, the modification of their production as compared to proton-proton collisions, where no significant QGP production is expected, is mostly due to their interaction with the rest of the medium partons. 

In this work, we deal with the heavy quark-antiquark systems, whose bound states are called quarkonia. More specifically, we focus on the key modeling of the interaction between the heavy quark and antiquark and of their interactions with the medium partons. Because the heavy quark masses are much larger than the typical Quantum Chromodynamic (QCD) scale $\Lambda_{\rm QCD} \sim 200$ MeV, the simplified description of their self interaction via a (non-relativistic) binding potential is possible. In the vacuum, the basic model is the so-called Cornell potential \cite{Eichten:1978tg}, 
\begin{eqnarray}\label{CornellPot}
V(r)=\sigma\,r-\frac{\alpha}{r},
\end{eqnarray}
where $r$ is the distance between the heavy quark and antiquark, and $\sigma$ and $\alpha$ two parameters. The first term describes the long distance non-perturbative confinement, whereas the second term corresponds to a short distance Coulombian like attraction. To take into account the quarkonium strong decays via fragmentation, the Cornell potential is often saturated to a certain value $V_{\rm SB}$ or at a certain distance $r_{\rm SB}$, in order to ``free" the heavy quark pairs with higher energies or larger separations. This simple modeling is close to what is obtained from lattice QCD calculations at the limit $T\rightarrow 0$ \cite{Kaczmarek:2005jy,Digal:2005ht,Mocsy:2007yj,Rothkopf:2011db,Burnier:2015tda}. The hierarchy of scales also ensures the interactions with the thermal medium to be weak enough to be described as a temperature dependent modification of the vacuum binding potential, which can be calculated via the hard-thermal loop (HTL) framework \cite{Laine:2006ns,Beraudo:2007ky,Brambilla:2008cx} or evaluated from lattice QCD results \cite{Kaczmarek:2005jy,Digal:2005ht,Mocsy:2007yj,Rothkopf:2011db,Burnier:2015tda,Lafferty:2019jpr}. The first main modification is an exponential damping of the self interaction due to a Debye-like screening from the color charges in the vicinity of the heavy quark pair. The second main modification is the addition of an imaginary contribution which translates the thermal effects -- i.e. the direct collisions between the medium partons and the heavy quarks -- and generates finite widths for the quarkonium states. It is important to note that this imaginary contribution is derived for a Schr\"odinger-type equation for a correlator and not for a wavefunction, i.e. it does not imply a disappearance of the heavy quarks but instead a decoherence of the pair over time.

In the last decade, a special effort has been made to study the real-time dynamics of heavy quarks pairs inside the QGP \cite{Young:2008he,Young:2010jq,Borghini:2011ms,Akamatsu:2011se,Akamatsu:2012vt,Katz:2013rpa,Akamatsu:2014qsa,Blaizot:2015hya,Katz:2015qja,Brambilla:2016wgg,DeBoni:2017ocl,Blaizot:2017ypk,Brambilla:2017zei,Kajimoto:2017rel,Yao:2018sgn,Blaizot:2018oev,Miura:2019ssi,Sharma:2019xum,Alund:2020ctu,Brambilla:2020qwo,Yao:2020xzw}. The purpose is to obtain a full quantum description of the quarkonium dynamics that can be implemented in a realistic QGP simulation and compared to experimental data. Inspired by the open quantum system framework and using the hierarchy of scales at play, several recent works have derived -- from first QCD principles -- some quantum master equations of the Lindblad type for the quantum Brownian motion of the heavy quarks \cite{Akamatsu:2012vt,Akamatsu:2014qsa,Blaizot:2015hya,Brambilla:2016wgg,Brambilla:2017zei,DeBoni:2017ocl,Blaizot:2017ypk,Blaizot:2018oev}. Various approaches have been proposed to resolve these equations at best: reduce them to their semi-classical limit and obtain a Langevin equation \cite{Blaizot:2015hya,Blaizot:2017ypk,Blaizot:2018oev}, expand them to a limited number of spherical harmonics \cite{Brambilla:2016wgg,Brambilla:2017zei}, turn the equations into a Schr\"odinger equation with a stochastic potential \cite{Kajimoto:2017rel,Sharma:2019xum} or to a stochastic Schr\"odinger equation in one or three dimensions \cite{Miura:2019ssi,Brambilla:2020qwo}, or perform a direct resolution in one dimension \cite{DeBoni:2017ocl,Alund:2020ctu}. The present authors have also performed a full resolution in one dimension (in preparation). The resolution of the full equations in three dimensions remains very challenging and will require a substantial computational effort. In the meantime, most of the studies restrict the resolution to one dimension, therefore limiting some aspects of the underlying physics (e.g. the transitions between states).

For all those approaches the potential is a one of the key ingredients of the dynamics. Several simple one-dimensional potentials have been proposed \cite{DeBoni:2017ocl,Kajimoto:2017rel,Miura:2019ssi,Alund:2020ctu}, mostly based on a single parameter tuned to explore the different regimes given by the master equations or related. However, these potentials are disconnected from the in-medium quarkonium physics and the three-dimensional potentials, making the connection to phenomenology very limited. A precise quantitative comparison to data would obviously require a three-dimensional description, but a well calibrated one-dimensional model could still be used as a good proxy. In this paper, we propose a one-dimensional potential more suited to quarkonium phenomenology. This potential is built to reproduce at best two key properties of a three-dimensional potential inspired by lattice data \cite{Burnier:2015tda,Lafferty:2019jpr}: the temperature-dependent mass and the decay width of the eigenstates. The equivalence of the temperature-dependent masses between the one- and three-dimensional cases ensures to reproduce the quarkonia spectrum at any temperatures as well as the dissociation temperatures. Combined with the decay widths equivalence, we ensure the dissociation to free states and the decoherence of the pair correlations to be approximately similar in the one- and three-dimensional models.

In Sec.\ \ref{pQCDHTL}, we first describe the three-dimensional complex potential obtained from the perturbative HTL framework. We then derive its bona fide reduction to one dimension and analyse its properties. In Sec.\ \ref{Rothkopf3D}, we introduce the three-dimensional lattice QCD inspired potential developed in \cite{Burnier:2015tda,Lafferty:2019jpr}. We analyse the behavior of its real part and the properties of the corresponding eigenstates. We compare the temperature-dependent masses obtained with the spectral functions in \cite{Burnier:2015tda,Lafferty:2019jpr} to the ones obtained with the expectation value methods. We then study the features of the imaginary part and compare the thermal decay widths obtained with the spectral function and expectation value method. We finally calculate the spectral decomposition of the imaginary part and check its compatibility with the positivity of the Lindblad equation. In Sec.\ \ref{3Dto1Dreal}, we describe a simple model for the real part of the one-dimensional potential which reproduces at best the experimental masses of the charmonium and bottomonium states and the temperature-dependent masses given by the three-dimensional potential described in Sec.\ \ref{Rothkopf3D}. The spatial features of the one- and three-dimensional eigenstates are also compared. In Sec.\ \ref{3Dto1Dim}, we propose an imaginary part for the one-dimensional potential, parameterized such as to reproduce at best the thermal decay widths of the three-dimensional potential. We then calculate the spectral decomposition of the one-dimensional imaginary part and analyse its compatibility with the positivity of the Lindblad equation. Finally, in Sec.\ \ref{Comparing1DModels}, the proposed one-dimensional potential is compared to the reduction of the HTL potential in one dimension given in Sec.\ \ref{pQCDHTL} and to a one-dimensional potential found in the literature \cite{Miura:2019ssi,Alund:2020ctu}.

\section{The HTL potential} \label{pQCDHTL}

\subsection{In three dimensions} \label{pQCDHTL3D}

A complex in-medium potential, appearing in a Schrödinger equation for the correlation function of two static heavy quarks, can be derived from the real-time evolution of the QCD Wilson loop using hard-thermal loop (HTL) resummed perturbation theory \cite{Laine:2006ns,Beraudo:2007ky}. The imaginary part of this potential originates from the scattering of the thermal medium particles with the gluons that mediates interaction between the two heavy quarks, similarly to Landau damping in electromagnetic plasma. With $r$ the distance between the two heavy quarks and $T$ the medium temperature, this complex potential reads
\begin{equation}
V ({\bf r},T) \equiv g^{2}C_f\int\frac{d^{3}\mathbf{q}}{(2\pi)^{3}}\Bigl(1 - e^{i \mathbf{q}\cdot {\bf r}}\Bigr)\Biggl[\frac{1}{\mathbf{q}^{2} + m_{D}^{2}} - i \frac{\pi m_{D}^{2}T}{|\mathbf{q}|(\mathbf{q}^{2} + m_{D}^{2})^{2}}\Biggr]. \label{eq:HTLpotential}
\end{equation}
After integration, this potential yields
\begin{equation}\label{eq:VHTL3D}
V(r,T) = -\frac{g^{2}C_f}{4\pi}\Biggl[m_D + \frac{e^{-m_{D}r}}{r}\Biggr]-i\frac{g^{2}T C_f}{4\pi}\phi(m_D r),             
\end{equation}
where 
\begin{equation}
\phi(x)=2\int_{0}^{\infty}\mathrm{d}z\;\frac{z}{\left(z^2+1\right)^2}\left(1-\frac{\sin(zx)}{zx}\right),
\end{equation} 
and $C_f=8/6$. In the framework of effective field theories (EFTs) for non-relativistic bound states, a complex potential can also be derived, but its (non-trivial) expression and the underlying physical phenomena depend on the considered separations of scales \cite{Brambilla:2008cx}. In addition to Landau damping, which dominates at high temperatures, the transitions between color singlet and octet states induced by a medium gluon may contribute as well to the imaginary part. In EFTs, the expression (\ref{eq:HTLpotential}) is only recovered when the temperature is larger than $1/r$ and the Debye mass of the order of $1/r$.

To illustrate the complex potential (\ref{eq:VHTL3D}), we use the QCD running coupling constant $\alpha_s$ determined with the one loop contribution and at the scale $2\pi T$, 
\begin{equation}\label{eq:g2HTL3D}
\alpha_s(T) = \frac{g^2(T)}{4\pi} = \frac{2\pi}{\left(11-\frac{2}{3}\,n_f\right)\,\log\left(\frac{2\pi T}{\Lambda_{\rm QCD}}\right)}, 
\end{equation} 
with $n_f = 3$ massless flavors and the QCD scale taken to be $\Lambda_{\rm QCD} = 0.250$ GeV \cite{Prosperi:2006hx,Escobedo:2019gzn}. For the Debye mass one can use its HTL approximation
\begin{equation}\label{eq:mDHTL3D}
m_D^2(T)=\frac{2\pi}{3}(6+n_f)\alpha_s T^2 = \frac{3}{2} g^2 T^2.
\end{equation} 

\begin{figure}[htb!]
    \centering
    \includegraphics[width=0.46\textwidth]{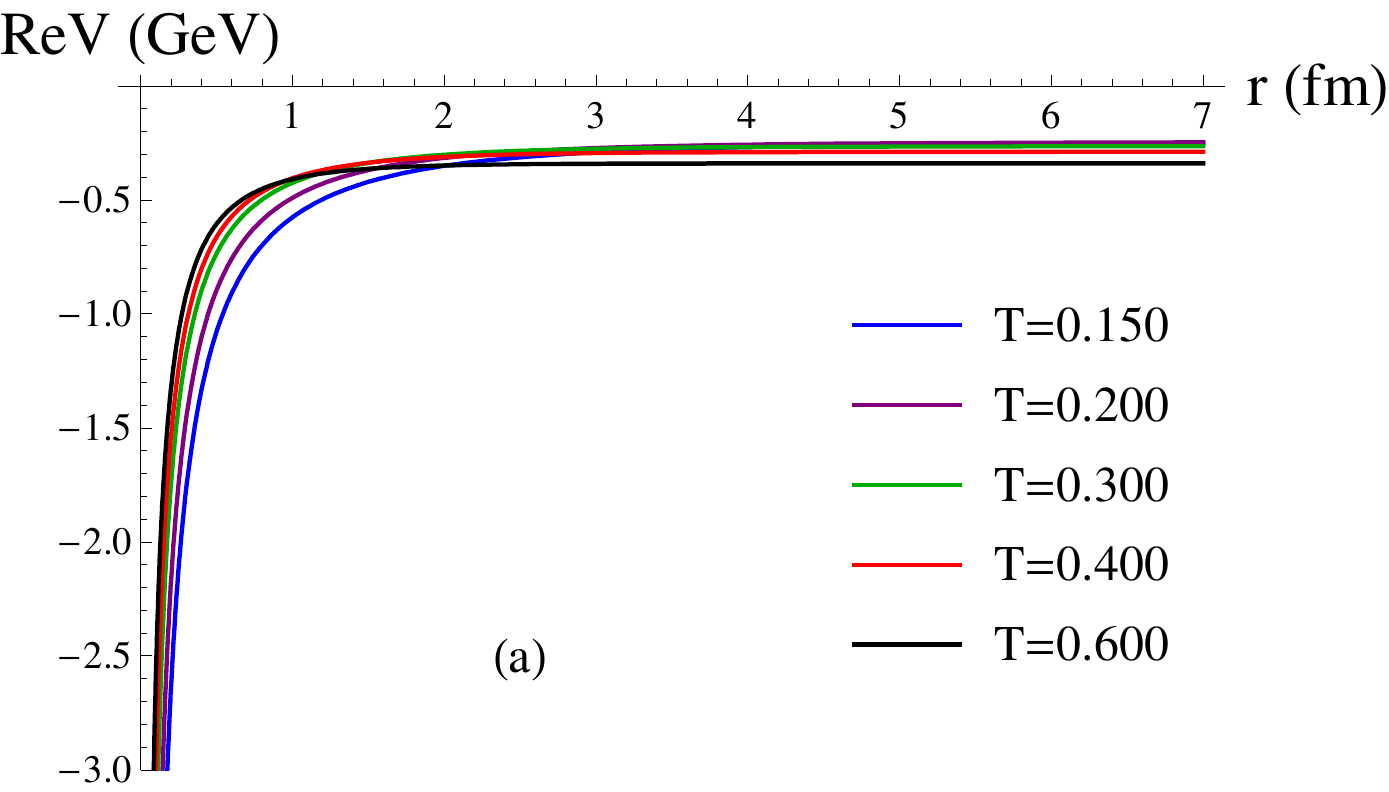}
    \hspace{2mm}
     \includegraphics[width=0.50\textwidth]{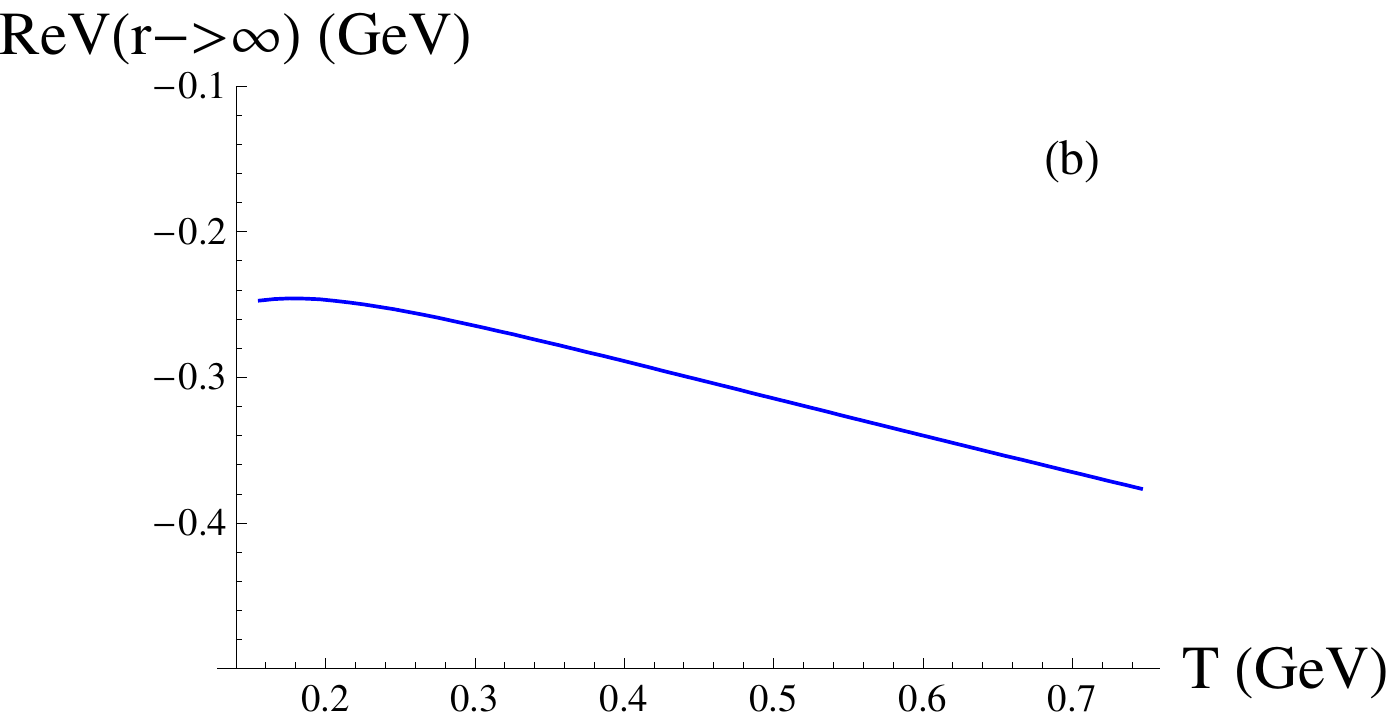}
    \caption{(a) Real part of the potential for different temperatures in GeV. (b) Values at large $r$ of the real part of the potential as a function of T.}
    \label{fig:3DVrealpQCDHTL}
\end{figure}

\begin{figure}[htb!]
    \centering
    \includegraphics[width=0.47\textwidth]{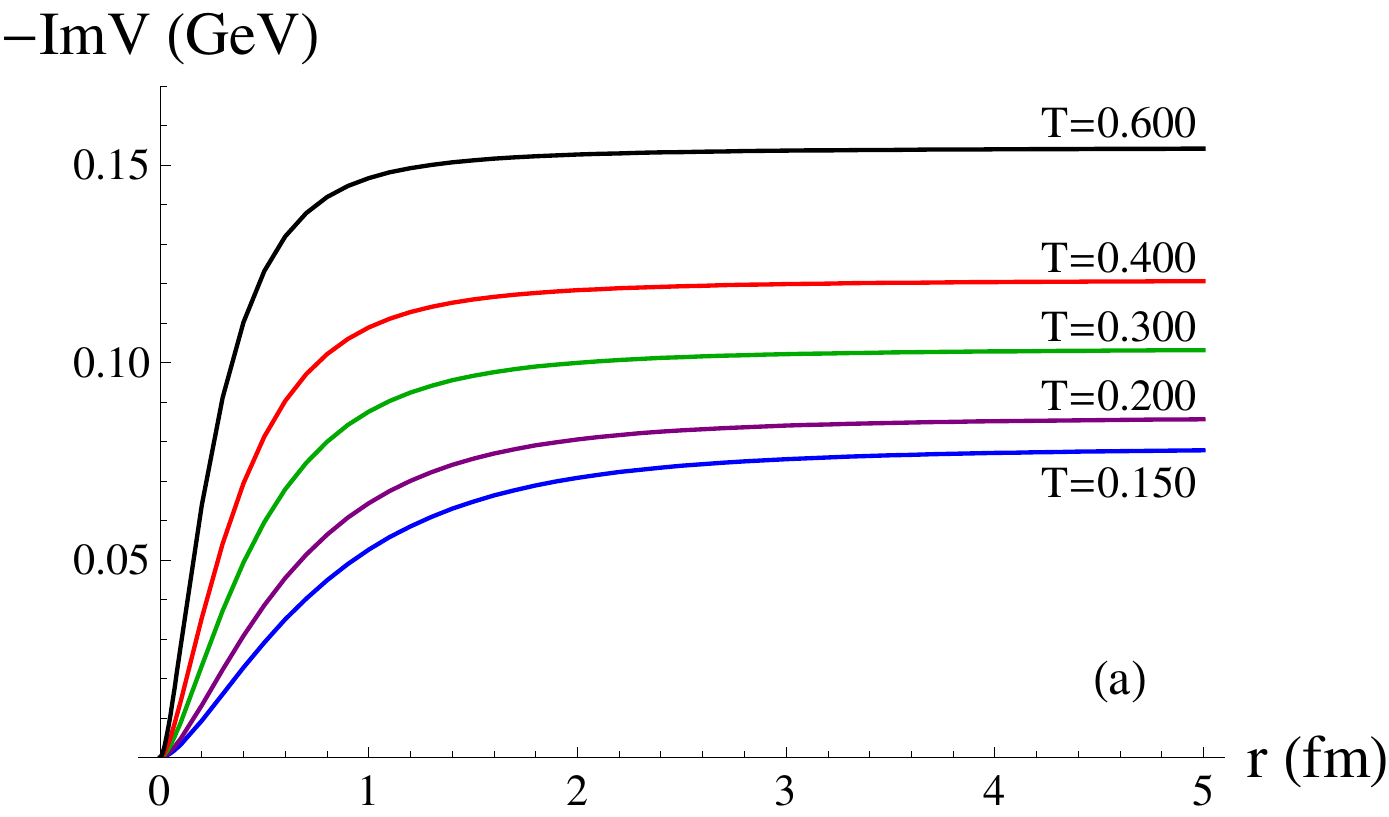}
    \hspace{1mm}
     \includegraphics[width=0.51\textwidth]{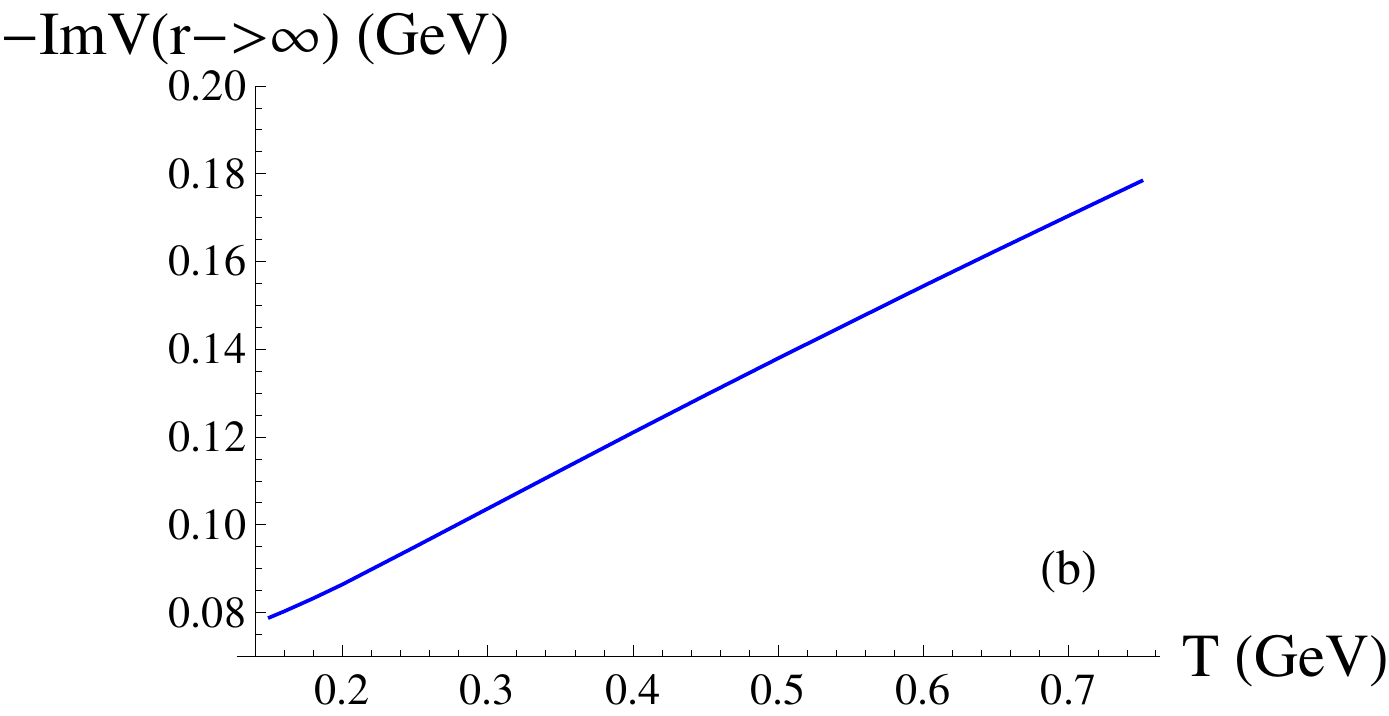}
    \caption{(a) Imaginary part of the potential for different temperatures in GeV. (b) Values at large r.}
    \label{fig:3DVimpQCDHTL}
\end{figure}

As shown in figures \ref{fig:3DVrealpQCDHTL} and \ref{fig:3DVimpQCDHTL}, the real part of (\ref{eq:VHTL3D}) is a screened Coulomb potential with a very small temperature dependence when $T>150$ MeV, and the imaginary part exhibits a small distance harmonic behavior and a large distance saturation that grows monotonously with temperature. The second derivative of the imaginary part at $r=0$ can then be related to the diffusion coefficients of the heavy quarks within the open quantum system framework \cite{Akamatsu:2012vt,Akamatsu:2014qsa,Blaizot:2015hya,DeBoni:2017ocl,Blaizot:2017ypk}.


The HTL potentials are only valid at high temperatures (from at least few hundred MeV) where the weak-coupling expansion is applicable. The Coulombic real part does not include the non-perturbative contributions, such as the string-like linear rise present at larger $r$ in lattice QCD results \cite{Mocsy:2007yj,Rothkopf:2011db,Burnier:2015tda} and in the Cornell form \cite{Eichten:1978tg}, which are necessary at lower temperatures to reproduce the QCD confinement, the quarkonium vacuum spectra and the gradual ``melting'' of the various states. Thus, these potentials are not adapted to the phenomenology of quarkonia in a QGP, which temperature ranges below few hundred MeV most of its lifetime. Nevertheless, their analytical expressions can easily be transposed to lower dimensions and it is thus temptating to use them in one dimension.

\subsection{A possible reduction of the HTL potential to one dimension} \label{pQCDHTL1D}

\begin{figure}[htb!]
    \centering
    \includegraphics[width=0.46\textwidth]{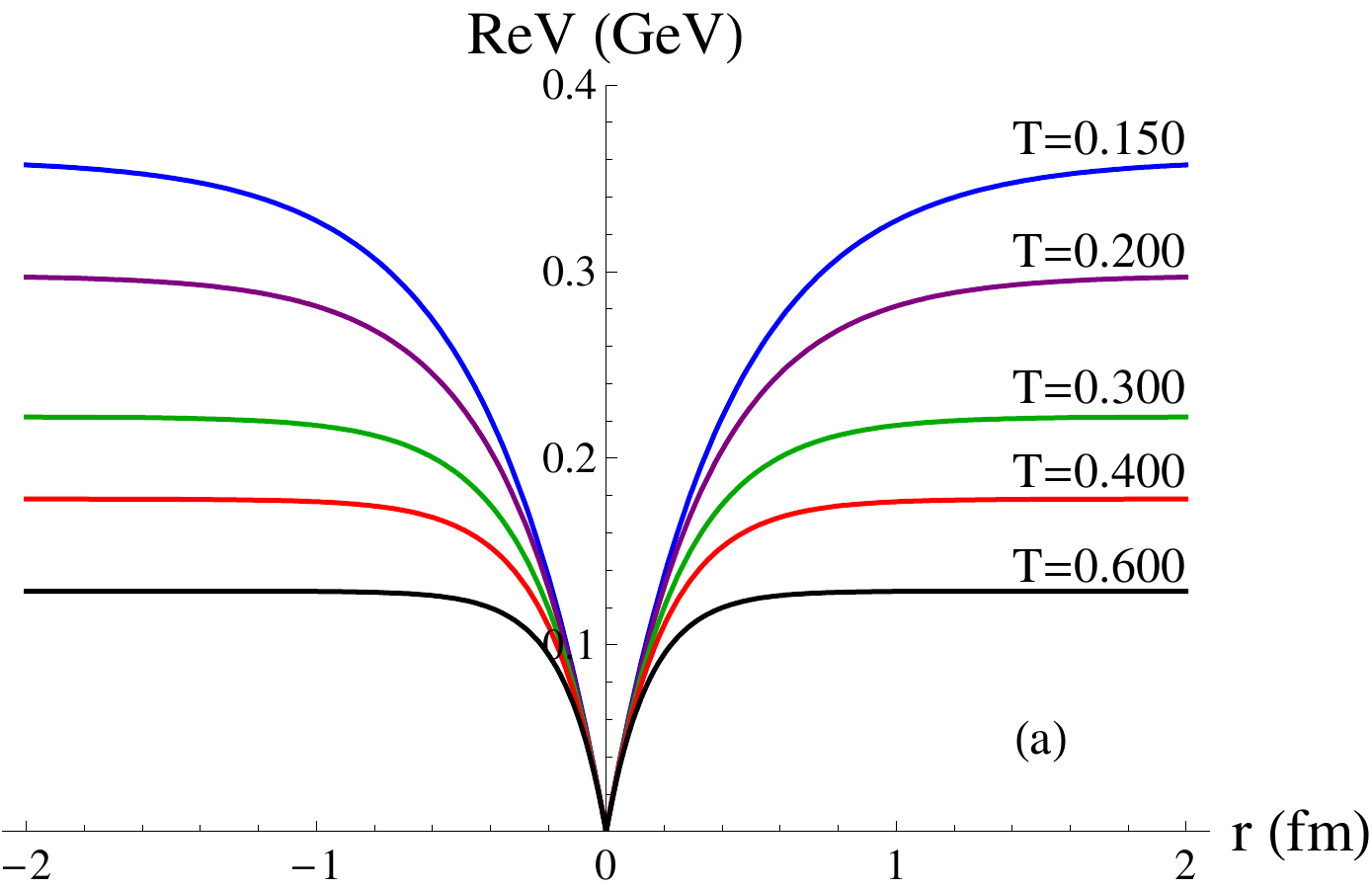}
    \hspace{2mm}
     \includegraphics[width=0.50\textwidth]{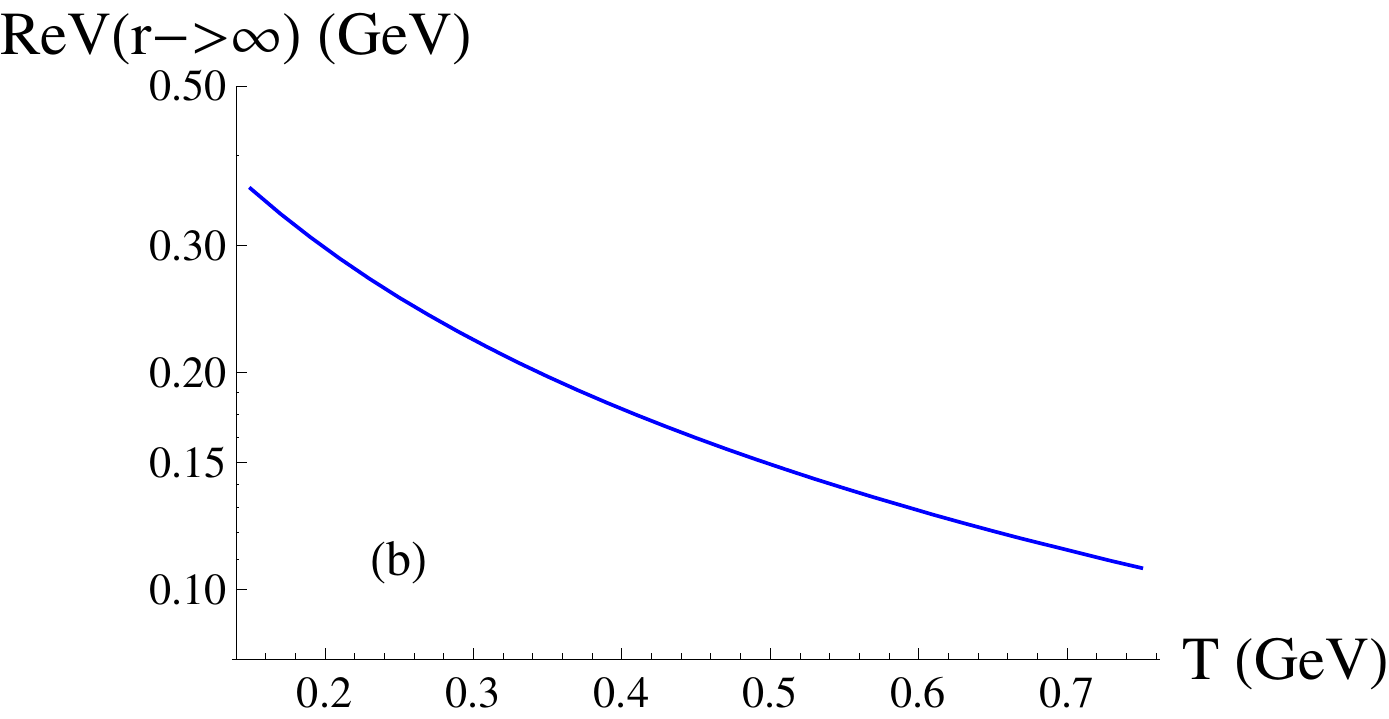}
    \caption{(a) Real part of the one-dimensional potential for different T in GeV. (b) Values at large $r$ of the real part of the potential as a function of the temperature. We use $\sigma=1.724$ GeV/fm for the charmonia taken from Sec.\ \ref{Sec:VacuumPotential1D}.}
    \label{fig:1DVrealpQCDHTL}
\end{figure}

In one dimension, a natural expression for the real part of the potential inspired by the three-dimensional equation (\ref{eq:HTLpotential}) could be taken as\footnote{Within the one dimensional reduction of the HTL framework, there is no screening of the potential per se.}
\begin{equation}
   {\rm Re}V(x,T) = g_{\rm 1D}^{2}C_f\int\frac{dq}{2\pi}\frac{1 - e^{i q|x|}}{q^{2} + m_{D}^{2}}, 
\end{equation}
with $x$ being the distance between the two heavy quarks. After integration, the expression becomes
\begin{equation}
   {\rm Re}V(x,T) = \frac{g_{\rm 1D}^{2}C_f}{2m_{D}}\Bigl(1 - e^{-m_{D}|x|}\Bigr).\label{eq:V1DHTL}
\end{equation}
As shown in Fig.~\ref{fig:1DVrealpQCDHTL}, the real part of this one-dimensional potential is not Colombic as in three dimensions. At small $m_{D}|x|$, it reduces to a linear potential $\propto |x|$, and at large $m_{D}|x|$ it asymptotes to a value strongly dependent on the temperature. In one dimension, $g^{2}$ is not anymore a dimensionless constant and one would need to derive its correct expression. It is more convenient to simplify the potential expression (\ref{eq:V1DHTL}) to
\begin{equation}
   {\rm Re}V(x,T) = \frac{\sigma}{m_{D}}\Bigl(1 - e^{-m_{D}|x|}\Bigr),\label{eq:V1DHTLsigma}
\end{equation}
by analogy with the linear part of the Cornell potential (\ref{CornellPot}), as the potential (\ref{eq:V1DHTLsigma}) now reduces to $\sigma |x|$ at small distances. One can evaluate the constant $\sigma$ such as to reproduce, for instance, the quarkonium spectra (as done in Sec.\ \ref{Sec:VacuumPotential1D}). Note that the plots in this section are only given on an indicative basis as we used the expression (\ref{eq:mDHTL3D}) of the Debye mass $m_{D}$ derived in three dimensions. Nonetheless, the main features of the one-dimensional potential discussed here should remain valid.

In one dimension, the imaginary part of the potential could be taken as
\begin{equation}
   {\rm Im}V(x,T) = - g_{\rm 1D}^{2}C_f\pi m_{D}^{2} T \int_{-\infty}^{\,\infty} \frac{dq}{2\pi} \frac{1 - e^{iq|x|}}{|q|(q^{2} + m_{D}^{2})^{2}},
\end{equation}
which also writes
\begin{equation}
   {\rm Im}V(x,T) = - g_{\rm 1D}^{2}C_f m_{D}^{2} T \int_{\,0}^{\,\infty} dq \frac{1 - \cos{q|x|}}{q(q^{2} + m_{D}^{2})^{2}}.
\end{equation}
At small distances $\big(|x|\ll \frac{1}{m_D}\big)$, the imaginary part reduces to a harmonic potential,
\begin{equation}
   {\rm Im}V(x,T) = - \frac{g_{\rm 1D}^{2}C_f T}{4} x^2,
\end{equation}
similarly to the three-dimensional case. At large distances $\big(|x|\gg \frac{1}{m_D}\big)$, the imaginary part diverges logarithmically,
\begin{equation}
   {\rm Im}V(x,T) \propto - \frac{g_{\rm 1D}^{2}C_f T}{m_{D}^{2}} \ln{(m_{D} |x|)},
\end{equation}
i.e.\ it does not saturate unlike the three-dimensional case. This logarithmic growth appears unphysical as a saturation of the imaginary part at large distances is expected from Landau damping \cite{Laine:2006ns,Beraudo:2007ky,Lafferty:2019jpr}. Additionally, at large distances, the physics of well separated heavy quarks should not influence the behaviour of the quarkonium states, which correspond to small distance correlations. 

\begin{figure}[htb!]
    \centering
    \includegraphics[width=0.48\textwidth]{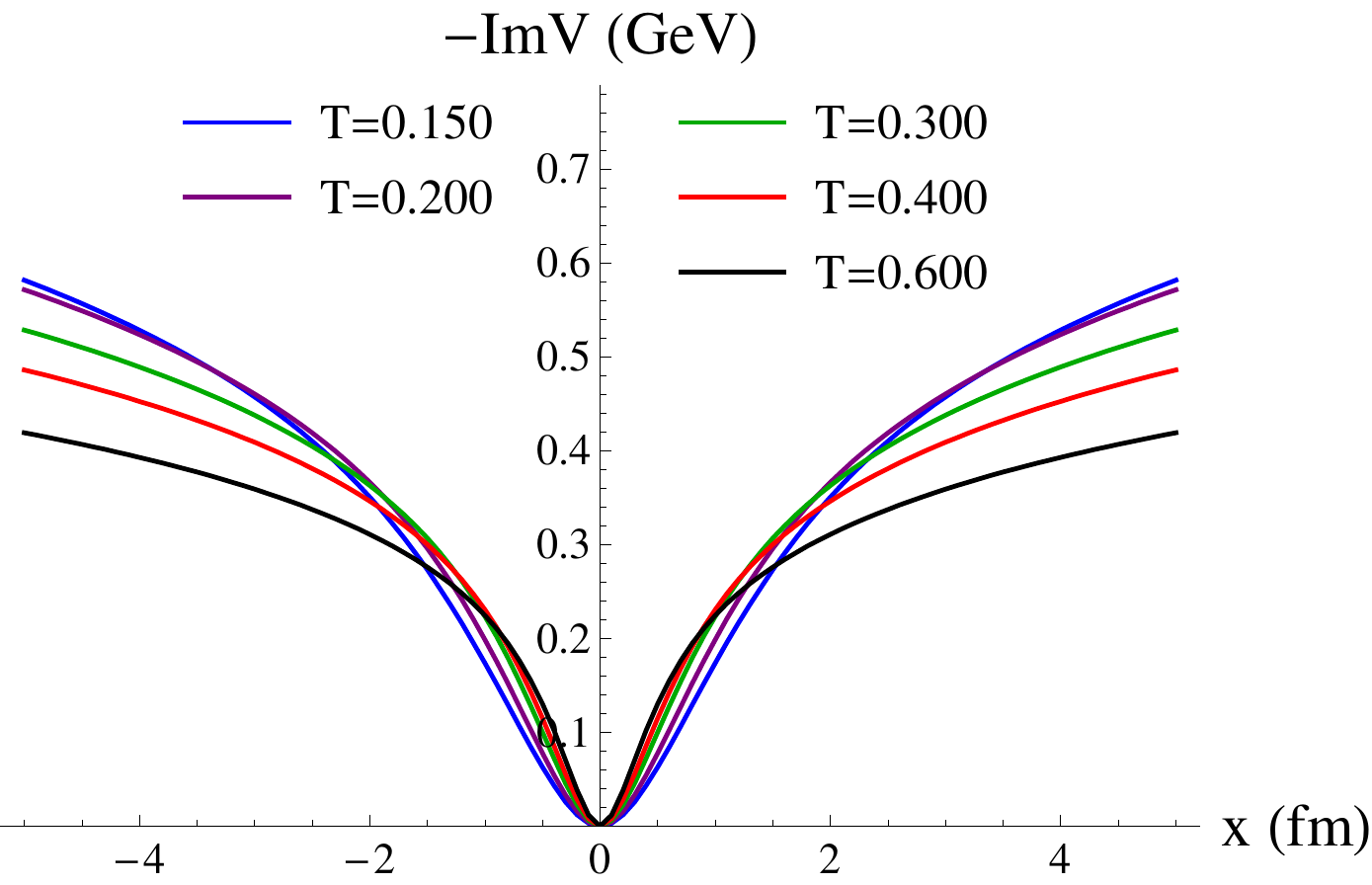}
    \caption{Imaginary part of the one-dimensional potential for different temperatures in GeV. We use $\sigma=1.724$ GeV/fm for the charmonia taken from Sec.\ \ref{Sec:VacuumPotential1D}.}
    \label{fig:1DVimpQCDHTL}
\end{figure}

In accordance with the real part of the potential (\ref{eq:V1DHTLsigma}), we can replace $g_{\rm 1D}^{2} C_f/2$ by $\sigma$: 
\begin{equation}
   {\rm Im}V(x,T) = - 2\sigma m_{D}^{2} T \int_{\,0}^{\,\infty} dq \frac{1 - \cos{q|x|}}{q(q^{2} + m_{D}^{2})^{2}}.
   \label{eq:1DVImpQCDHTL}
\end{equation}
Using the Debye mass derived in three dimensions as for the real part, the imaginary part of the potential (\ref{eq:1DVImpQCDHTL}) is shown for different temperatures in Fig.~\ref{fig:1DVimpQCDHTL}.

Although this one-dimensional potential does not fit our basic physical expectations - because of the logarithmic growth of the imaginary part - its properties will be further studied in Sec.\ \ref{Comparing1DModels} and compared to other models.


\section{The three-dimensional lattice QCD inspired complex potential} \label{Rothkopf3D}

To find a three dimension in-medium quarkonium potential valid in a broader range of temperature and relevant in terms of phenomenology, one can turn to lattice QCD inspired potentials \cite{Kaczmarek:2005jy,Digal:2005ht,Mocsy:2007yj,Burnier:2015tda,Lafferty:2019jpr}. The real parts of these potentials are generally parameterizations of Cornell forms supplemented by exponential damping factors \cite{Kaczmarek:2005jy,Digal:2005ht,Mocsy:2007yj} or parameterizations of more advanced analytical expressions derived from linear response theory \cite{Thakur:2013nia,Burnier:2015tda,Lafferty:2019jpr}. 
The parameters are then generally determined by a fit to lattice QCD numerical calculations at different distances and temperatures. We focus in this section on the only complex potential model inspired by lattice QCD results developed so far \cite{Burnier:2015tda,Lafferty:2019jpr}. For the sake of completeness, we first sum up the model further described in \cite{Burnier:2015tda,Lafferty:2019jpr}.

\subsection{The model from  Y.\ Burnier, O.\ Kaczmarek, A.\ Rothkopf and  D.\ Lafferty \cite{Burnier:2015tda,Lafferty:2019jpr}} \label{Rothkopf3DModel}

\begin{figure}[htb!]
    \centering
    \includegraphics[width=0.50\textwidth]{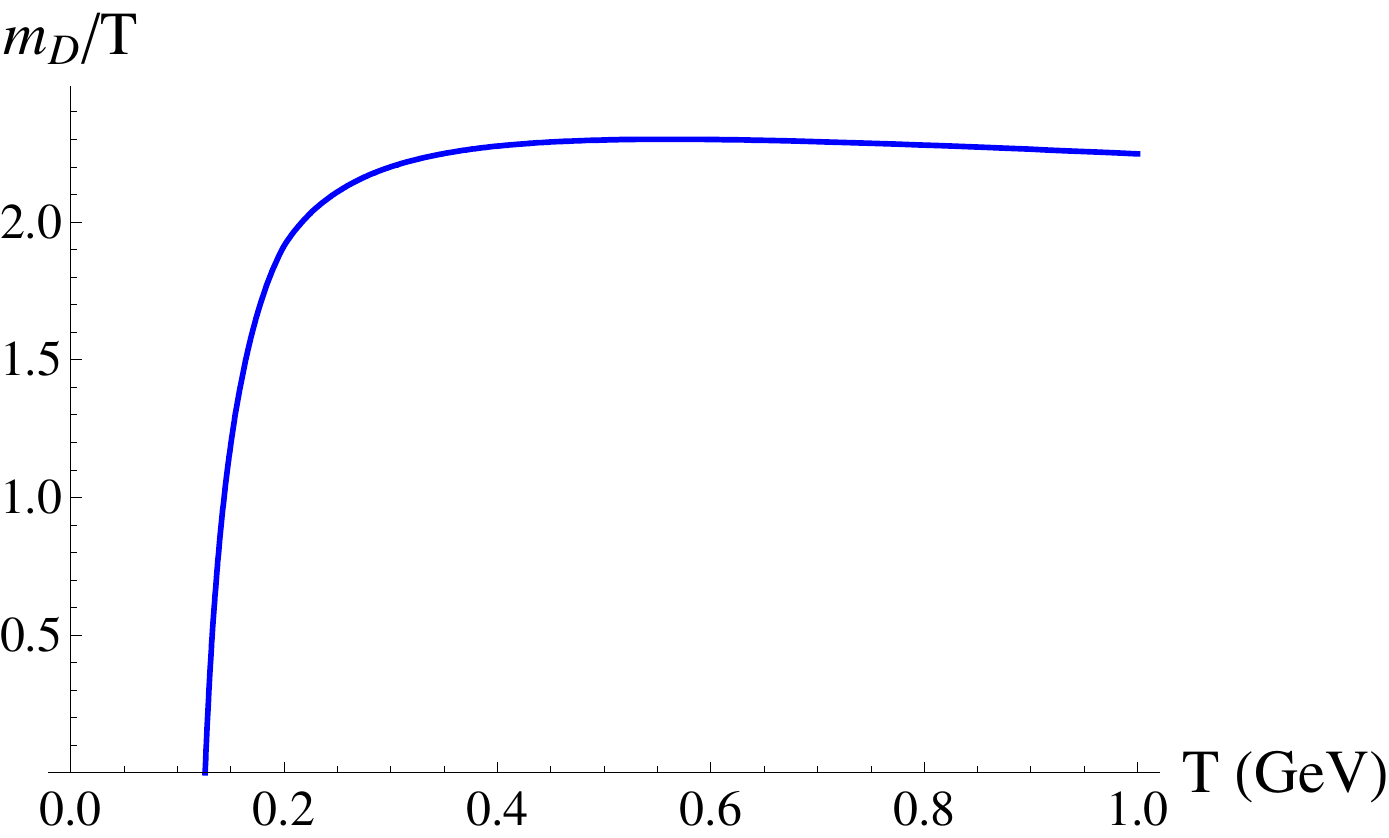}    
    \caption{Debye mass dependence on the temperature \cite{Burnier:2015tda,Lafferty:2019jpr}.}
    \label{fig:mDRothkopf}
\end{figure}

Based on the linear response theory framework and a generalized Gauss law ansatz, the in-medium complex potential is derived from the vacuum Cornell potential using the in-medium complex permittivity obtained from a HTL calculation. Similarly to the Cornell form, the resulting analytic expressions can be decomposed in a Coulombic (``$V_C$'') and a linear string-like part (``$V_S$'') which both depend on a single temperature dependent parameter, the Debye mass $m_D$.
\begin{equation}\label{eq:3DPottot}
\mathrm{Re}V = \mathrm{Re}V_C + \mathrm{Re}V_S + c \,\,\,\,\,\,{\rm and}\,\,\,\,\,\,  \mathrm{Im}V = \mathrm{Im}V_C + \mathrm{Im}V_S,
\end{equation}
where $c$ is a constant. The Coulombic part writes:
\begin{equation}\label{eq:Coulombic3D}
\mathrm{Re}V_C\!\left(r\right)=-\alpha_s\left[m_D+\frac{e^{-m_Dr}}{r} \right], \quad
\mathrm{Im}V_C\!\left(r\right)=\alpha_s T\phi(m_D r),
\end{equation}
where \begin{equation}\label{eq:CoulombicInt}
\phi(x)=2\int_{0}^{\infty}\mathrm{d}z\;\frac{z}{\left(z^2+1\right)^2}\left(1-\frac{\sin(zx)}{zx}\right).
\end{equation} 
$V_C$ is perfectly equivalent to the HTL potential (\ref{eq:VHTL3D}) seen in Sec.\ \ref{pQCDHTL3D}. The string part writes:
\begin{align}\label{eq:String3D}
\mathrm{Re}V_S\!\left(r\right)=\frac{2\sigma}{m_D}-\frac{e^{-m_D r}\left(2+m_Dr\right)\sigma}{m_D}, \quad \mathrm{Im}V_S(r)=\frac{\sigma T}{m_D^2} \chi(m_D r),
\end{align}
where 
\begin{equation}\label{eq:StringInt}
\chi(x)=2\int_0^{\infty}\mathrm{d}p\;\frac{2-2\cos(px)-px\sin(px)}{\sqrt{p^2+\Delta_D^2}\left(p^2+1\right)^2},
\end{equation}
and \(\Delta_D=\Delta/m_D\) $\approx$ 3.0369 with \(\Delta\) being a suitably chosen regularization scale. The latter has been introduced to avoid the logarithmic divergence of the imaginary string part at large r, which can be traced back to the absence of string breaking in the initial Cornell potential. The choice in the formulation of this regularization scale has been checked to not influence the end results. The set of phenomenological parameters coming from the vacuum Cornell potential, 
\begin{equation}
\label{eq:vac_param}
\alpha_s =0.513\pm 0.0024, \quad \sqrt{\sigma}=0.412\pm 0.0041~\mathrm{GeV},\quad c=-0.161\pm 0.0025~\mathrm{GeV} ,
\end{equation}
is fixed to reproduce the energy spectrum of the bottomonia with the bottom quark mass taken to be the so-called renormalon subtracted mass \cite{Pineda:2001zq},
\begin{equation}\label{massBottom}
 m_b=m_b^{RS}=4.882~\mathrm{GeV}. 
 \end{equation}
The charm quark mass \(m_c=m_c^{\mathrm{fit}}=1.4692~\mathrm{GeV}\) is then determined to reproduce the energy spectrum of the charmonia while keeping the same vacuum parameters (\ref{eq:vac_param}). The Debye mass parameter is obtained via the next-to-leading order HTL expression supplemented by two terms (quadratic and cubic in the coupling constant $g$) accounting for non-perturbative contribution \cite{Burnier:2015tda,Lafferty:2019jpr,Vermaseren:1997fq}. The two constants appearing in the latter are fixed using continuum corrected lattice results. The resulting Debye mass\footnote{A good fit to the resulting Debye mass is given by: $m_D/T=1.275 \ln{(170.813\,T - 19.4086)} - 7.593\,T$  for $T_0<T<0.2$ GeV and $m_D/T=0.6724\,{\rm Re}\left[\ln{(95347.824\,T - 12617.627)}\right] - 
    7.864\,T^{0.12} + 2.4975$ for $0.2<T<1$ GeV.} dependence on temperature is shown in Fig.\ \ref{fig:mDRothkopf}. 
The zero of the Debye mass defines the ``vacuum'' temperature $T_0 \approx 0.126025$ GeV of this potential model. The resulting curves for the real part of the potential fit nicely the data extracted from the lattice at any temperature \cite{Burnier:2014ssa,Burnier:2015tda}. The curves for the imaginary part also agree quite well with the tentative lattice QCD results up to $r \approx 1$ fm and down to $T \sim 160$ MeV (see figure 2 in \cite{Lafferty:2019jpr}). At high temperatures (i.e.\ large $m_D$), the string part $V_S$ becomes negligible and the potential reduces to the HTL result $V_C$. At low temperatures (i.e.\ when $m_D \rightarrow 0$ and $T\rightarrow T_0$), the model approaches the vacuum Cornell form. To include string breaking to the real part of the potential, a flat asymptotics is enforced from $r_{SB} = 1.25$ fm at $T=T_0$, which corresponds to a saturation of the potential at $V_{\rm SB}\approx 0.8352$ GeV. The latter is enforced to be the maximum value of the real part for $T>T_0$ as well. The value of $V_{\rm SB}$ is consistently close to the fragmentation threshold $2m_B - 2 m_b \approx 2m_{D^0} - 2 m_c \approx 0.8$ GeV. We do not consider the complex potential extension to the finite baryon density and finite velocity regimes developed in \cite{Lafferty:2019jpr}. 

\subsection{Features of the real part of the potential} \label{Rothkopf3DFeaturesReal}

\begin{figure}[htb!]
    \centering
    \includegraphics[width=0.47\textwidth]{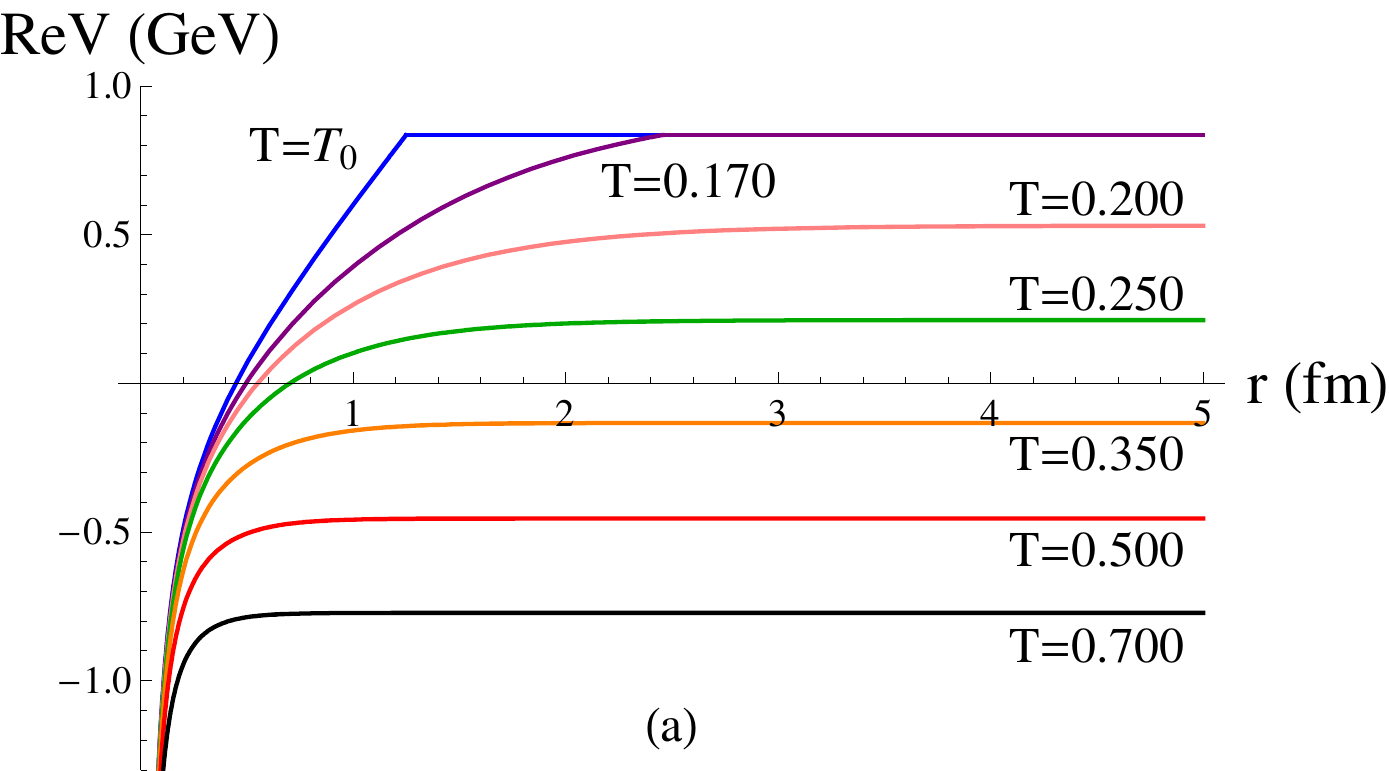}
     \includegraphics[width=0.51\textwidth]{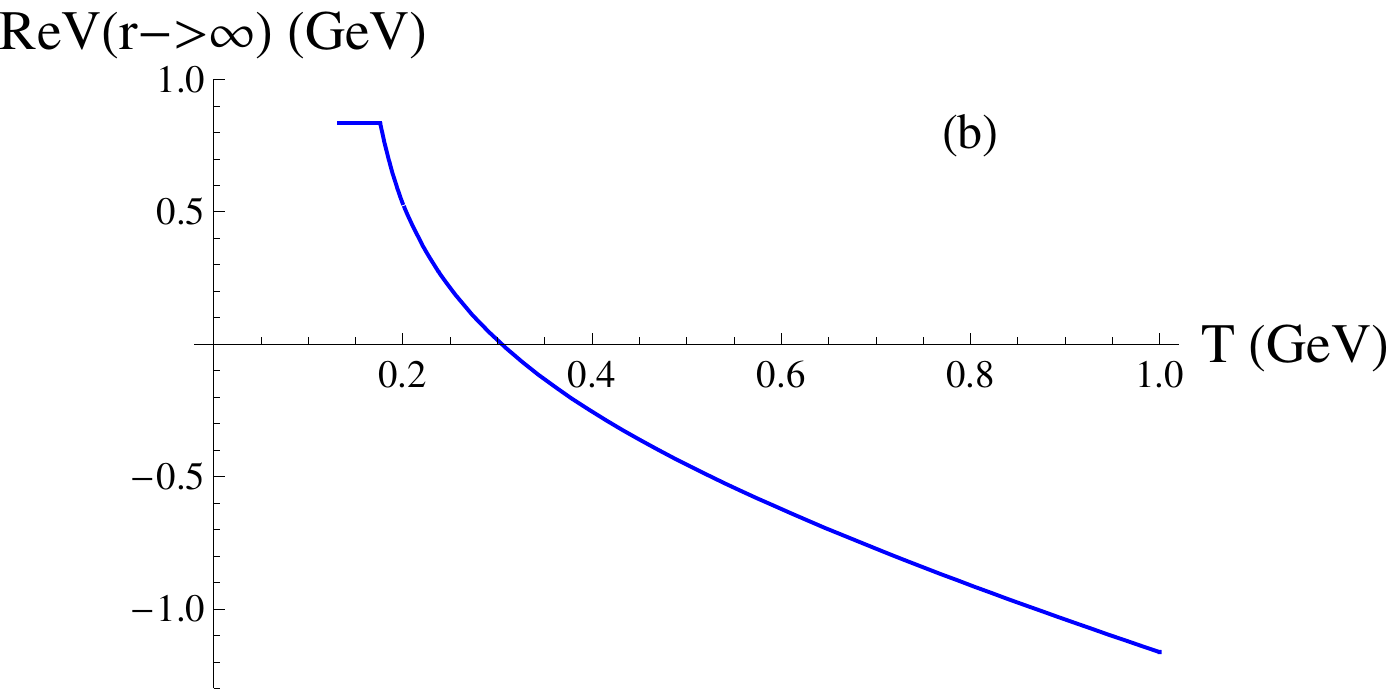}
    \caption{(a) Real part of the potential for different temperatures in GeV. (b) Values at large distances of the real part of the potential as a function of the temperature.}
    \label{fig:3DVreal}
\end{figure}

In Fig.\ \ref{fig:3DVreal}, we show the resulting real part of the potential at different temperatures and its temperature dependence at large distances. At small distances $\big(r\ll \frac{1}{m_D}\big)$, the potential becomes purely Coulombic and is independent on the temperature. At large distances $\big(r\gg \frac{1}{m_D}\big)$, due to the string part, the potential decrease with the temperature is much steeper than it was in Fig.\ \ref{fig:3DVrealpQCDHTL} with the HTL potential. 

\begin{figure}[htb!]
    \centering
    \includegraphics[width=0.47\textwidth]{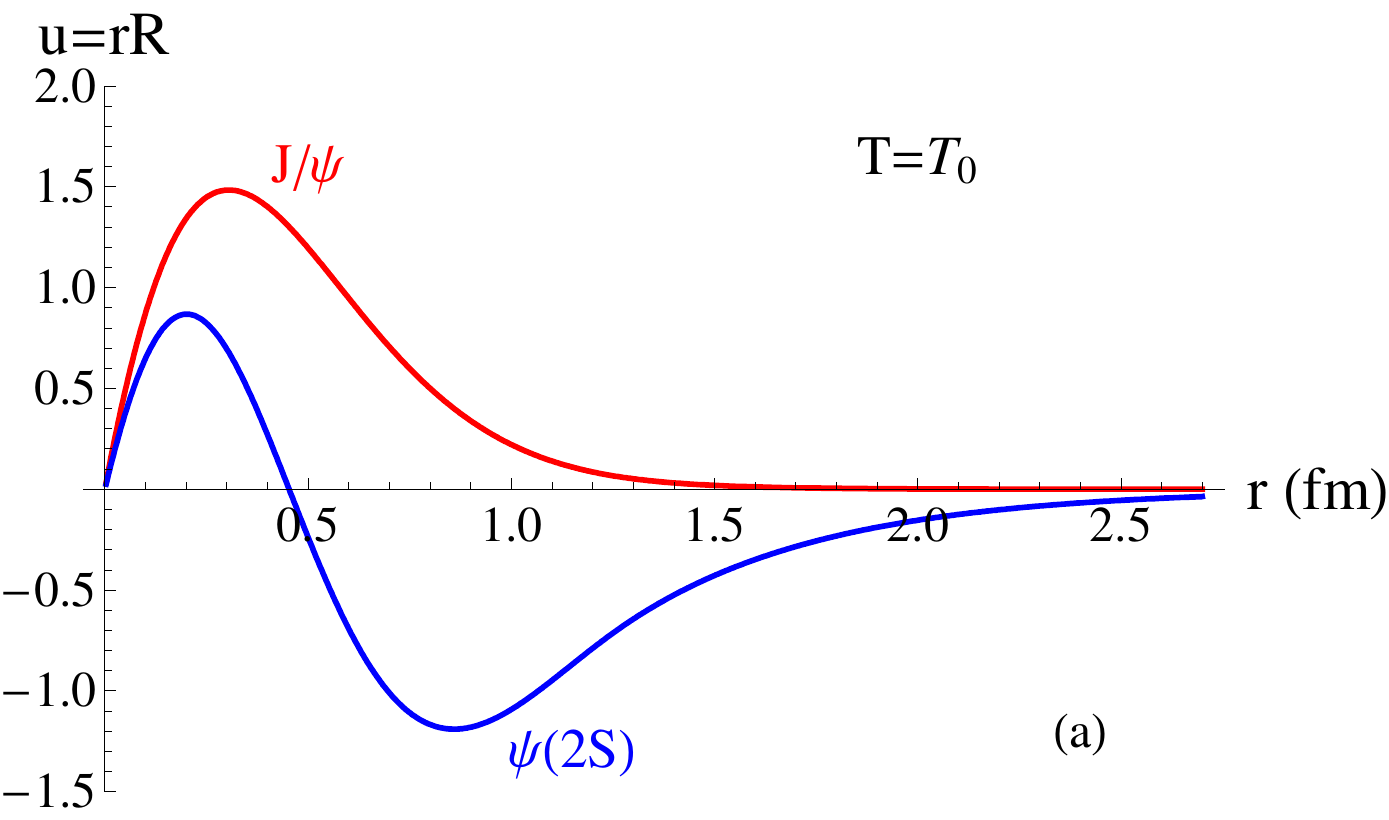}
     \includegraphics[width=0.47\textwidth]{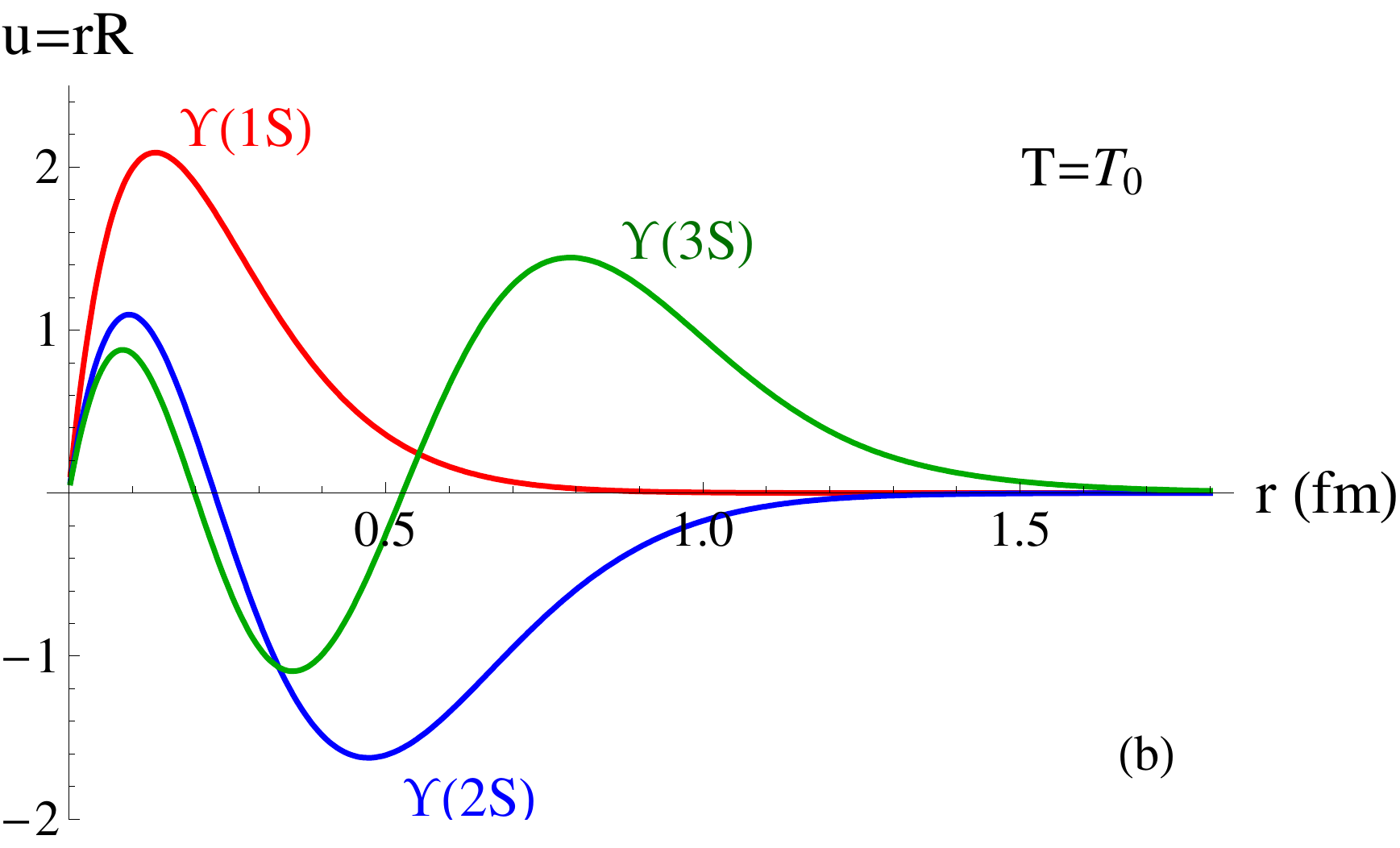}
    \caption{(a) Reduced radial wavefunctions $u(r)=rR(r)$ of the charmonium S states at the vacuum temperature $T_0$ (b) Same for bottomonia. }
    \label{fig:eigenfunctions3D}
\end{figure}

To further analyse the properties of this potential, we use the Hamiltonian of a non-relativistic two particles system in the $Q{\bar Q}$ pair center of mass frame, given by :
\begin{eqnarray}\label{Hamiltonian}
H=2m_Q-\frac{1}{m_Q}\nabla^2+{\rm Re}V(r,T),
\end{eqnarray}
where $\nabla$ is the nabla operator and $m_Q$ is the heavy quark mass\footnote{The value of ${\rm Re}V(r\rightarrow\infty,T)$ can be interpreted as a temperature dependent ``mean field" contribution to the heavy quark mass \cite{Beraudo:2010tw,Riek:2010fk,Liu:2017qah} : $m_Q(T)= m_Q^0+{\rm Re}V(r\rightarrow\infty, T)/2$, where $m_Q^0$ is the heavy quark bare mass. If one would consider $m_Q(T)$ in the Hamiltonian, then the potential would be rescaled to ${\rm Re}V(r, T)-{\rm Re}V(r\rightarrow\infty, T)$ and one may use $m_Q(T)$ in the kinetic term as well. To our knowledge, the effect of a temperature dependent inertial mass has never been studied within the quarkonium framework. In the present work, this temperature dependence of the heavy quark mass is not considered.} (given by Eq.~\ref{massBottom} and beneath).
The reduced radial wavefunctions $u(r)=rR(r)$ (with $R(r)$ the radial part of the wavefunction) for the charmonium and bottomonium S states at the vacuum temperature $T_0$ are shown in Fig. \ref{fig:eigenfunctions3D}. In Tab.\ \ref{tab:Eandrvacuum3D}, the masses $m^{\rm EV}$ of the different quarkonium S states obtained from the expectation values of the Hamiltonian at the vacuum temperature $T_0$ are compared to the masses obtained with the spectral functions $m^{\rm SF}$ in \cite{Lafferty:2019jpr} and to the experimental values. The differences between the results of the two methods are negligible and the values are consistent with the data. The mean radiuses of the ``vacuum'' states obtained from the expectation values $\langle r \rangle$ are also given in Tab.\ \ref{tab:Eandrvacuum3D}. 

\begin{table}[h!]
\begin{center}
    \begin{tabular}{|C{2.5cm}||C{1.5cm}|C{1.5cm}||C{1.5cm}|C{1.5cm}|C{1.5cm}|}
    \hline
At $T=T_0$ & $J/\psi$ & $\psi$(2S) & $\Upsilon$(1S) & $\Upsilon$(2S) & $\Upsilon$(3S) \\
    \hline
    \hline
$m^{\rm PDG}$ \cite{Zyla:2020zbs} & 3.0969 & 3.6861 & 9.4603 & 10.023  & 10.355  \\
    \hline
$m^{\rm SF}$ \cite{Lafferty:2019jpr} & 3.0969 & 3.6632 & 9.4603 & 10.023  & 10.355  \\ 
    \hline
$m^{\rm EV}$ & 3.0964 & 3.6642 & 9.4611 & 10.023  & 10.355  \\ 
    \hline
    \hline
$r^{\rm EV}$ & 0.387 & 0.856 & 0.188 & 0.463  & 0.687  \\ 
    \hline
    \end{tabular}
\caption {\label{tab:Eandrvacuum3D} 
\small Masses (in GeV) and mean radiuses (in fm) for the ``vacuum'' states (at $m_D=0$ $\Leftrightarrow T_0 \approx 0.126025$ GeV). $m^{\rm PDG}$ is the experimental mass given by the particle data group, $m^{\rm SF}$ corresponds to the mass obtained from the spectral functions \cite{Lafferty:2019jpr} and $m^{\rm EV}$ ($r^{\rm EV}$) is the mass (mean radius respectively) calculated from the expectation values of the Hamiltonian (of the radius operator respectively).}
\end{center}
\end {table}

\begin{figure}[htb!]
    \centering
    \includegraphics[width=0.49\textwidth]{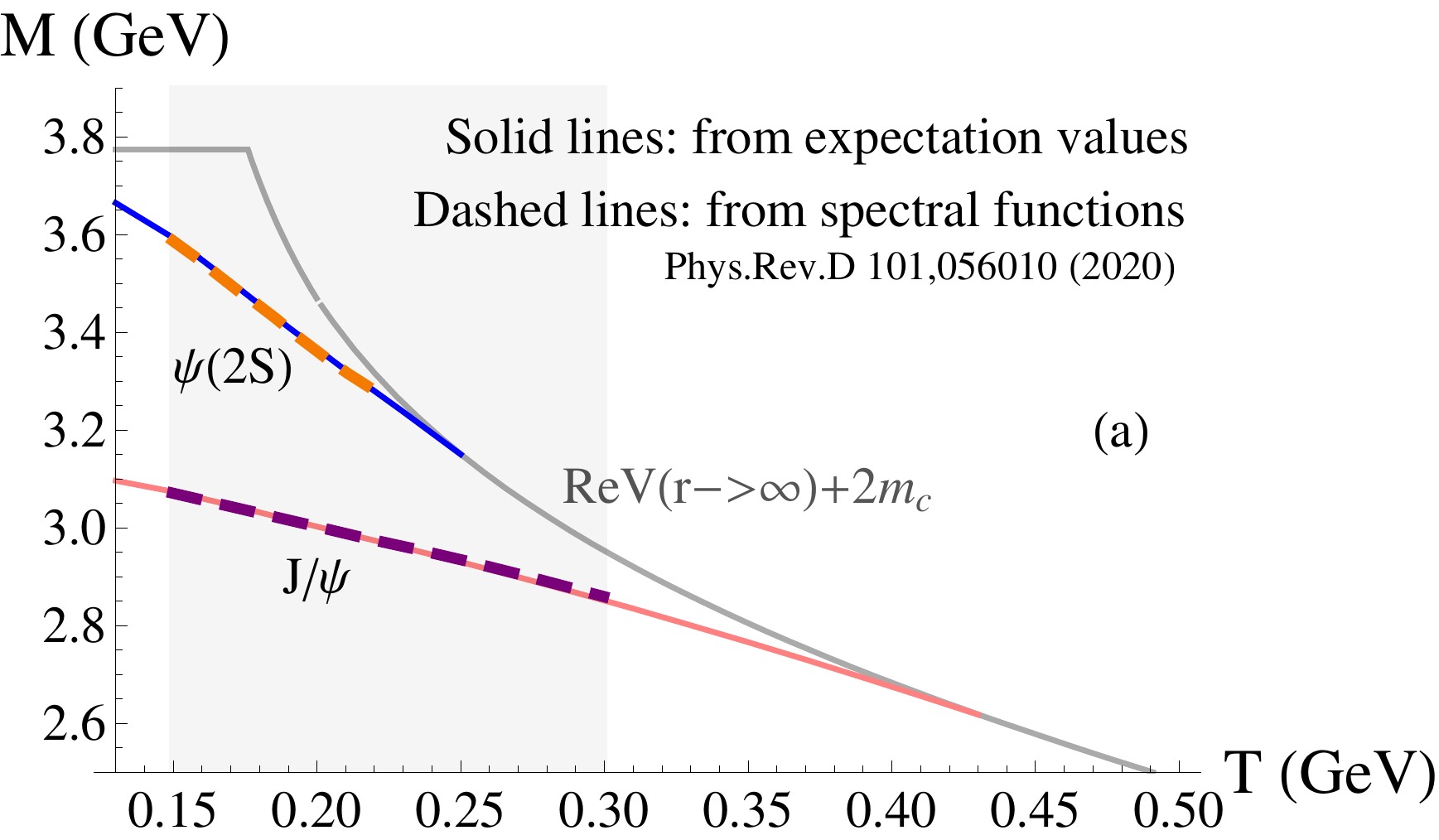}
     \includegraphics[width=0.49\textwidth]{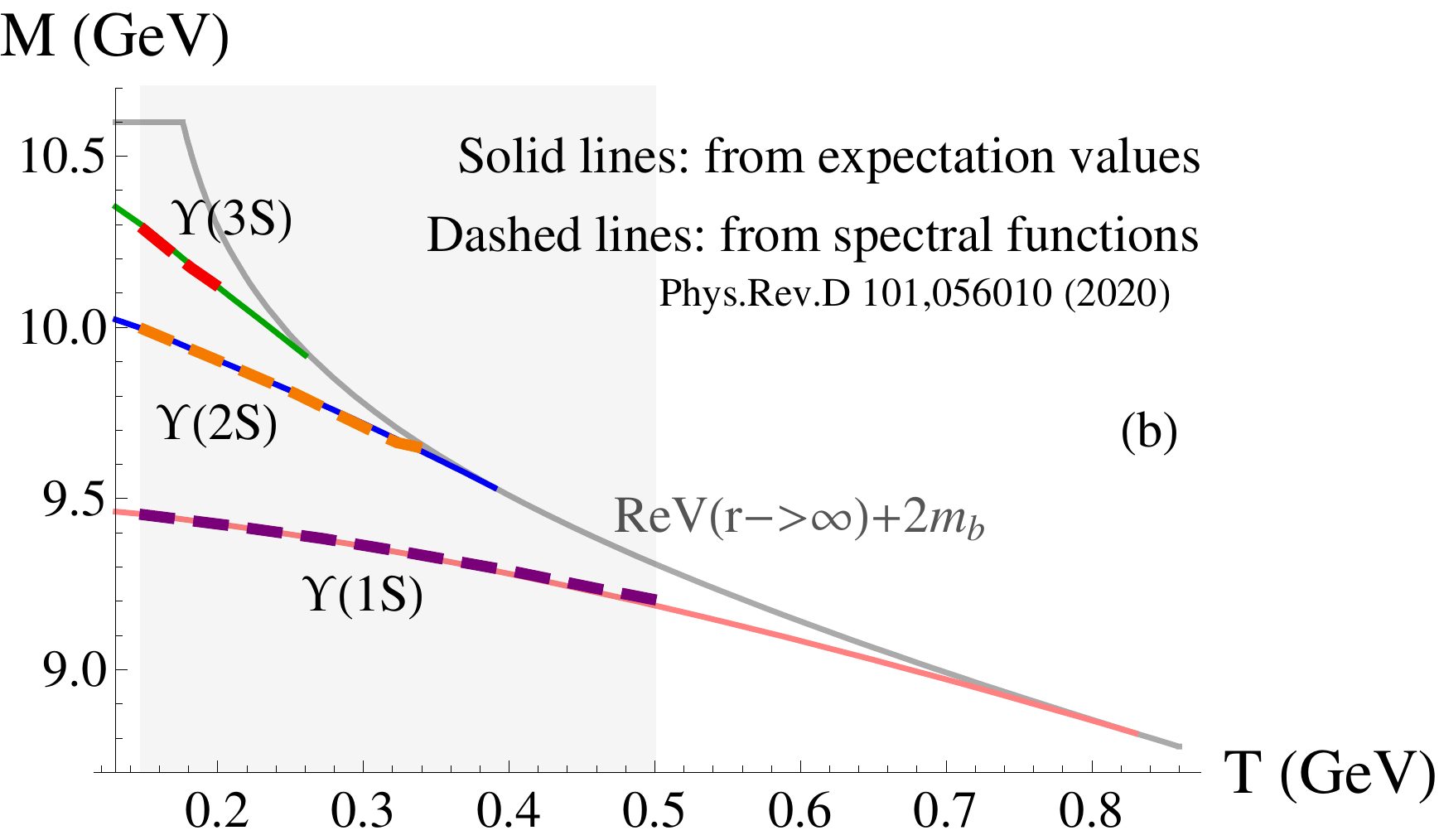}
    \caption{(a) Thermal masses $M(T)=E_n(T)+2m_c$ of the charmonium states and (b) Same for bottomonia. Gray areas corresponds to the results available from spectral functions in \cite{Lafferty:2019jpr}.}
    \label{fig:3DEstatesT}
\end{figure}

In Fig.\ \ref{fig:3DEstatesT}, the temperature-dependent masses of the states $M(T)=E_n(T)+2m_Q$, where $E_n$ are the expectation values of the Hamiltonian, are compared to the temperature-dependent masses obtained in \cite{Lafferty:2019jpr} with the spectral functions via the fit of the spectral peaks with (skewed) Breit-Wigner distributions. The masses obtained with both methods are similar and globally decrease with the temperature, but the dissociation temperatures are observed to be larger with the expectation value method -- they correspond to the crossing of $E_n(T)$ and ${\rm Re}V(r\rightarrow\infty)$ -- than with the spectral function method. This difference in behavior close to dissociation might be explained by the difficulty of fitting large spectral peaks with skewed Breit-Wigner distributions. As noted in \cite{Lafferty:2019jpr}, the variation with temperature is stronger for excited states with smaller binding energies $E^{\rm bind}_n={\rm Re}V(r\rightarrow\infty,T_0)-E_n(T_0)$, which corresponds to the intuition that tightly bound states are less sensitive to medium effects.

\subsection{Features of the imaginary part of the potential} \label{Rothkopf3DFeaturesIm}

\begin{figure}[htb!]
    \centering
    \includegraphics[width=0.47\textwidth]{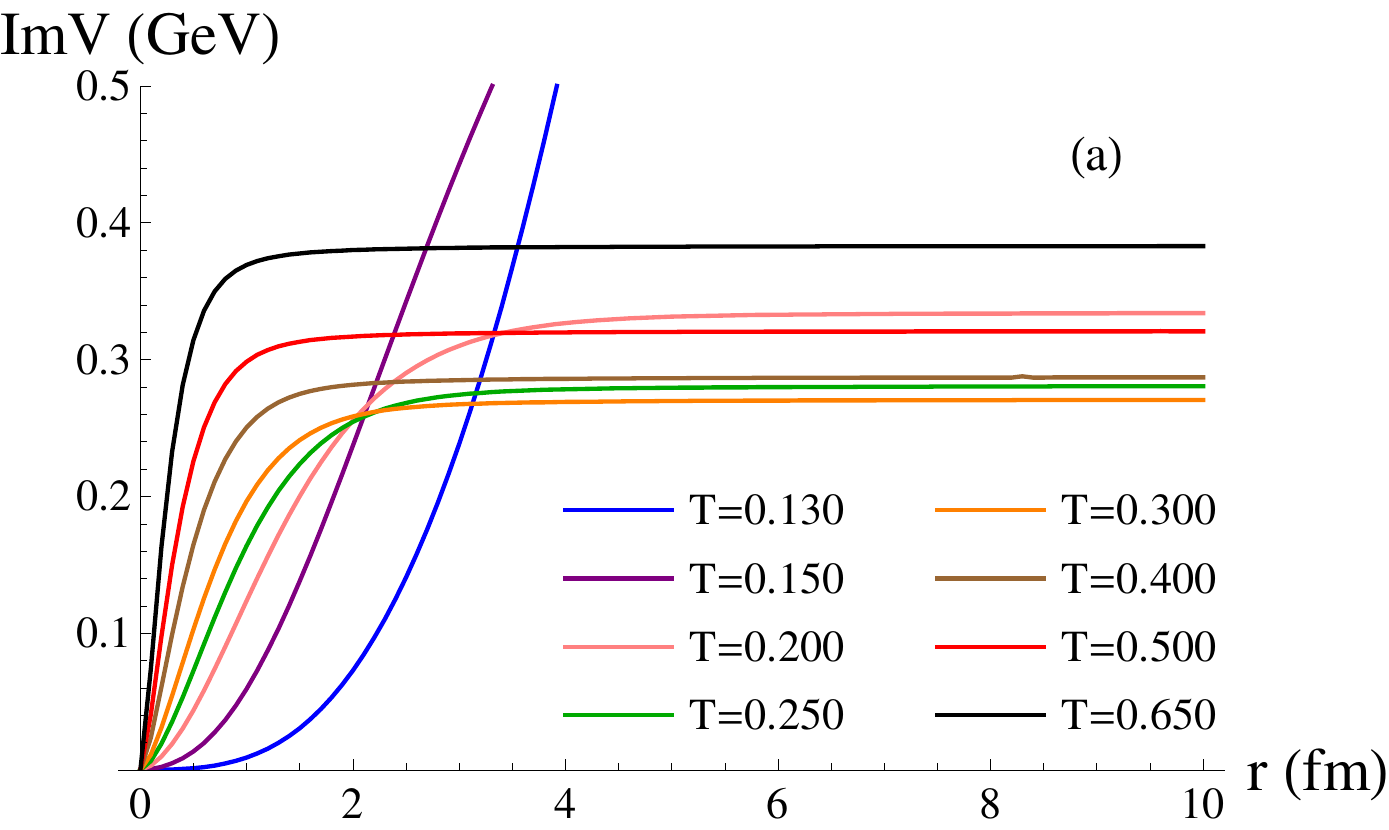}
     \includegraphics[width=0.51\textwidth]{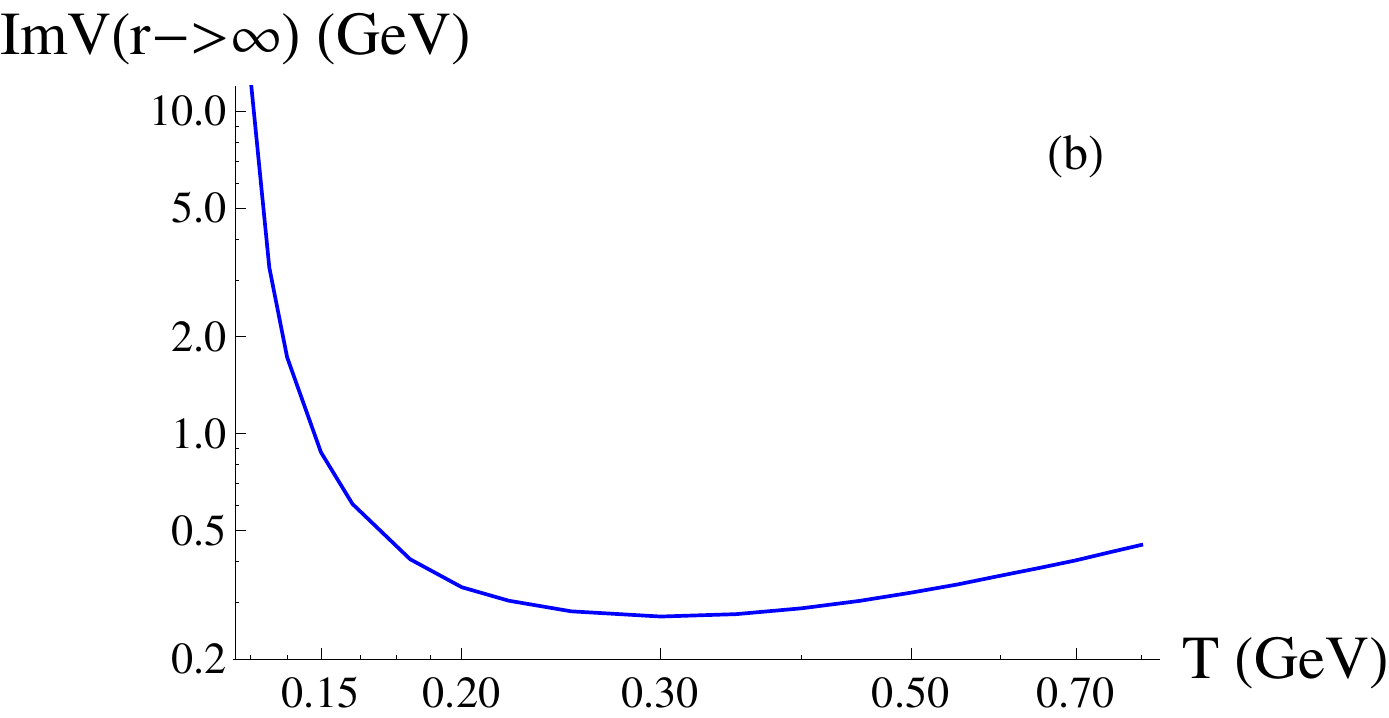}
    \caption{(a) Imaginary part of the potential for different temperatures in GeV. (b) The temperature dependence of the imaginary part at large distances.}
    \label{fig:3DVim}
\end{figure}

\begin{figure}[htb!]
    \centering
    \includegraphics[width=0.49\textwidth]{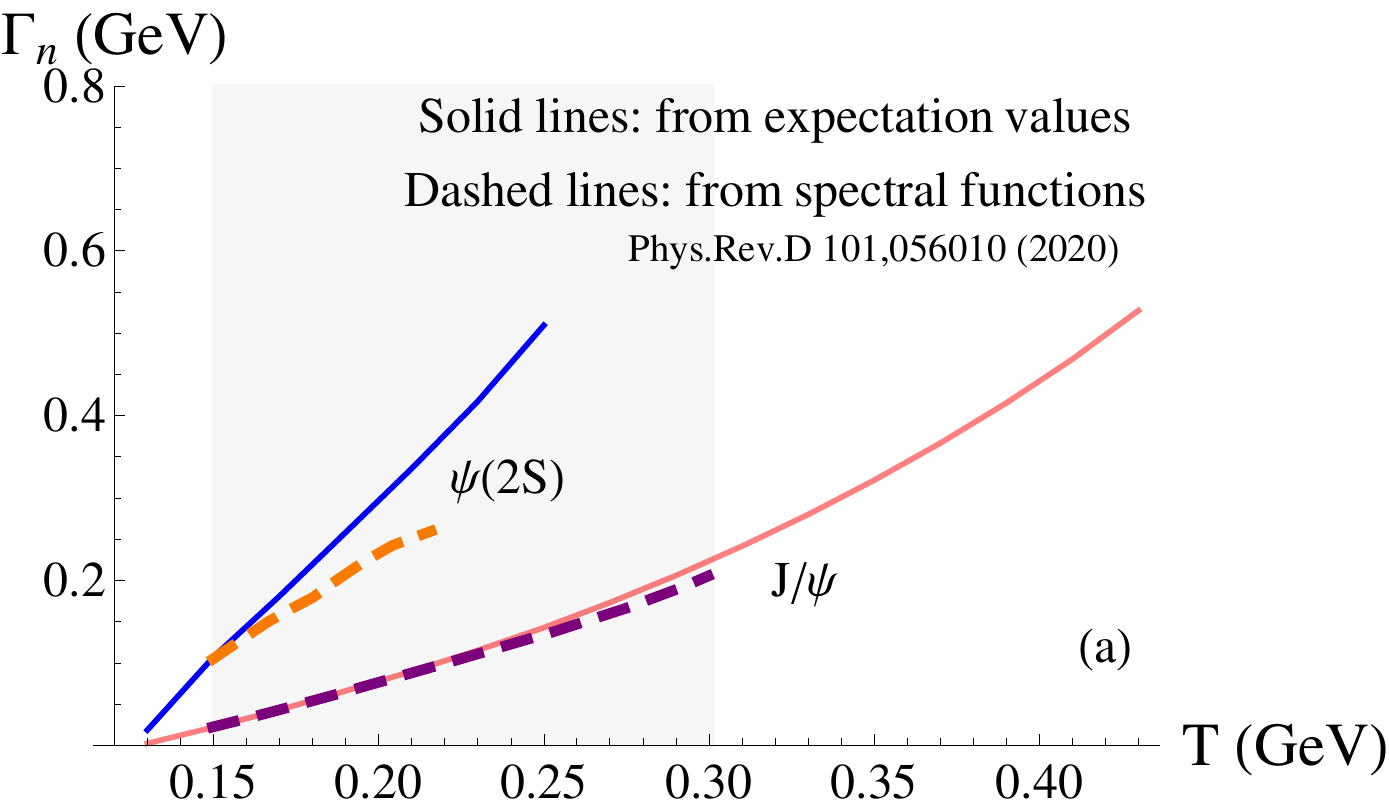}
     \includegraphics[width=0.49\textwidth]{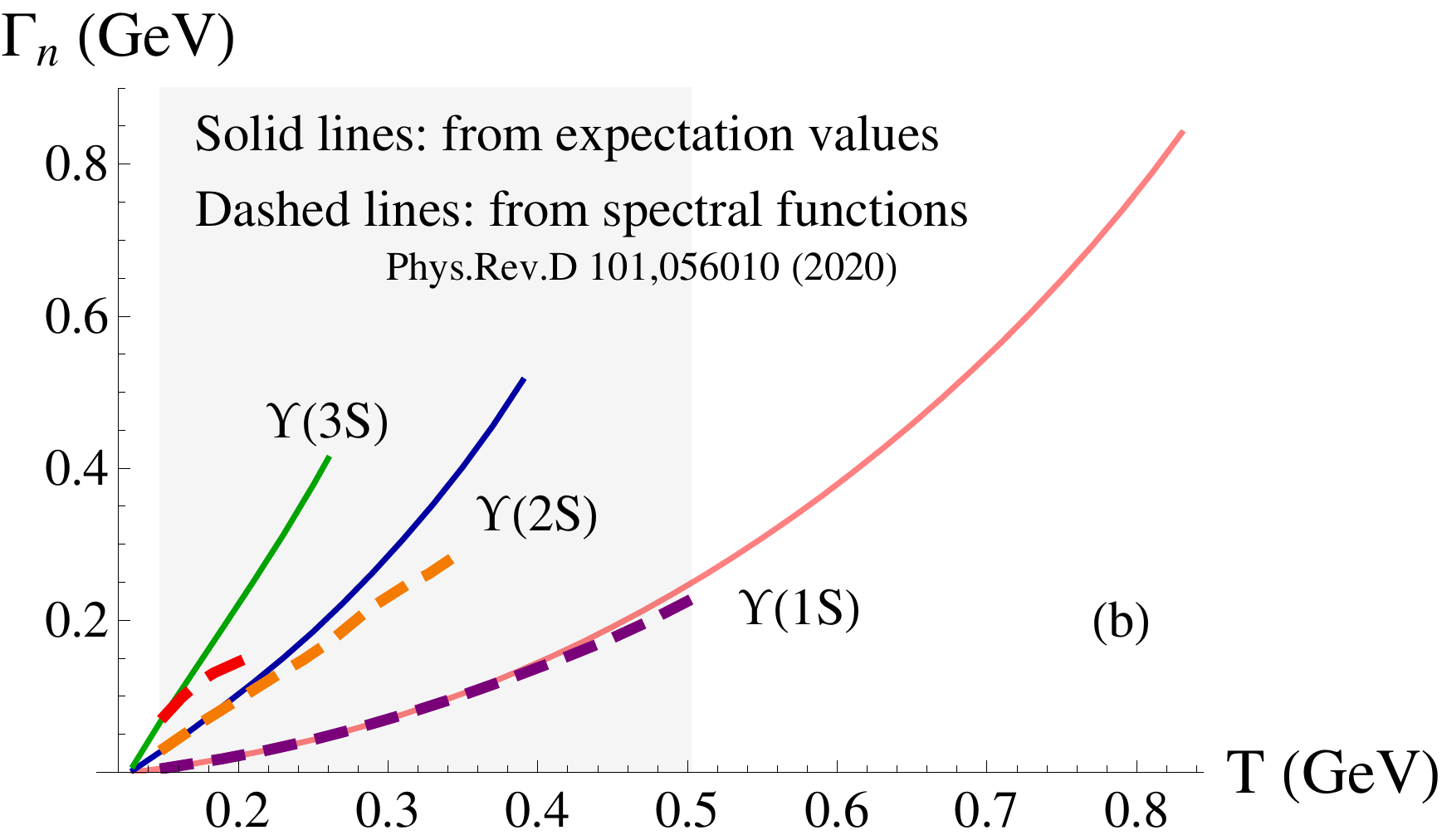}
    \caption{(a) $\Gamma_n(T)$ of the charmonia and (b) $\Gamma_n(T)$ of the bottomonia. Gray areas corresponds to the results available from spectral functions in \cite{Lafferty:2019jpr}.}
    \label{fig:3DDecayWidthStatesT}
\end{figure}

In Fig.\ \ref{fig:3DVim}, we show the imaginary part of the potential at different temperatures and its temperature dependence at large distances. As with the HTL potential (Fig.\ \ref{fig:3DVimpQCDHTL}), the imaginary part exhibits a small distance harmonic behavior with a harmonic coefficient that gets larger with the temperature. However, the large distance saturation is less trivial than it was in Fig.\ \ref{fig:3DVimpQCDHTL} with the HTL potential: the string part has a strong influence up to at least T=400 MeV. To evaluate the impact of the imaginary part on the heavy quark pair, one can compute the thermal decay widths $\Gamma_n$ of the bound states. Within the spectral function method, the $\Gamma_n$ correspond to the widths of the bound state peaks at finite temperature and are calculated in \cite{Lafferty:2019jpr} via a fit of the peaks with (skewed) Breit-Wigner distributions. The thermal decay width of the state $n$ can also be evaluated via the expectation value of the imaginary part
\begin{align}\label{eq:DecayWidths}
\Gamma_n=2 {\rm Tr}(\rho_n{\rm Im}V)=2 \langle {\rm Im}V\rangle_n,
\end{align}
where $\rho_n$ is the density matrix of the state $n$ and $\langle ... \rangle_n$ the related expectation value. As shown in Fig.\ \ref{fig:3DDecayWidthStatesT}, the thermal decay widths increase with temperature and are larger for excited states, which corresponds to the intuition that more compact states are less sensitive to medium effects. The two methods give similar results when the temperature is much smaller than the dissociation temperature of the considered state, but increasing discrepancies appear as this temperature is approached. These differences might originate once again from the difficulty of fitting large spectral peaks with skewed Breit-Wigner distributions.
%
%

\subsection{Spectral decomposition} \label{Rothkopf3DSpectralDecompo}

The spectral decomposition ${\rm Im}{\tilde V}({\vec q})={\rm Im}{\tilde V}_c+{\rm Im}{\tilde V}_s$ of the imaginary part of the potential $\mathrm{Im}V({\vec r})={\rm Im}V_c+{\rm Im}V_s$ is given by its three-dimensional Fourier transform:
\begin{equation}\label{Eq:FourierTransfo}
{\rm Im}{\tilde V}({\bf {\vec q}})=\frac{1}{(2\pi)^3}\int_{-\infty}^{\infty}d{\vec r}\,\mathrm{Im}V({\vec r})\,e^{-i{\vec q}{\vec r}}.
\end{equation}
Within the open quantum system framework, it can be shown that the spectral decomposition of the imaginary part of the potential must be globally positive (or negative depending on the chosen convention) to ensure the positivity of the Lindblad equation \cite{Akamatsu:2012vt,Akamatsu:2014qsa,Blaizot:2015hya,DeBoni:2017ocl,Blaizot:2017ypk,Blaizot:2018oev}. In this section, we will check whether the spectral decomposition of the QCD inspired potential \ref{eq:3DPottot} satisfies this condition. 

For the Coulombic part, the angular integration of the Fourier transform yield
\begin{eqnarray}
{\rm Im}{\tilde V}_c =\frac{\alpha_s T}{2\pi q} \int_0^\infty dr\, r \sin(q r)\big(\phi(m_D r)-\phi_{\rm asympt}\big),
\end{eqnarray}
where $\phi_{\rm asympt}=1$ is subtracted by hand to remove the asymptotic value of the complex potential that would lead to a Dirac delta $\delta(q)$. Similarly, for the string part we get
\begin{eqnarray}
{\rm Im}{\tilde V}_s =\frac{\sigma T}{2\pi^2 q\, m_D^2} \int_0^\infty dr\, r \sin(q r)\big(\chi(m_D r)-\chi_{\rm asympt}\big),
\end{eqnarray}
where $\chi_{\rm asympt}$ is the asymptotic value of the function $\chi$ and is given by
\begin{eqnarray}
\chi_{\rm asympt}=4  \int_0^\infty dk \frac{1}{(k^2 + 1)^2\sqrt{k^2 + \Delta_D^2}}.
\end{eqnarray}

\begin{figure}[htb!]
    \centering
    \includegraphics[width=0.49\textwidth]{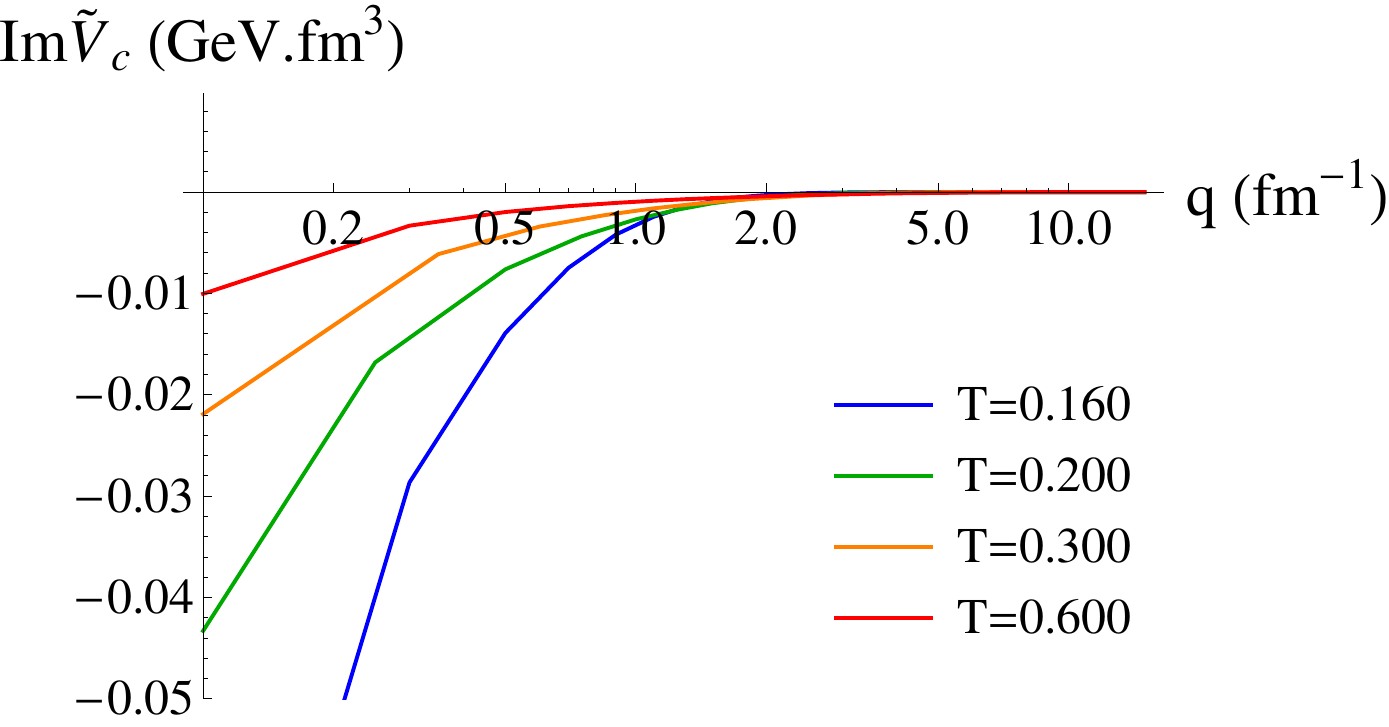}
    \caption{Spectral decomposition of the Coulombic-like imaginary part of the potential ${\rm Im}V_c$} at different temperatures (in GeV).
    \label{fig:3DspectraldecompoVc}
\end{figure}
\begin{figure}[htb!]
    \centering
    \includegraphics[width=0.48\textwidth]{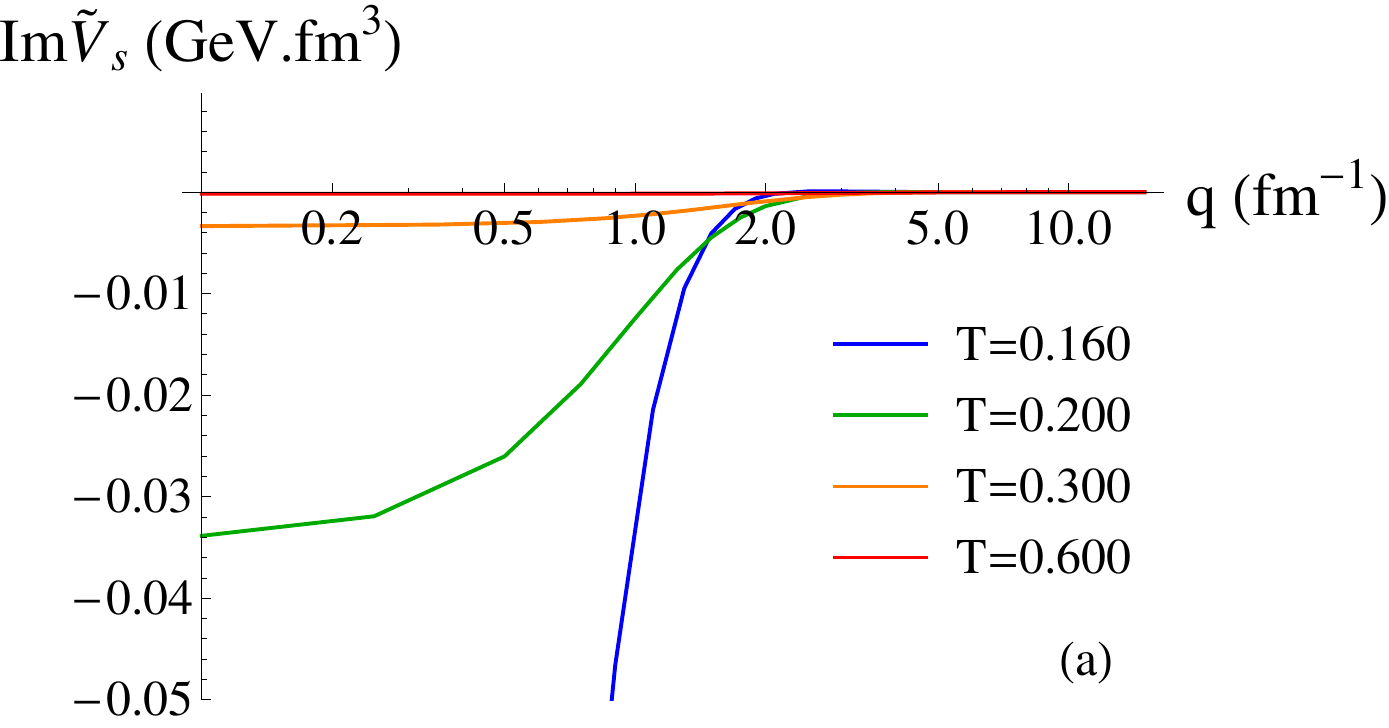}
    \hspace{4mm}
     \includegraphics[width=0.48\textwidth]{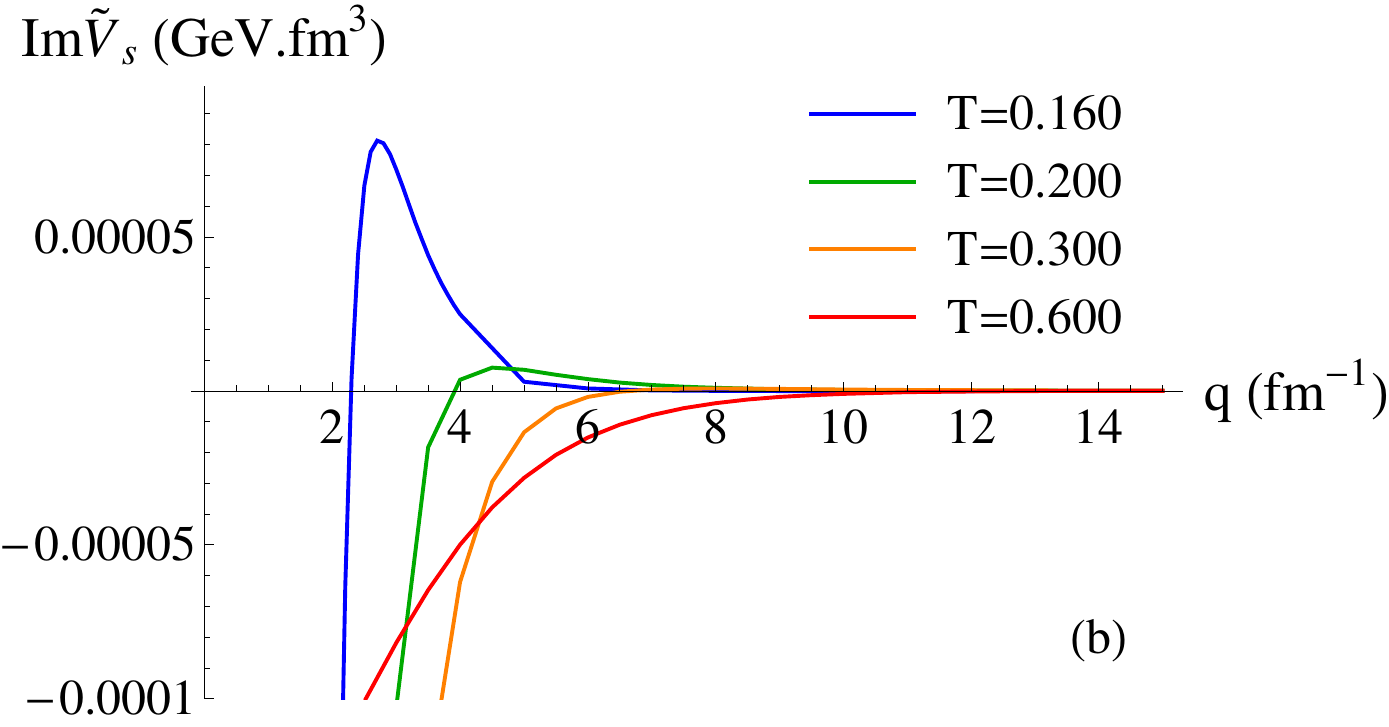}
    \caption{Spectral decomposition of the String-like imaginary part of the potential ${\rm Im}V_s$} at different temperatures (in GeV). (a) General behaviour. (b) Very small positive bumps at low temperatures.
    \label{fig:3DspectraldecompoVs}
\end{figure}

In Fig.\ \ref{fig:3DspectraldecompoVc}, the spectral decomposition of the Coulombic-like part (equivalent to the HTL potential of Sec.\ \ref{pQCDHTL3D}) is observed to be negative for all $q$, while in Fig.\ \ref{fig:3DspectraldecompoVs} for the string-like part the spectral decomposition is negative at small $q$ but exhibits a very small positive bump at larger q and smaller temperatures. 

\begin{figure}[htb!]
    \centering
    \includegraphics[width=0.48\textwidth]{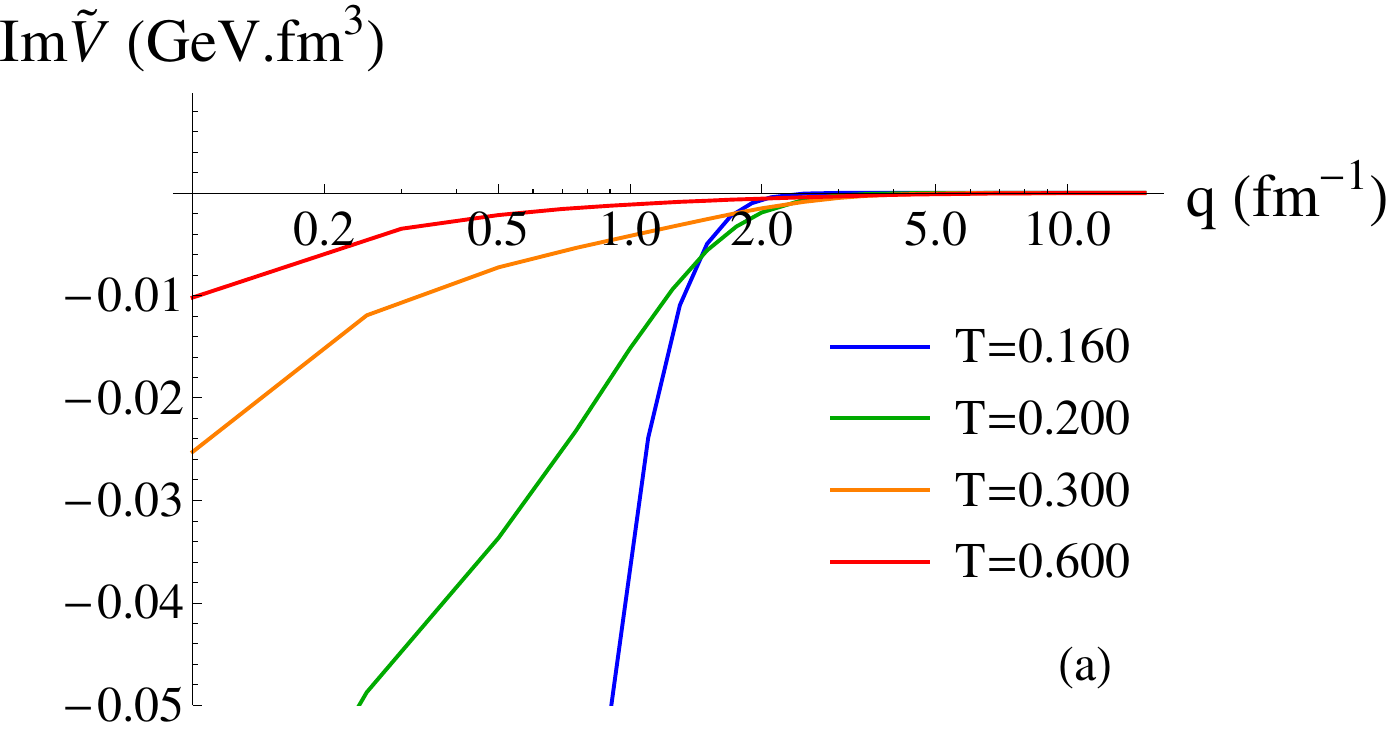}
    \hspace{4mm}
     \includegraphics[width=0.48\textwidth]{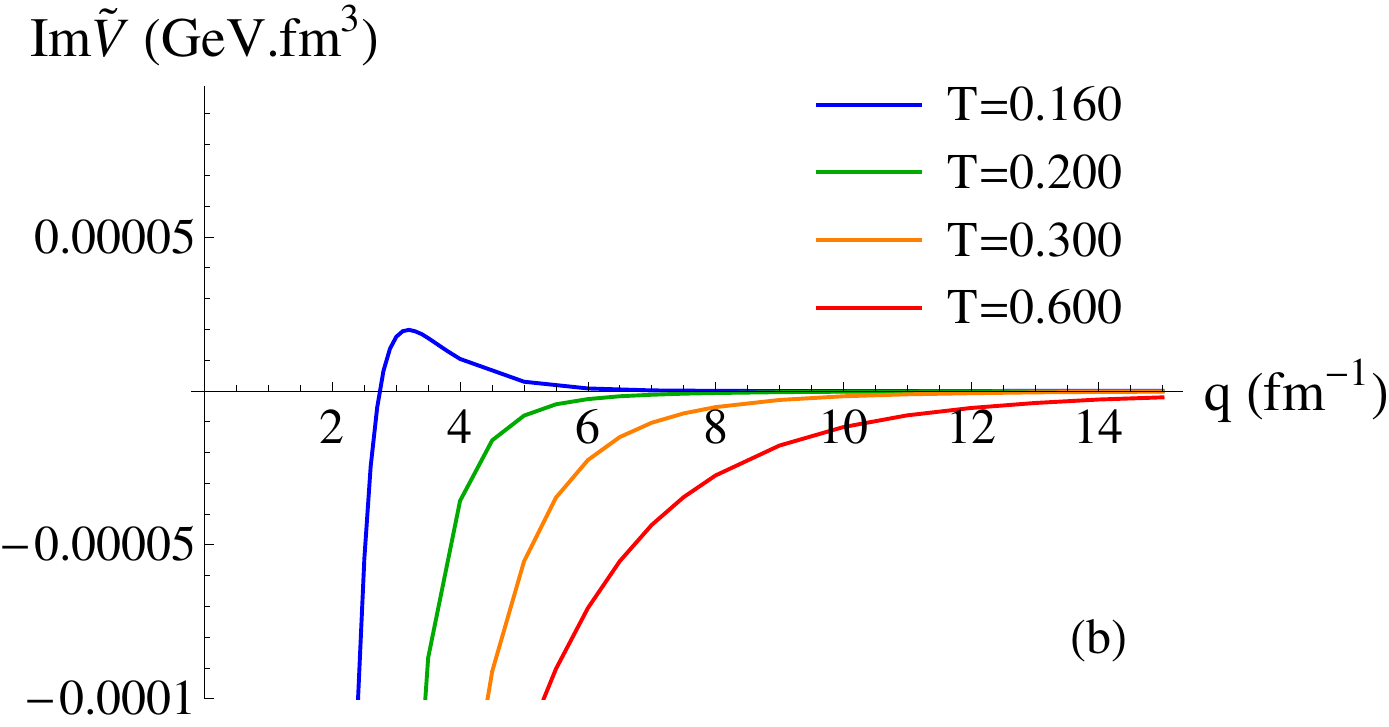}
    \caption{Spectral decomposition of the imaginary part of the potential ${\rm Im}V={\rm Im}V_c+{\rm Im}V_s$} at different temperatures (in GeV). (a) General behaviour. (b) Very small positive bumps at low temperatures.
    \label{fig:3DspectraldecompoVtot}
\end{figure}

Finally, as shown in Fig.\ \ref{fig:3DspectraldecompoVtot}, when summing the Coulomb and string like parts, the small positive bump is only observed for temperatures $T\lesssim$ 160 MeV. We can therefore conclude that the spectral decomposition is globally negative (with the chosen convention) and that this three-dimensional potential will globally satisfies the positivity requirement of the Lindblad equations.  

\newpage

\section{A one-dimensional potential: the real part}\label{3Dto1Dreal}

In this section, we propose a one-dimensional potential parameterized to reproduce at best the properties of the three-dimensional potential from the generalized Gauss law model described in Sec.\ \ref{Rothkopf3D}. We first focus on the real part of the potential and the corresponding energy spectra. To study the one-dimensional potential, we use the Hamiltonian :
\begin{eqnarray}\label{Hamiltonian1D}
H=2m_Q-\frac{1}{m_Q}\frac{\partial^2}{\partial x^2}+{\rm Re}V(x,T).
\end{eqnarray}


\subsection{Vacuum potential} \label{Sec:VacuumPotential1D}

In the ``vacuum'', a simple form is proposed and tuned to reproduce at best the experimental masses. Inspired by the Cornell potential and by the one-dimensional version of the pQCD+HTL potential in Sec.\ \ref{pQCDHTL} -- which are mainly linear around the considered eigenenergies --, we approximate the heavy quark/anti-quark self interaction to a 1D symmetrical linear potential $1/2\,K|x|+C$ saturated at $V_{\rm SB}\approx 0.8352$ GeV for the string breaking, i.e.
\begin{eqnarray}\label{V1Dvac}
\mathrm{Re}V_{\rm 1D}(x,T=T_0)=\left\{ \begin{array}{lcl}
1/2\,K|x|+C & \mbox{   when   } & 1/2\,K|x|+C < V_{\rm SB}  \\
V_{\rm SB}  &  \mbox{   when   } & 1/2\,K|x|+C \geq V_{\rm SB}.  \,
\end{array}\right.  
\end{eqnarray}
We keep the same quark masses as in Sec.\ \ref{Rothkopf3D}, i.e.\ $m_c=1.4692$ GeV and $m_b=4.882$ GeV. The string parameters $K_c$ for the charmonia and $K_b$ for the bottomonia are chosen such as to obtain an energy difference between the first two even eigenstates (subscripted by $n=0$ and $n=2$) given by $E_2-E_0=E(\psi')-E(J/\psi)=589$ MeV for charmonia and an intermediate situation between $E_2-E_0\approx E(\Upsilon')-E(\Upsilon)=563$ MeV and $E_4-E_0\approx E(\Upsilon'')-E(\Upsilon)=895$ MeV for bottomonia (with subscript $n=4$ for the third even eigenstate); it leads respectively to $K_c=1.724$ GeV.fm$^{-1}$ and $K_b=2.692$ GeV.fm$^{-1}$. The overall constants $C_c$ for the charmonia and $C_b$ for the bottomonia are then fixed to get $E_0+2m_c=m_{J/\psi}$ for the charmonia and $E_0+2m_b=m_\Upsilon$ for the bottomonia, i.e.\ $C_c=-0.115$ GeV and $C_b=-0.55$ GeV respectively. 

\begin{figure}[htb!]
    \centering
    \includegraphics[width=0.45\textwidth]{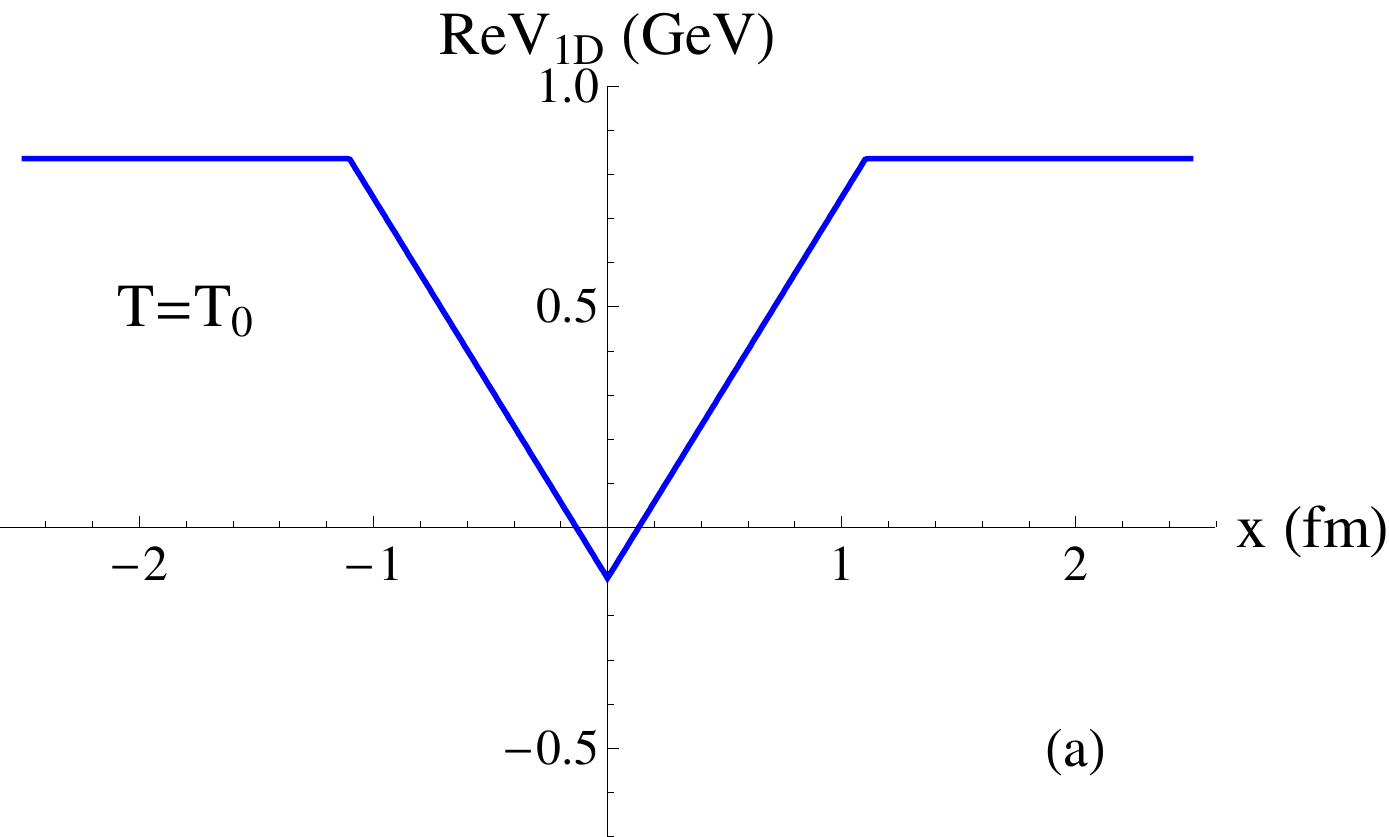}
     \includegraphics[width=0.45\textwidth]{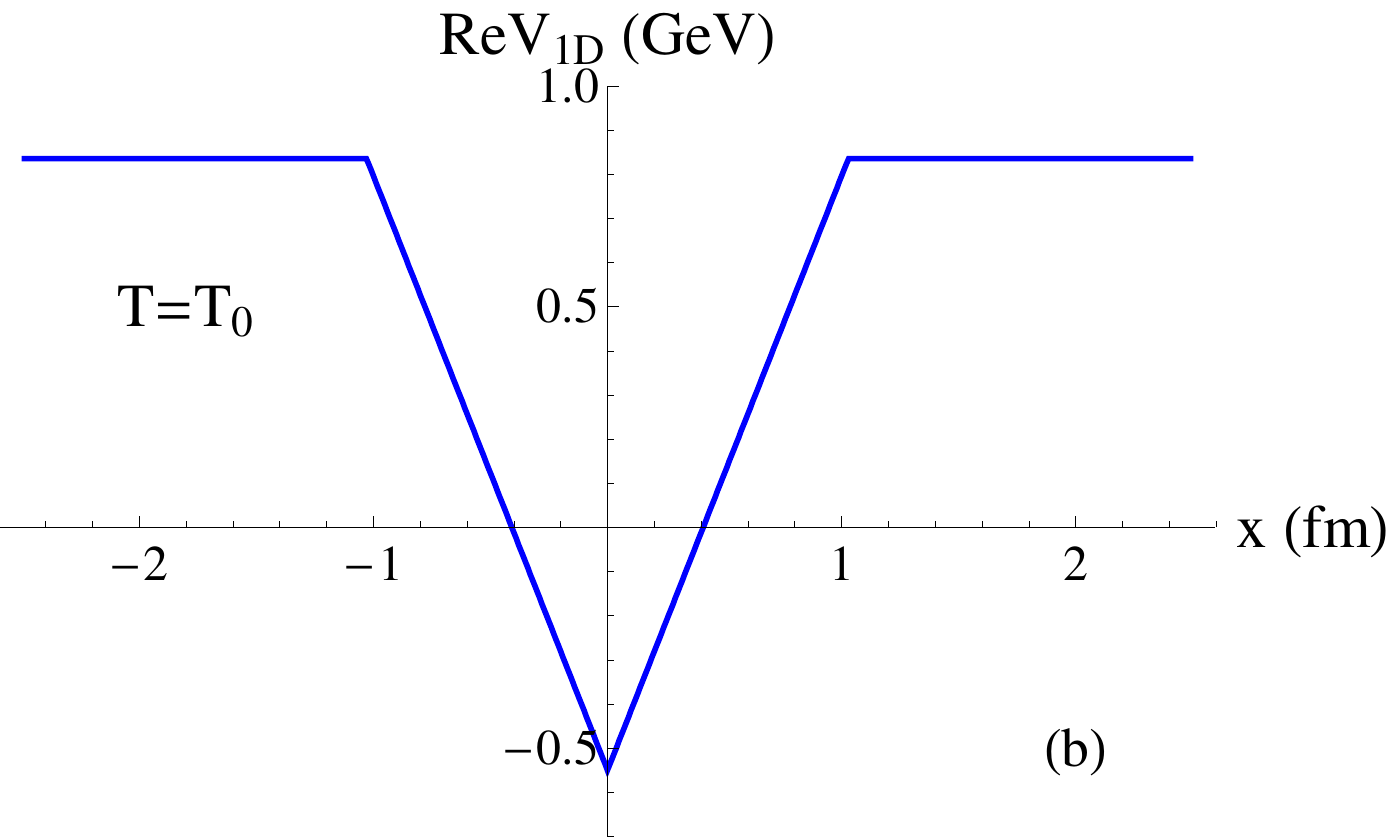}
    \caption{Vacuum one-dimensional linear potential for (a) charmonia and (b) bottomonia.}
    \label{fig:1DVacRePot}
\end{figure}

\begin{figure}[htb!]
    \centering
    \includegraphics[width=0.46\textwidth]{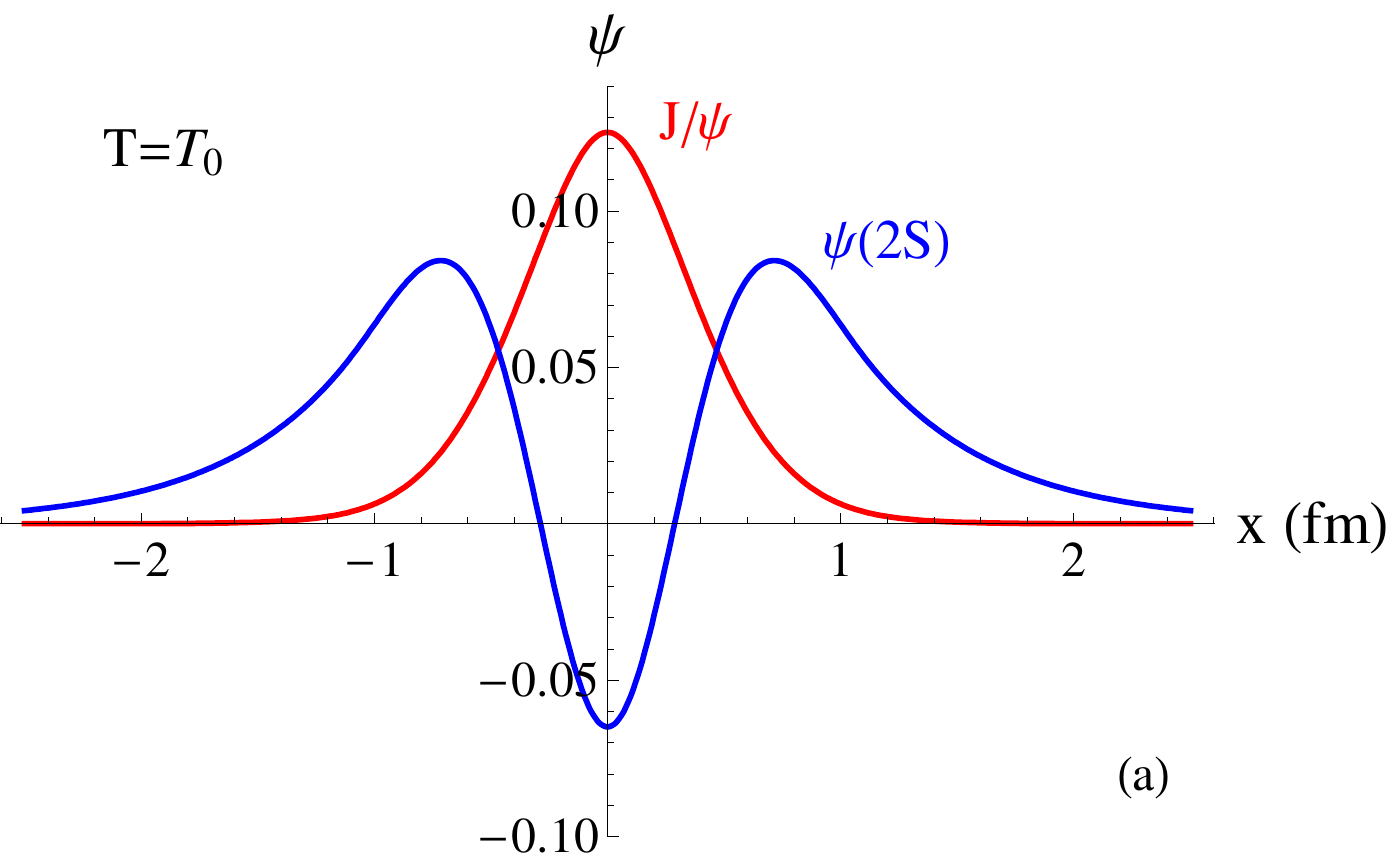}
    \hspace{3mm}
     \includegraphics[width=0.46\textwidth]{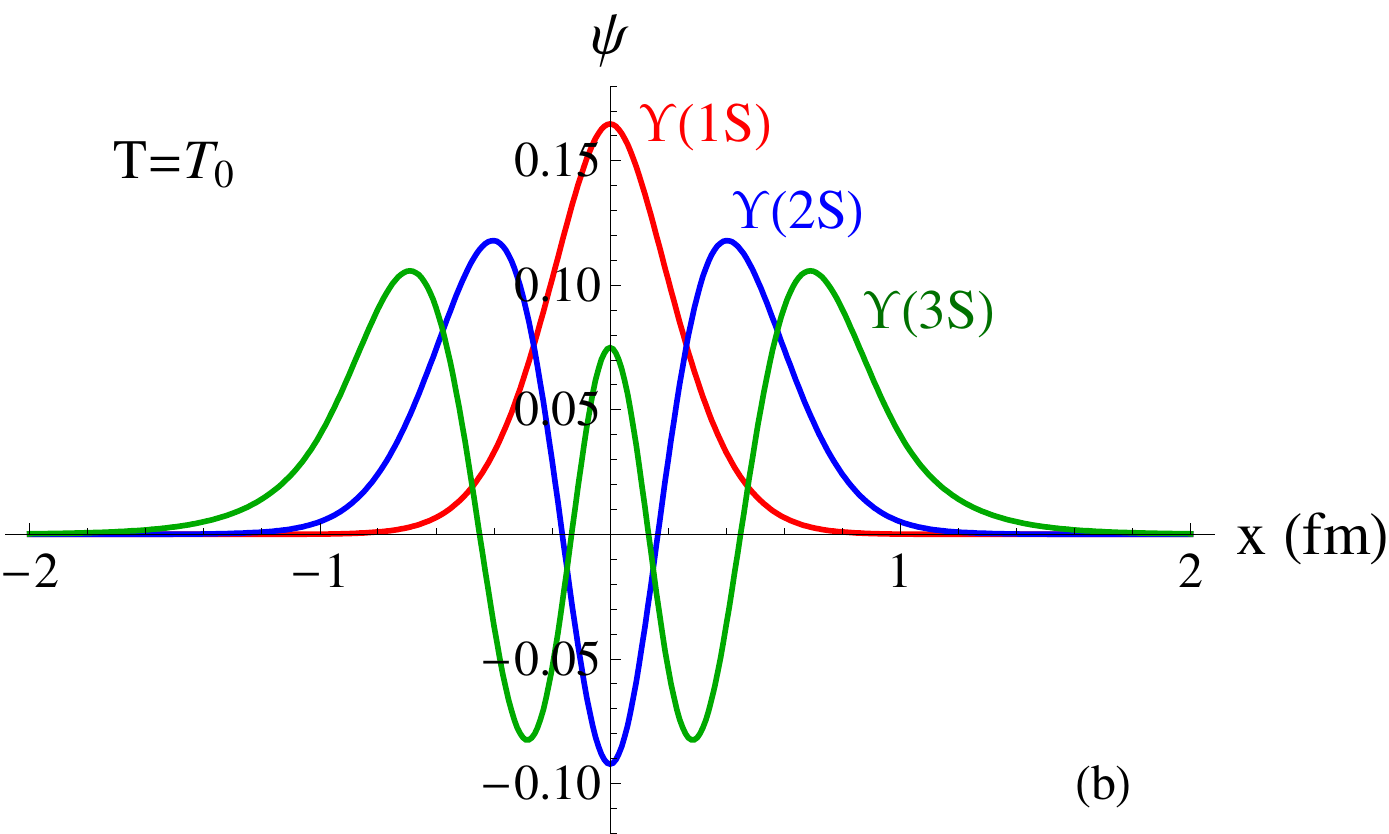}
    \caption{(a) Charmonium S-like states obtained with the real part of the one-dimensional potential in the ``vacuum'' (at $T_0 \approx 0.126025$ GeV). (b) Same but for bottomonia.}
    \label{fig:1Deigenfunctions}
\end{figure}

The corresponding vacuum potentials are shown in Fig.\ \ref{fig:1DVacRePot} and feature three charmonium and six bottomonium eigenstates. The masses (and root mean square radiuses) of the even ``S-like'' and odd ``P-like'' states are summarized in Tab.\ \ref{tab:Eandrvacuum1D} and \ref{tab:Eandrvacuum1DPstates} and compared to the three-dimensional and experimental values. With errors of maximum $\approx 30$ MeV for the S-like states and $\approx 100$ MeV for the P-like states, the one dimension linear model reproduces in good approximation the mass spectra. In Fig.\ \ref{fig:1Deigenfunctions}, the charmonium and bottomonium S-like states obtained with the one-dimensional potential are shown, and can be compared to the states obtained with the three-dimensional potential (Fig.\ \ref{fig:eigenfunctions3D}). As expected, the shapes of the one- and three-dimensional 1S-states are different : the radial parts of the three-dimensional wavefunctions are typically exponential for Coulombic potentials, whereas they are approximately Gaussian for linear potentials. Despite this difference in nature, some similar features can be observed : the typical sizes of the states (as confirmed by the root mean square radiuses in Tab. \ref{tab:Eandrvacuum1D}), the large distance behaviour, the wave nodes and antinodes positions and relative amplitudes... 

\begin{table}[h!]
\begin{center}
    \begin{tabular}{|C{3.5cm}||C{1.5cm}|C{1.5cm}||C{1.5cm}|C{1.5cm}|C{1.5cm}|}
    \hline
At $T=T_0$ & $J/\psi$ & $\psi$(2S) & $\Upsilon$(1S) & $\Upsilon$(2S) & $\Upsilon$(3S) \\
    \hline
    \hline
$m^{\rm PDG}$ \cite{Zyla:2020zbs} & 3.0969 & 3.6861 & 9.4603 & 10.023  & 10.355  \\
    \hline
$m^{\rm EV}_{\rm 3D}$ & 3.0964 & 3.6642 & 9.4611 & 10.023  & 10.355  \\
    \hline 
$m^{\rm EV}_{\rm 1D}$ & 3.0981 & 3.6858 & 9.4621 & 10.005  & 10.386  \\ 
    \hline 
$|m^{\rm PDG}-m^{\rm EV}_{\rm 1D}|$ (MeV) & 1.2 & 0.3 & 2.1 & 18  & 31  \\ 
    \hline
    \hline
${\rm 3D} \,\,\, \sqrt{\langle r^2 \rangle}$ & 0.431 & 0.943 & 0.213 & 0.501 & 0.7449  \\ 
    \hline
$ {\rm 1D} \,\,\, \sqrt{\langle x^2 \rangle}$ & 0.387 & 0.856 & 0.157 & 0.432  & 0.648  \\ 
    \hline
    \end{tabular}
\caption {\label{tab:Eandrvacuum1D} 
\small Masses (in GeV) and root mean square radiuses (in fm) for the ``vacuum'' S-like states (at $T_0 \approx 0.126025$ GeV). $m^{\rm PDG}$ is the experimental mass given by the particle data group, $m^{\rm EV}_{\rm 3D}$ ($\sqrt{\langle r^2 \rangle}$) is the mass (root mean square radius respectively) calculated from the expectation values of the Hamiltonian (of the square radius operator respectively) with the three-dimensional potential (see Sec.\ \ref{Rothkopf3D}). $m^{\rm EV}_{\rm 1D}$ and $\sqrt{\langle x^2 \rangle}$ are the corresponding values obtained with the one-dimensional potential.}
\end{center}
\end {table}

\begin{table}[h!]
\begin{center}
    \begin{tabular}{|C{3.5cm}||C{1.5cm}||C{1.5cm}|C{1.5cm}|C{1.5cm}|}
    \hline
At $T=T_0$ & $\chi_c$(1P) & $\chi_b$(1P) & $\chi_b$(2P) & $\chi_b$(3P) \\
    \hline
    \hline
$m^{\rm PDG}$ \cite{Zyla:2020zbs} & 3.494 & 9.888 & 10.252 & 10.534    \\
    \hline
$m^{\rm EV}_{\rm 3D}$ & 3.509 & 9.932 & 10.273 & 10.540   \\ 
    \hline
$m^{\rm EV}_{\rm 1D}$ & 3.453 & 9.783 & 10.209 & 10.544  \\ 
    \hline
$|m^{\rm PDG}-m^{\rm EV}_{\rm 1D}|$ (MeV) & 41 & 105 & 43 & 10    \\
    \hline
    \end{tabular}
\caption {\label{tab:Eandrvacuum1DPstates} 
\small Masses (in GeV) for the ``vacuum'' P-like states (at $T_0 \approx 0.126025$ GeV). $m^{\rm PDG}$ is the experimental mass given by the particle data group \cite{Zyla:2020zbs}. $m^{\rm EV}_{\rm 3D}$ is the mass calculated from the expectation values of the Hamiltonian with the three-dimensional potential and $m^{\rm EV}_{\rm 1D}$ is the mass obtained with the one-dimensional potential.}
\end{center}
\end {table}

\newpage

\subsection{Temperature dependence} 

When $m_D > 0$ (for $T>T_0 \approx 0.126025$ GeV), the one-dimensional potential is parameterized such as to reproduce the temperature dependence of the mass spectrum obtained from the three-dimensional potential, i.e.
\begin{eqnarray}\label{Eq:M1DM3Dcondition}
M^{\rm 1D}(T) = \langle {\hat H}^{\rm 1D}(T) \rangle \approx \langle {\hat H}^{\rm 3D}(T) \rangle = M^{\rm 3D}(T),
\end{eqnarray}
where $M^{\rm 1D}$ and ${\hat H}^{\rm 1D}$ (respectively $M^{\rm 3D}$ and ${\hat H}^{\rm 3D}$) are the mass spectrum and Hamiltonian operator associated with the real part of the one-dimensional (respectively three-dimensional) potential. Additionally, the effect of the temperature dependent Debye screening is implemented by saturating the one-dimensional potential $\mathrm{Re}V_{\rm 1D}$ to the value of the three-dimensional potential at large distances,
\begin{equation}\label{Eq:saturationReV}
\mathrm{Re}V^{x\rightarrow\infty}_{\rm 1D}(T)\equiv\mathrm{Re}V(r\rightarrow\infty,T)={\rm Minimum}\,[\,2\sigma/m_D(T)-\alpha_sm_D(T)+c, \,V_{\rm SB}].
\end{equation}
This combination of similar mass spectra and equal large distance values of the potential implies that the binding energies of the states -- which are key features of the real part of the potential in the open quantum system framework -- are equivalent in both cases. In other words, the energy gaps between bound and free states are then similar in the one- and three-dimensional cases. 

Inspired by the exponential ``$e^{-m_D r}$'' terms present in $\mathrm{Re}V_C$ and $\mathrm{Re}V_{S}$, we propose the following parameterization for the one-dimensional potential\footnote{Many other parametrizations were tested, using for instance the the expression of the string part of the potential \ref{eq:String3D}. Nevertheless, none of those was able to correctly reproduce the three-dimensional mass spectra (condition \ref{Eq:M1DM3Dcondition}).}: 
\begin{eqnarray}\label{V1D}
\mathrm{Re}V_{\rm 1D}(x,T) = \left\{ \begin{array}{lcl}
1/2\,K|x|\,e^{-\mu_1|x|} + \mu_2 & \mbox{   when   } & 1/2\,K|x|\,e^{-\mu_1|x|} + \mu_2 < \mathrm{Re}V^{x\rightarrow\infty}_{\rm 1D}(T)  \\
\mathrm{Re}V^{x\rightarrow\infty}_{\rm 1D}(T)  &  \mbox{   when   } & 1/2\,K|x|\,e^{-\mu_1|x|} + \mu_2  \geq \mathrm{Re}V^{x\rightarrow\infty}_{\rm 1D}(T),  \,
\end{array}\right.  
\end{eqnarray}
where
\begin{equation}\label{V1Dmu}
\mu_1(x,T) = a_1\,(T - T_0)^{b_1} \, e^{-\lambda (T - T_0)\,|x|} \,\,\,\,\,\,{\rm and}\,\,\,\,\,\, \mu_2(T) = -a_2\,(T-T_0)^{b_2} + C.
\end{equation}
The $\mu_2$ term is a global temperature-dependent shift of the potential leading to a global decrease of the mass spectrum values as the temperature increases. The $e^{-\mu_1|x|}$ term symmetrically curves the shape of the potential at intermediate distances, resulting in different (temperature-dependent) variations of the mass spectrum values around the global decrease given by $\mu_2$. The $e^{-\lambda (T - T_0)\,|x|}$ term present in $\mu_1$ is introduced to smoothly cancel $\mu_1$ at large distances. Without this term, the potential would not reach or would not remain at the saturation value at large distances given by Eq.\ \ref{Eq:saturationReV}, but would rather tend to the value $\mu_2$. The values of the parameters are summed up in Tab.\ \ref{tab:1DRePotParameters} and were tuned to reproduce at best the mass spectra of the three-dimensional potential shown in Fig.\ \ref{fig:3DEstatesT}. We keep the values of the K and C constants determined in the vacuum case. As $\mu_1=0$ and $\mu_2=0$ when $T = T_0$, the potential $\mathrm{Re}V_{\rm 1D}(x,T)$ reduces to the vacuum linear potential described in the previous Sec.\ \ref{Sec:VacuumPotential1D}.

\vspace{0.2cm}
\begin{table}[h!]
\begin{center}
    \begin{tabular}{|C{3.5cm}||C{1.5cm}|C{1.5cm}|C{1.5cm}|C{1.5cm}|C{1.5cm}|C{1.5cm}|C{1.5cm}|}
    \hline
$\mathrm{Re}V_{\rm 1D}$ parameters & $K$ & $C$ & $a_1$ & $b_1$ & $\lambda$ &  $a_2$ & $b_2$ \\
    \hline
    \hline
Charmonia & 1.724 & -0.115 & 7.5 & 0.98 & \multirow{2}{*}{1.5} & 0.96 & 0.97 \\
    \cline{1-5} \cline{7-8}
Bottomonia & 2.692 & -0.55 & 7.7 & 1.15 &  & 0.58 & 1.15 \\ 
    \hline
    \end{tabular}
\caption {\label{tab:1DRePotParameters} 
\small Parameters for the real part of the 1D potential $\mathrm{Re}V_{\rm 1D}$. $K$ is in GeV.fm$^{-1}$, $C$ in GeV, $a_1$, $b_1$ and $\lambda$ dimensions are such as $\mu_1$ is in fm$^{-1}$, $a_2$ and $b_2$ dimensions are such as $\mu_2$ is in GeV.}
\end{center}
\vspace{-0.4cm}
\end {table}
\begin{figure}[htb!]
    \centering
    \includegraphics[width=0.49\textwidth]{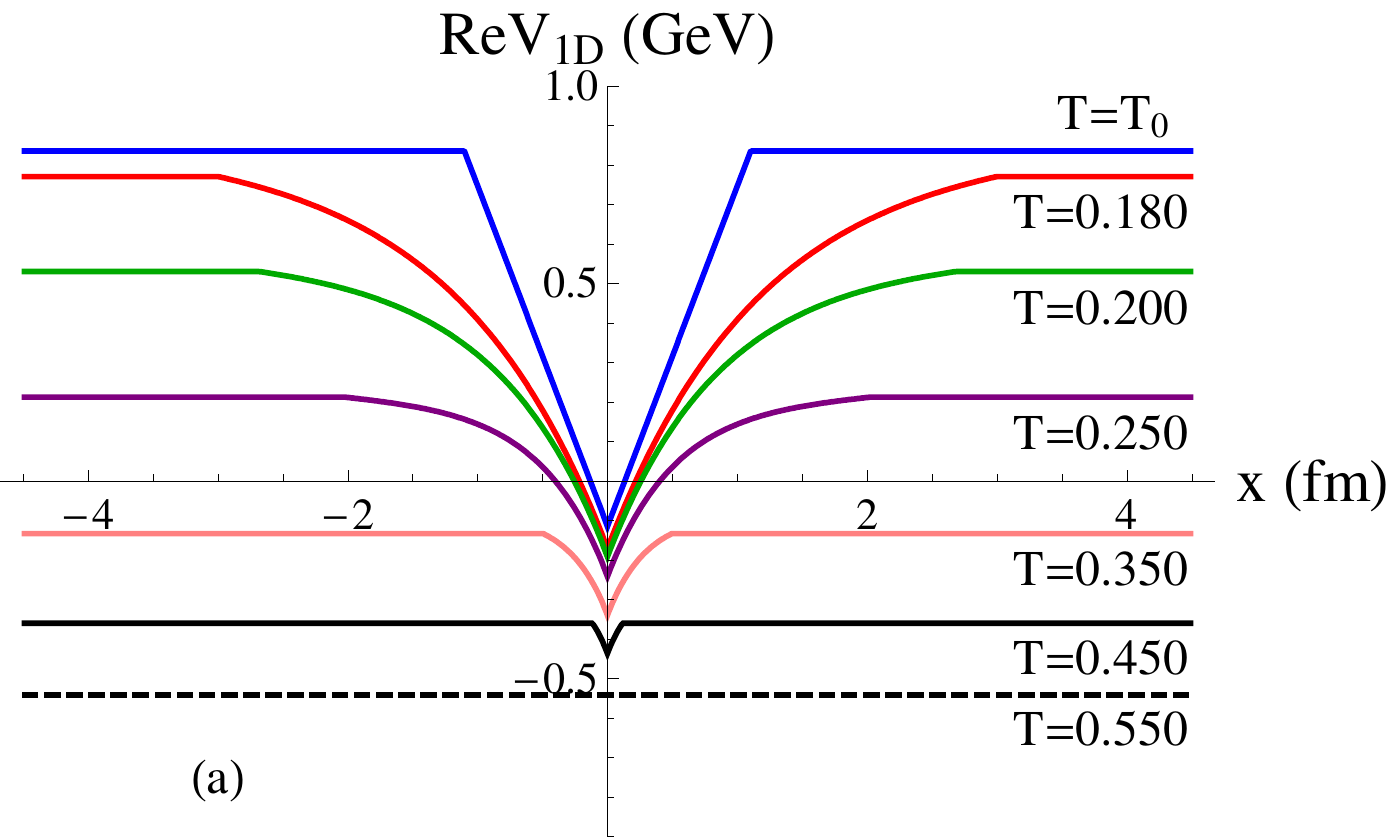}
     \includegraphics[width=0.49\textwidth]{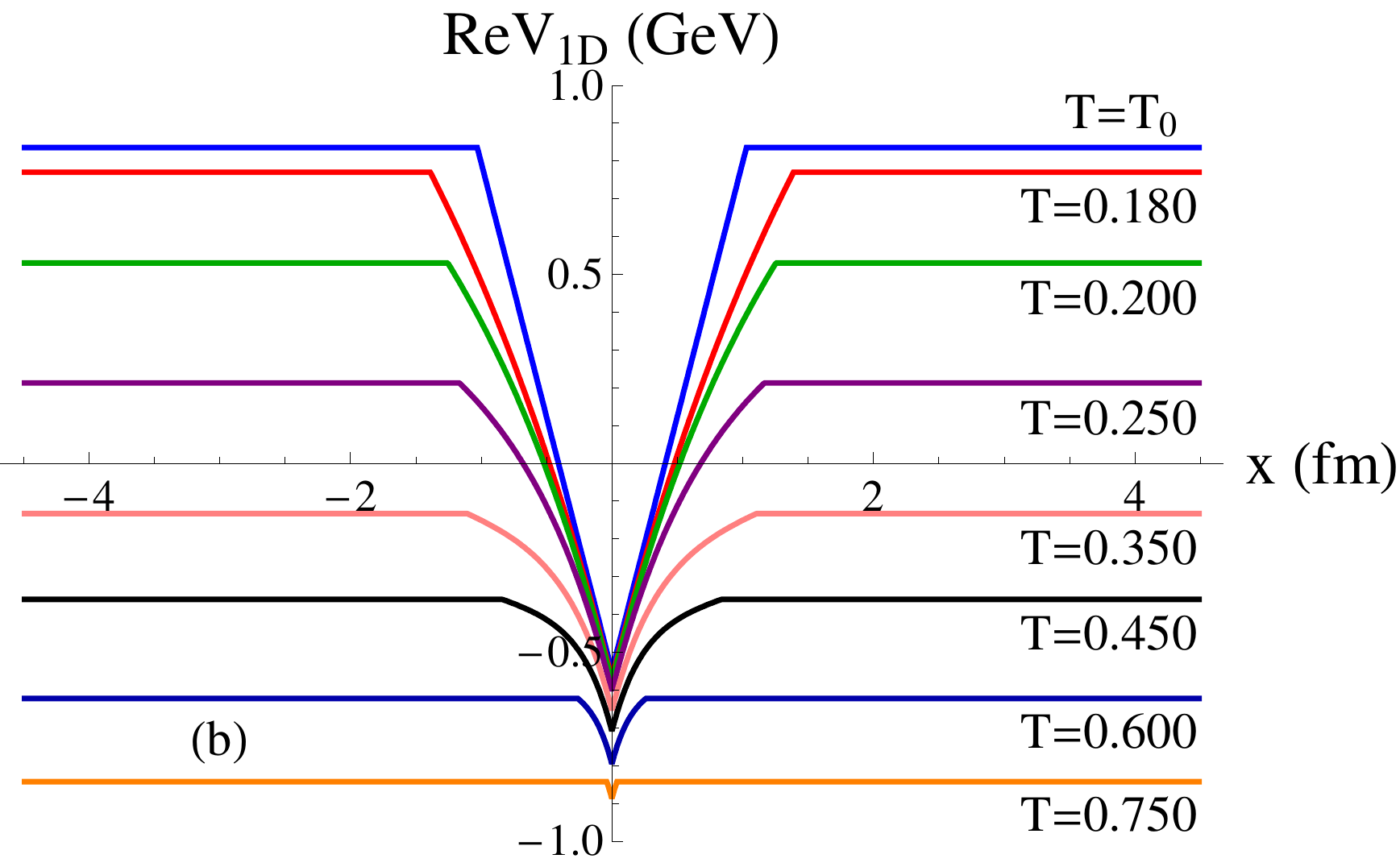}
        \caption{(a) One-dimensional potential at different temperatures (in GeV) for the charmonia and (b) for the bottomonia.}
    \label{fig:RePot1D}
\end{figure}

The resulting potentials $\mathrm{Re}V_{\rm 1D}$ for the charmonia and bottomonia are shown in Fig.\ \ref{fig:RePot1D} at different temperatures. Although the three-dimensional quarkonium potentials are independent of the heavy quark mass, it is not the case within this one-dimensional model which rather prioritizes the phenomenological features. Due to a larger $K$ and a smaller $C$, the potential well for the bottomonia is narrower and deeper than for the charmonia. Larger values of $\mu_1(x,T)$ in the bottomonium case also lead to a less curved and narrower potential well close to the asymptotes. The potentials become completely non-bonding (i.e.~constant) at $T\approx 530$ MeV for the charmonia and $T\approx 800$ MeV for the bottomonia.

\begin{figure}[htb!]
    \centering
    \includegraphics[width=0.49\textwidth]{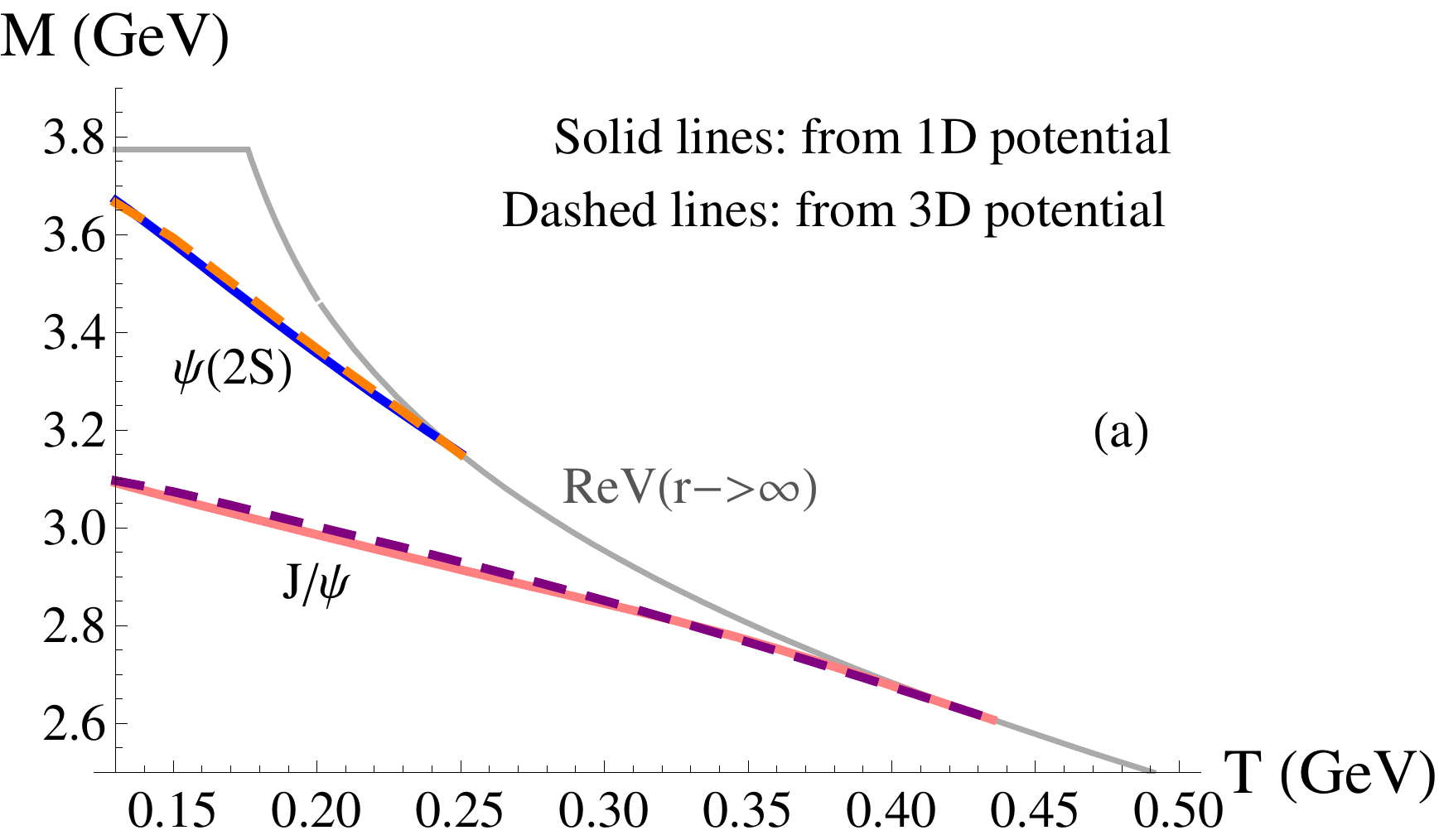}
     \includegraphics[width=0.49\textwidth]{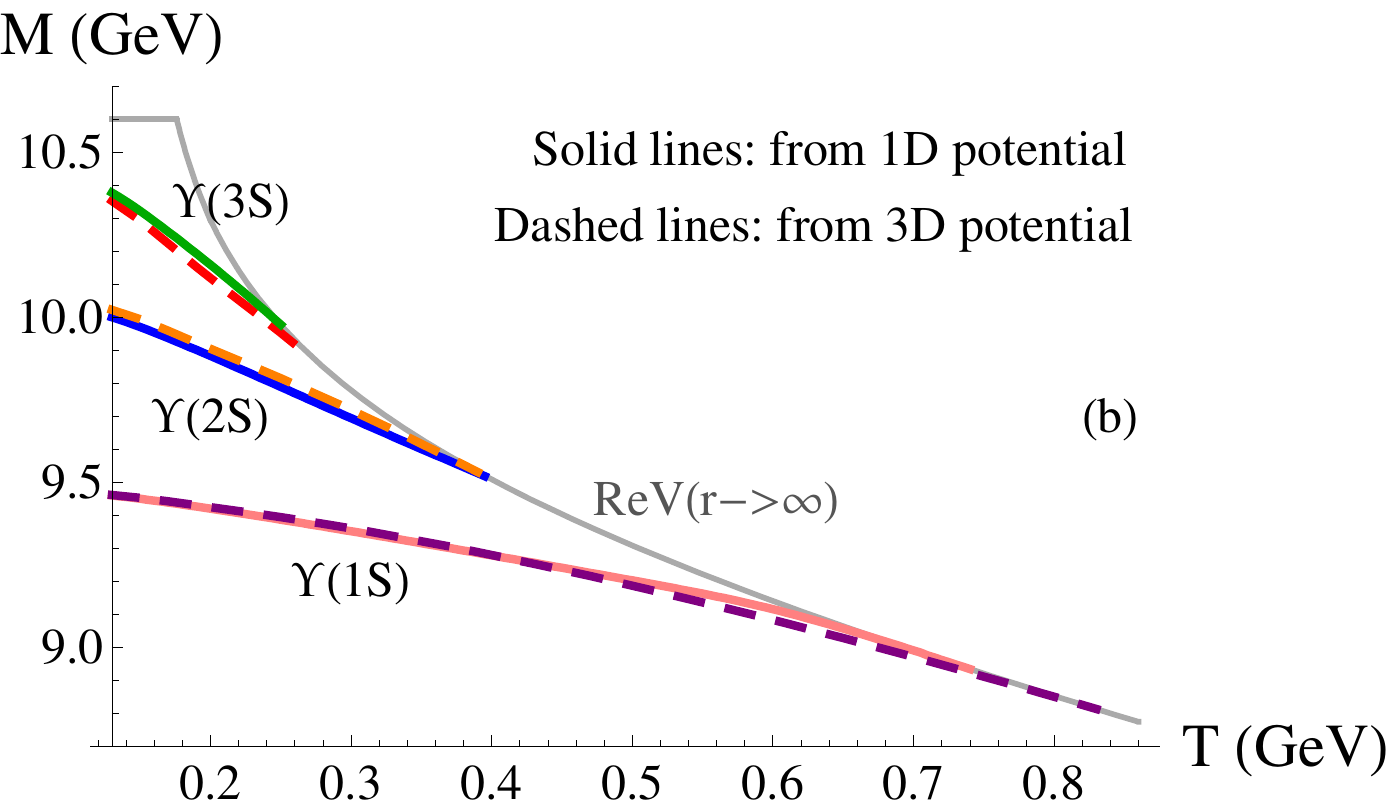}
    \caption{(a) Charmonium mass spectrum of the S states for the one-dimensional potential (solid lines) obtained from the expectation values of the Hamiltonian compared to the three-dimensional case (dashed lines). (b) Same for the bottomonia.}
    \label{fig:MassSpectra1D}
\end{figure}

In Fig.\ \ref{fig:MassSpectra1D}, the temperature dependence of the mass spectra -- using the expectation values of the Hamiltonian -- for the charmonium and bottomonium S-like states obtained from the one-dimensional potential are compared to the three-dimensional case. The mass spectra are in good agreement at any temperature for the charmonia and up to $T\approx 550$ MeV for the bottomonia (which includes the typical range of temperature reached in heavy-ion collisions). The maximum mass difference between the one- and the three-dimensional cases is $\approx$ 40 MeV for the S states (mainly for the $\Upsilon({\rm 1S})$ above $T=550$ MeV and for the $\Upsilon({\rm 3S})$). Additionally, the dissociation temperature of the $\Upsilon({\rm 1S})$ is approximately $100$ MeV smaller with the one-dimensional potential ($T_{\rm diss}^{\Upsilon({\rm 1S})}\approx 740$ MeV). 

\begin{figure}[htb!]
    \centering
    \includegraphics[width=0.50\textwidth]{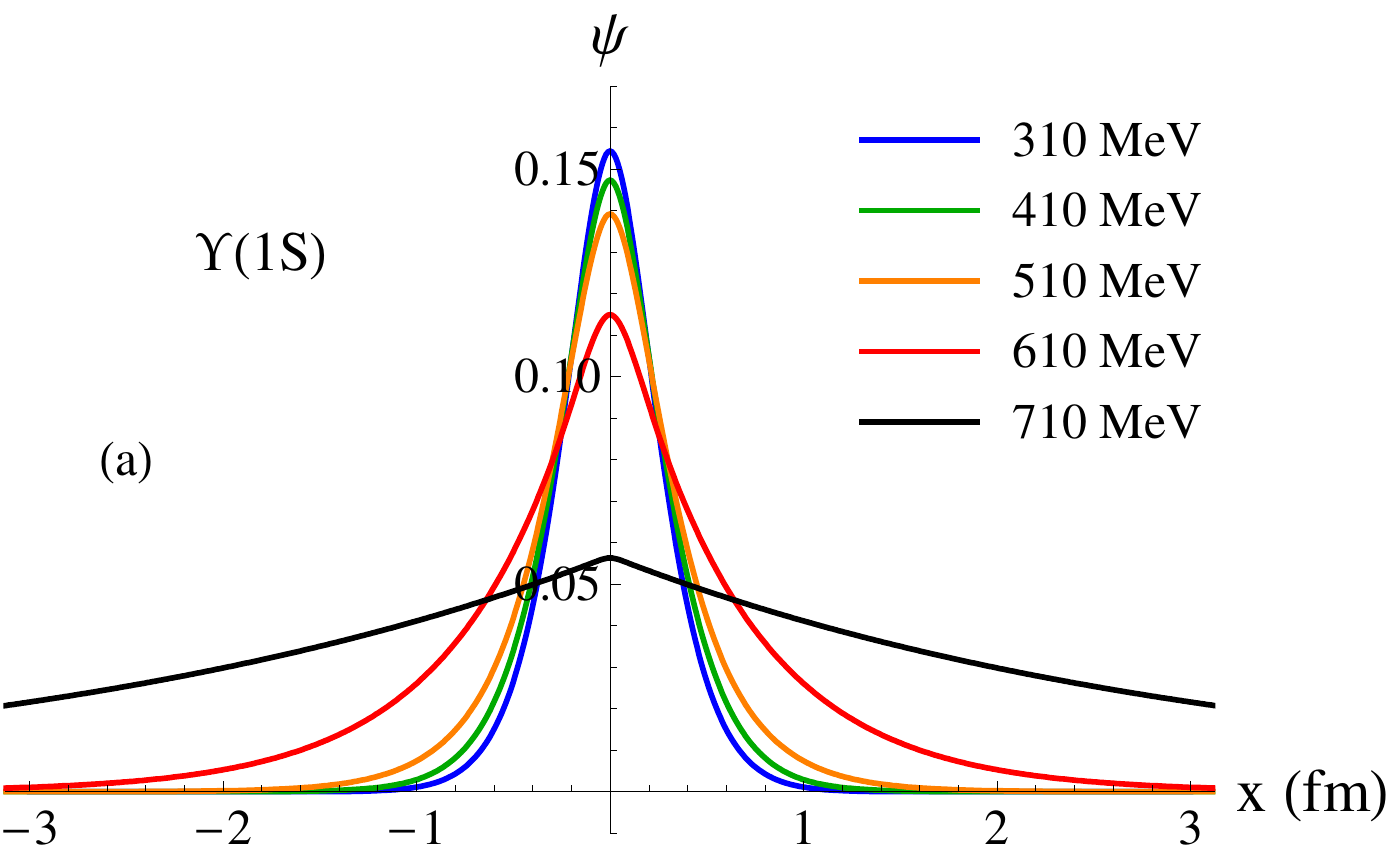}
     \includegraphics[width=0.48\textwidth]{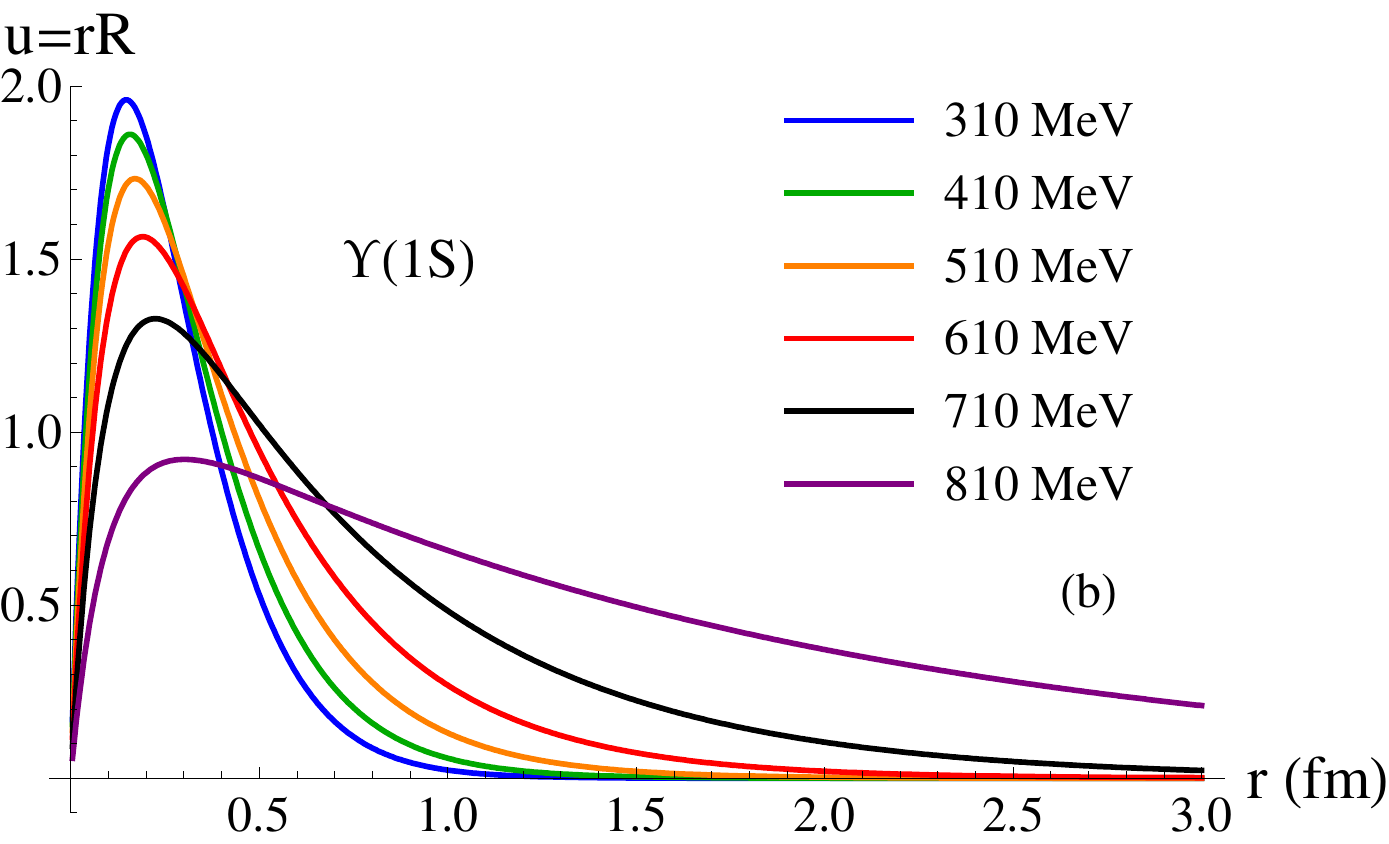}
    \caption{(a) Bottomonium 1S-like state obtained with the real part of the one-dimensional potential at various temperatures. (b) Reduced radial part of the bottomonium 1S state obtained with the three-dimensional potential at various temperatures.}
    \label{fig:eigenfunctionshighT}
\end{figure}

The bottomonium 1S states obtained with the one- and three-dimensional potentials are shown at different temperatures in Fig.\ \ref{fig:eigenfunctionshighT}. As in the vacuum case (see Sec.~\ref{Sec:VacuumPotential1D}), the features of the one- and three-dimensional eigenstates remain roughly similar throughout the temperature range. Nevertheless, the spatial distributions of the one-dimensional 1S-like states show more deviations from Gaussianity at temperatures close to the state dissociation temperature (see for instance $\Upsilon(1S)$ at $T=710$ MeV in Fig.\ \ref{fig:eigenfunctionshighT}). These deviations can be explained by the one-dimensional potential being almost flat at these high temperatures (see Fig~\ref{fig:RePot1D}).

\begin{figure}[htb!]
    \centering
    \includegraphics[width=0.49\textwidth]{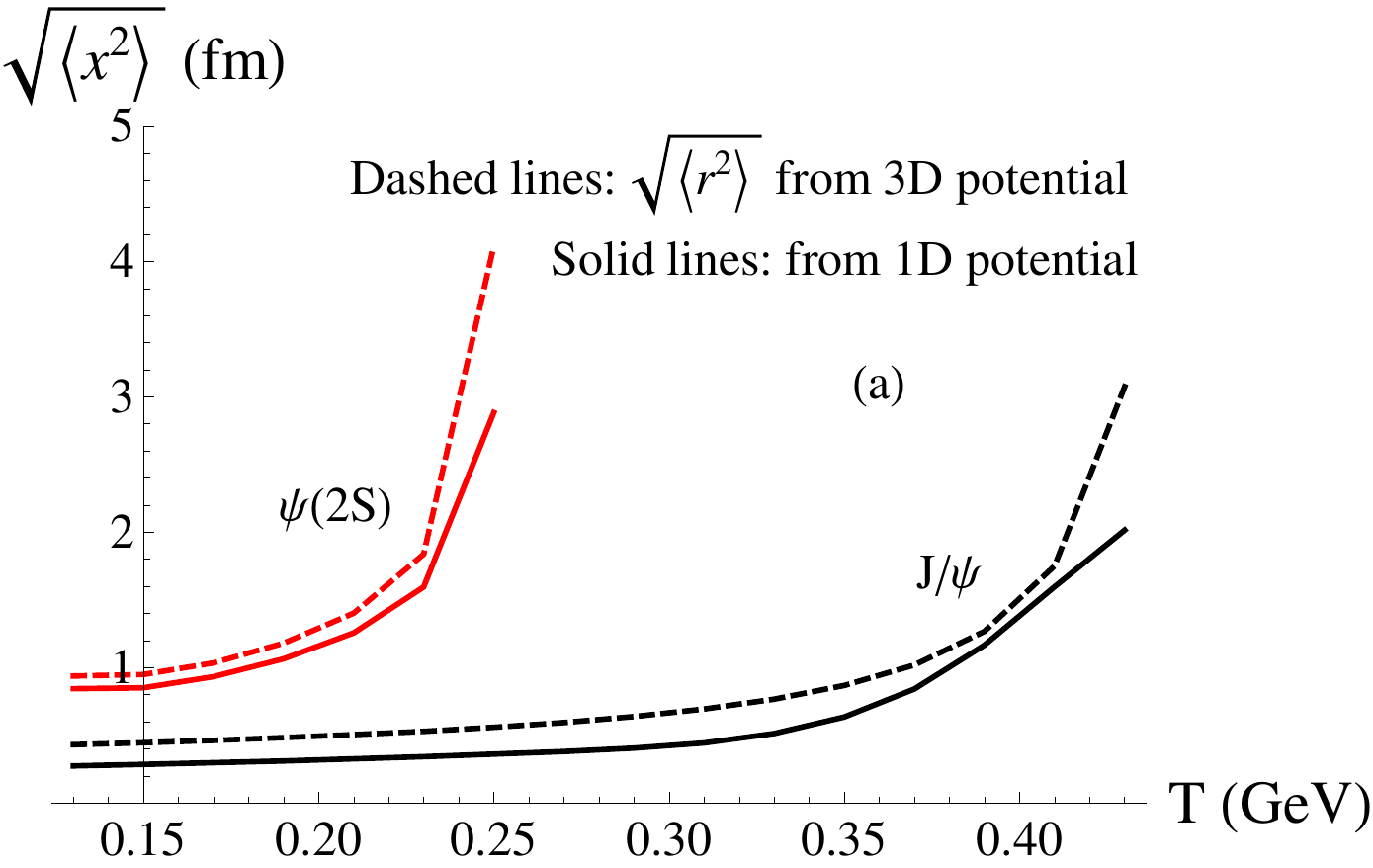}
     \includegraphics[width=0.50\textwidth]{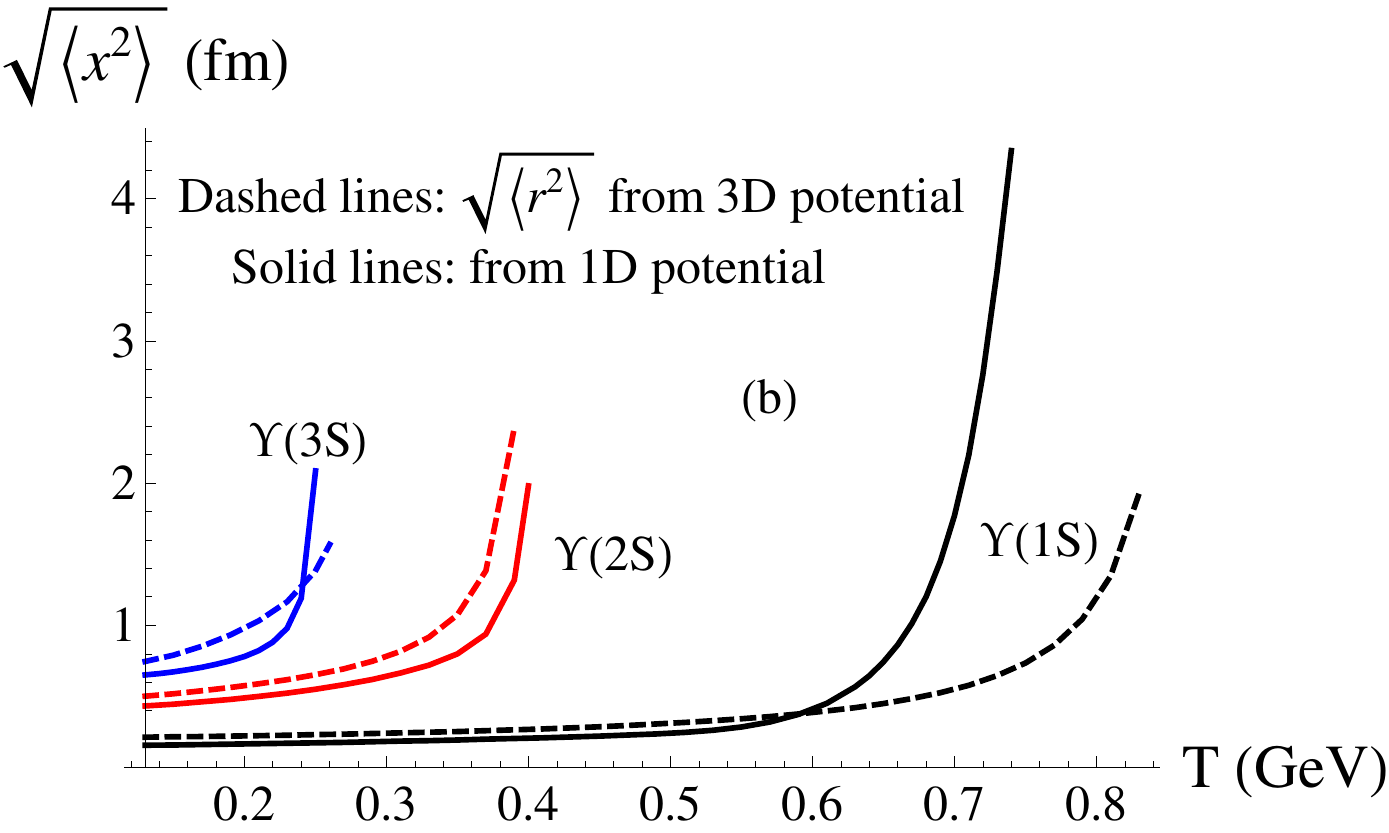}
    \caption{(a) Comparison of the root-mean-square radiuses of the charmonium S states obtained with the one- and three-dimensional potentials. (b) Same for the bottomonia.}
    \label{fig:MeanSquaredRadius1S}
\end{figure}

In Fig \ref {fig:MeanSquaredRadius1S}, the root-mean-square radiuses of the S states obtained with the one- and three-dimensional potentials are compared. The 1S states are observed to be $\approx 35\%$ larger with the three-dimensional potential and $\approx 10-15\%$ larger for the 2S states. For most of the states, the disagreements increase close to their dissociation temperatures : a large extension of the state spatial distributions exacerbates the differences between the two potentials. Nevertheless, as the imaginary part of the potential dominates the evolution of a states close to its dissociation temperature, these particular discrepancies are expected to be negligible.

\section{A one-dimensional potential: the imaginary part} \label{3Dto1Dim}

In this section, we propose an imaginary part for the one-dimensional potential parameterized to reproduce at best the temperature dependence of the decay widths of the three-dimensional potential described in Sec.\ \ref{Rothkopf3D}.

\subsection{One dimension model and decay widths} 

A simple possibility is to extend the radial imaginary part of the three-dimensional potential $\mathrm{Im}V = \mathrm{Im}V_C + \mathrm{Im}V_S$ (see Eqs.\ \ref{eq:String3D} and \ref{eq:Coulombic3D}) to ``$-x$'' (i.e. the symmetry to the $x=0$ axis) and to include by hand two coefficients $\alpha$ and $\beta$ to be tuned,

\begin{equation}\label{eq:1DImV}
\mathrm{Im}V^{\rm 1D}(x,T) = \alpha\,\mathrm{Im}V_C\left(|x|,T\right) + \beta\,\mathrm{Im}V_S\left(|x|,T\right).
\end{equation}

\begin{figure}[htb!]
    \centering
    \includegraphics[width=0.47\textwidth]{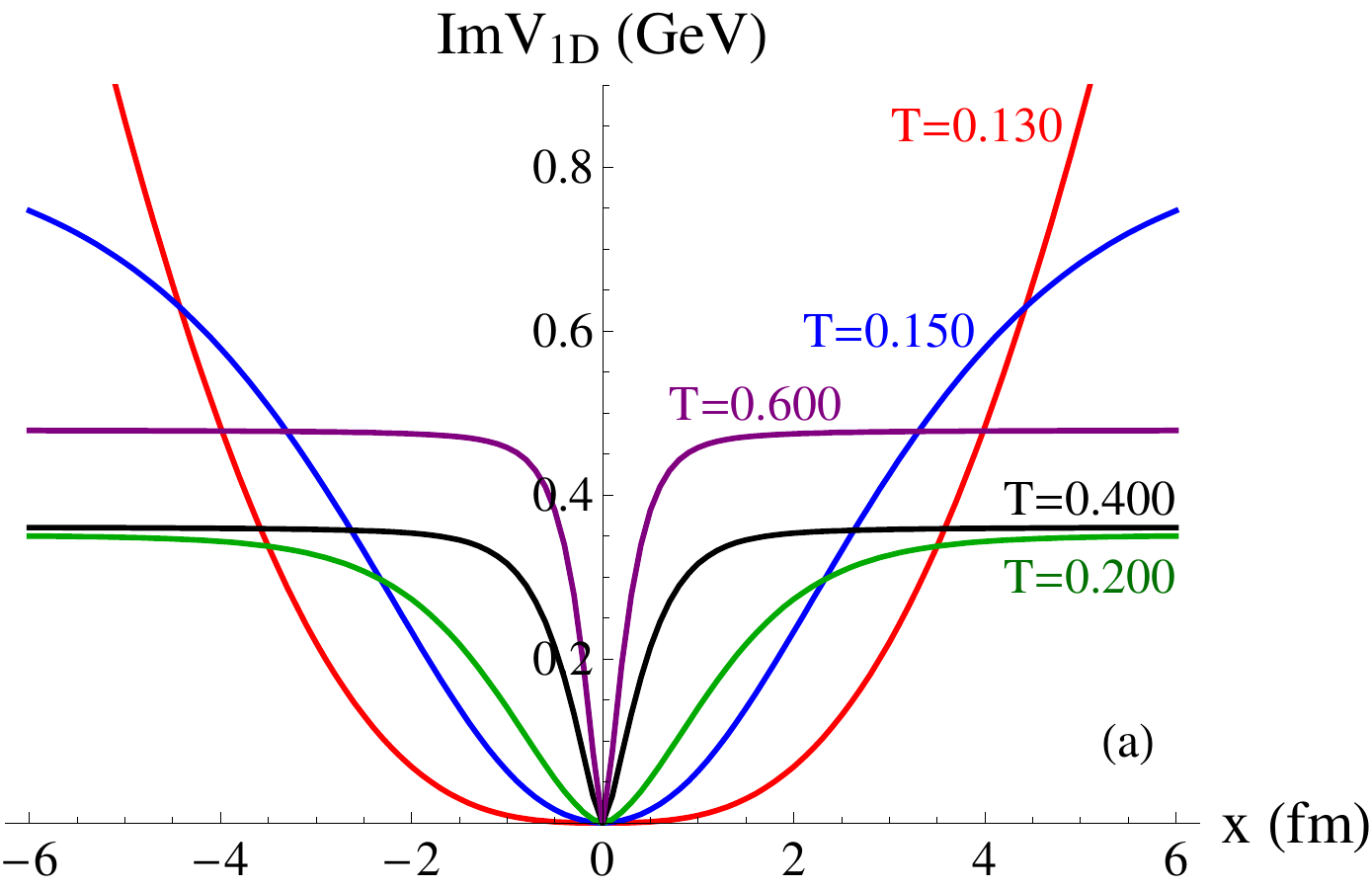}  
     \includegraphics[width=0.50\textwidth]{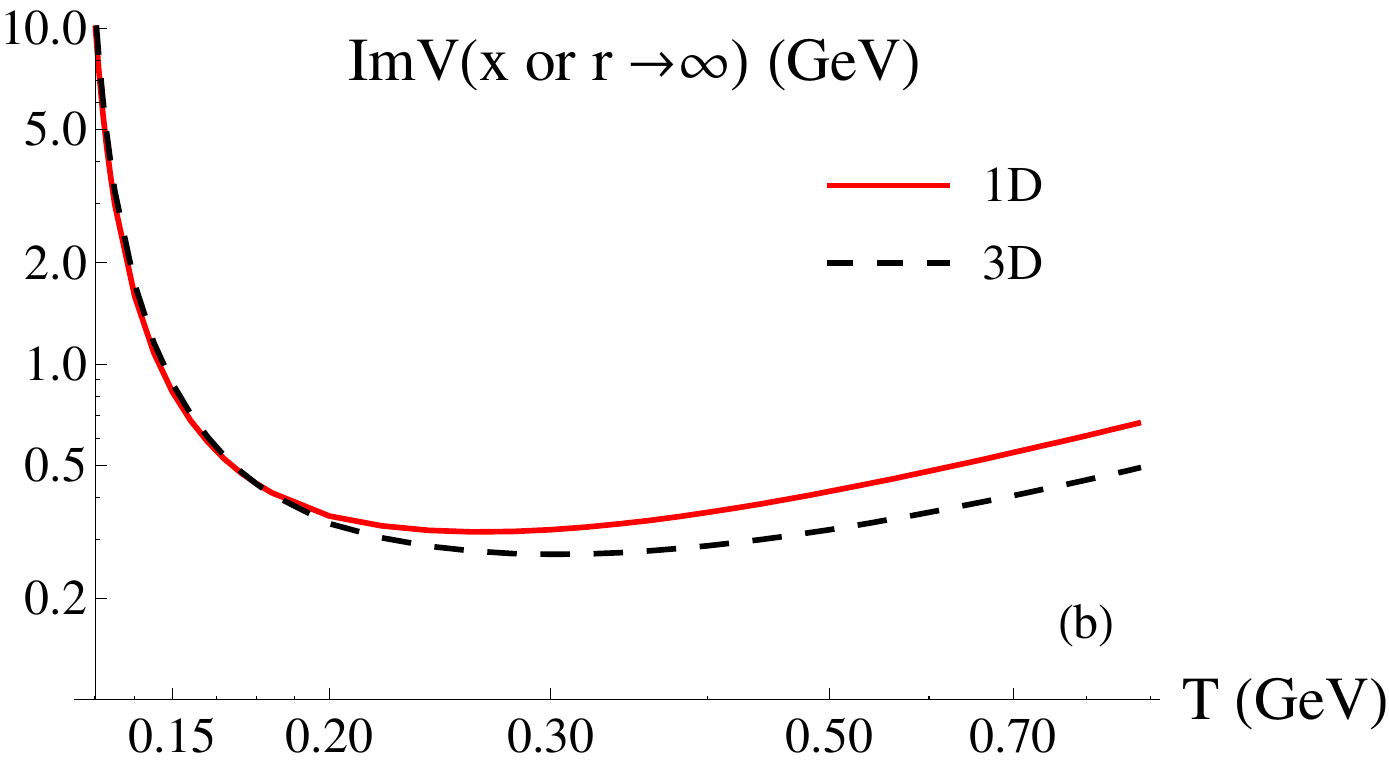}
    \caption{(a) Imaginary part of the one-dimensional potential for the bottomonia. (b) Large distance behaviour of the imaginary part of the one- and three-dimensional potentials for the bottomonia.}
    \label{fig:1DImPot}
\end{figure}

Then, like in three dimensions, the one-dimensional imaginary part has a harmonic behaviour at small distances -- that can be related to the diffusion coefficients of the heavy quarks -- and it saturates at large distances -- which is expected from Landau damping --. The values of $\alpha$ and $\beta$ which lead to the best similarity between the one- and three-dimensional decay widths are summed up in Tab.\ \ref{tab:1DImPotParameters}. The resulting potential $\mathrm{Im}V_{\rm 1D}$ is shown in Fig.\ \ref{fig:1DImPot} (a) at different temperatures. By giving more weights to the Coulombic part, the coefficients tend to narrow the potential well at short distances and to uplift the large distance values at high temperatures as compared to the three-dimensional case (as shown in Fig.\ \ref{fig:1DImPot} (b)). 

\begin{table}[h!]
\begin{center}
    \begin{tabular}{|C{3.5cm}||C{1.5cm}|C{1.5cm}|}
    \hline
$\mathrm{Im}V_{\rm 1D}$ coefficients & $\alpha$ & $\beta$  \\
    \hline
    \hline
Charmonia &  1.7 & 0.8  \\   
   \hline
Bottomonia &  1.4 & 0.9  \\ 
    \hline
    \end{tabular}
\caption {\label{tab:1DImPotParameters} 
\small Coefficients for the imaginary part of the one-dimensional potential $\mathrm{Im}V_{\rm 1D}$. }
\end{center}
\end {table}

\begin{figure}[htb!]
    \centering
    \includegraphics[width=0.49\textwidth]{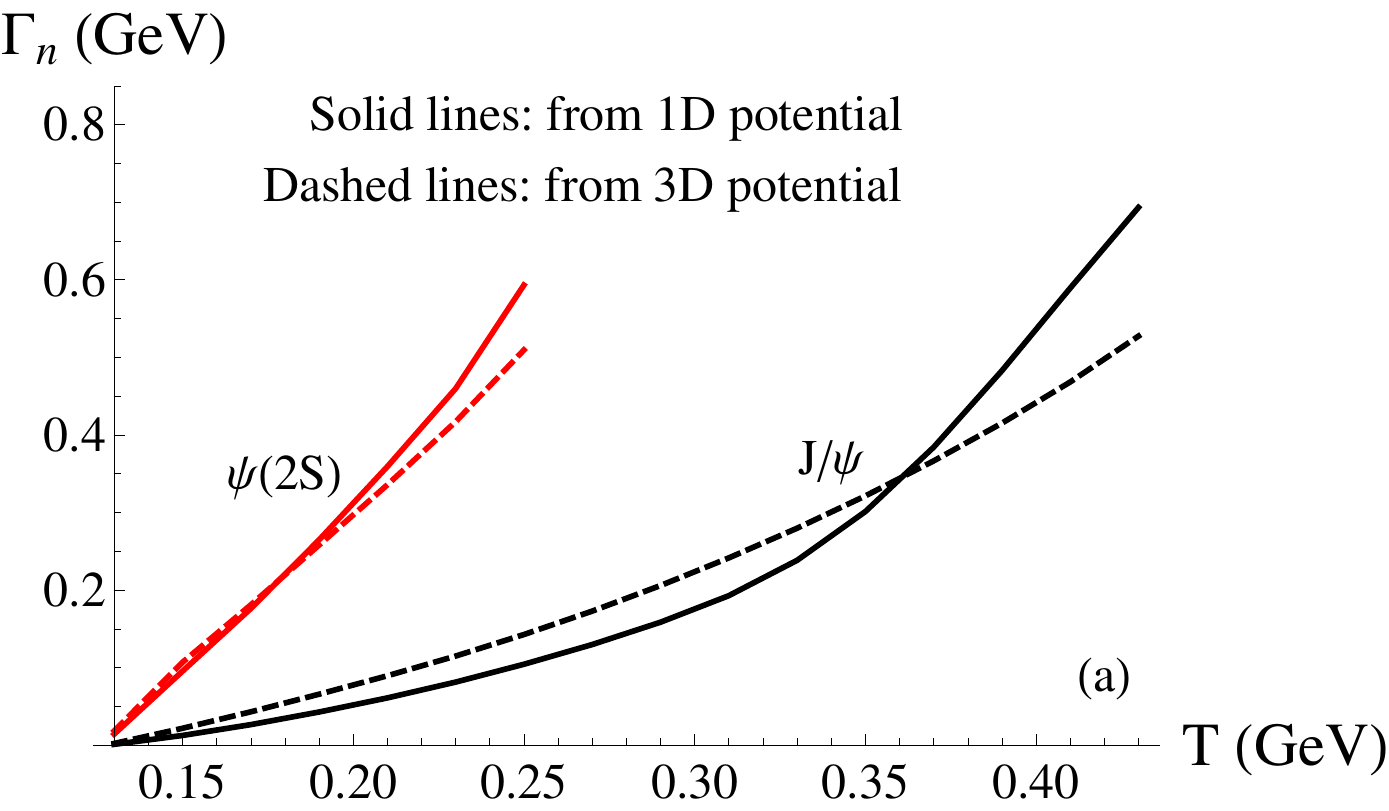}
        \includegraphics[width=0.49\textwidth]{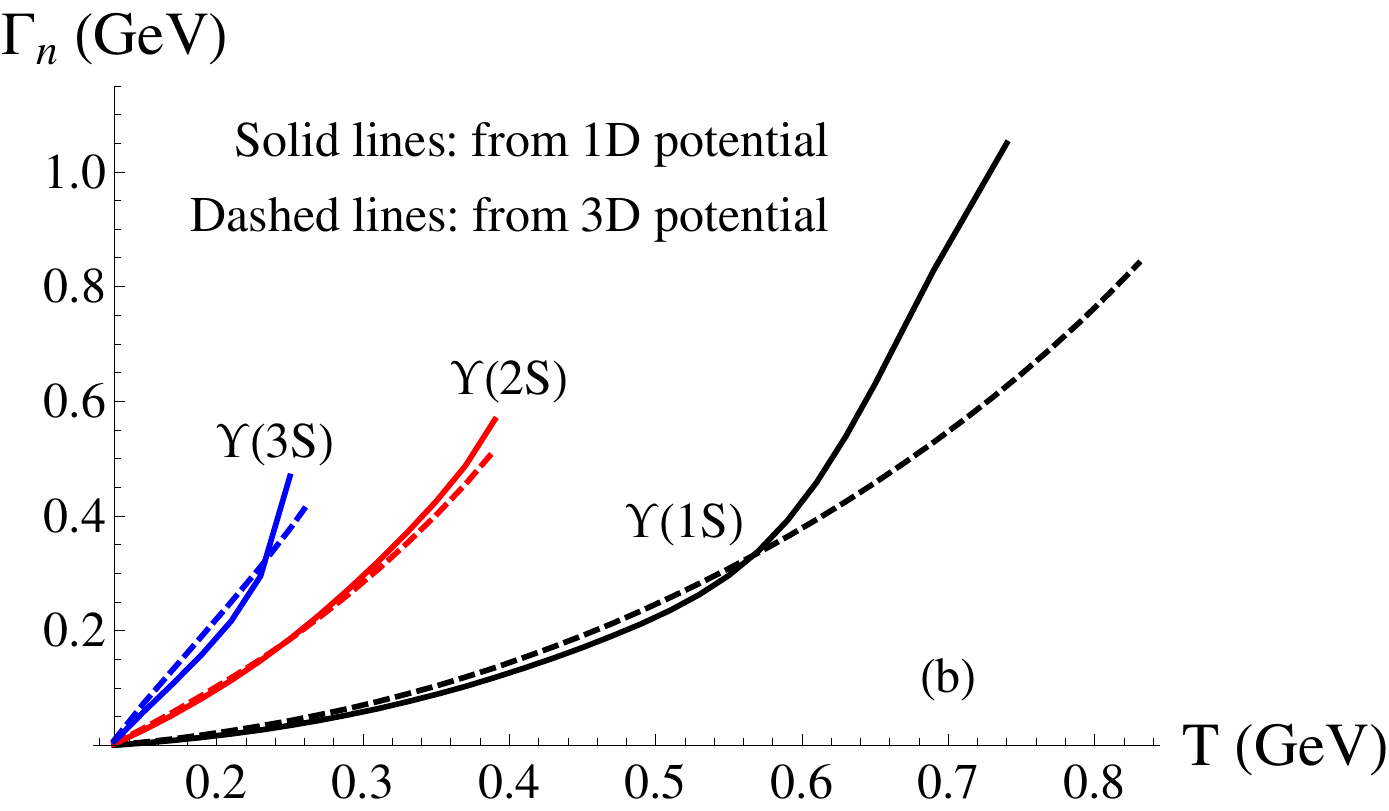}
        \caption{(a) Charmonium decay widths obtained with the one- and three-dimensional potentials. (b) Same for bottomonia.}
    \label{fig:DecayWitdthsCharmoBotto1D}
\end{figure}

The imaginary part of the one-dimensional potential must reproduce at best the decay widths of the three-dimensional potential from Sec.\ \ref{Rothkopf3D}. To compute the decay widths $\Gamma^{\rm 1D}_n$ of the one-dimensional potential we use the eigenstates of $\mathrm{Re}V_{\rm 1D}$ determined in Sec.\ \ref{3Dto1Dreal} and calculate the expectation value of $\mathrm{Im}V^{\rm 1D}$. They are then compared to the decay widths $\Gamma^{\rm 3D}_n$ of the three-dimensional potential calculated with the expectation values of $\mathrm{Im}V$, and we expect that for a state $n$:
\begin{eqnarray}
\Gamma^{\rm 1D}_n(T) = 2 \left\langle {\rm Im}V^{\rm 1D}(T) \right\rangle_n \approx 2 \left\langle {\rm Im}V(T) \right\rangle_n = \Gamma^{\rm 3D}_n(T).
\end{eqnarray}
The temperature dependent decay widths for the charmonium and bottomonium S states obtained with the one-dimensional potential (\ref{eq:1DImV}) and the set of parameters given in Tab.\ \ref{tab:1DImPotParameters} are shown in Fig.\ \ref{fig:DecayWitdthsCharmoBotto1D}. They are in a good agreement with the three-dimensional decay widths up to $T=600$ MeV, at the exception of the $J/\psi$ state which is underestimated for $T<350$ MeV and overestimated for $T>380$ MeV to a maximum of $\approx 30\%$. The large distance saturation of the imaginary part is observed to be important to obtain decay widths that are nearly linear with the temperature. At high temperatures, the discrepancies between the one- and three-dimensional 1S state decay widths can be traced back to the differences in eigenstate spatial features (see Fig.\ \ref{fig:eigenfunctionshighT}) and to the differences in complex potentials at large distances (see Fig.\ \ref{fig:1DImPot}).

\vspace{-3mm}

\subsection{Spectral decomposition} \label{1DmodelSpectralDecompo}

%
\begin{figure}[htb!]
    \centering
    \includegraphics[width=0.48\textwidth]{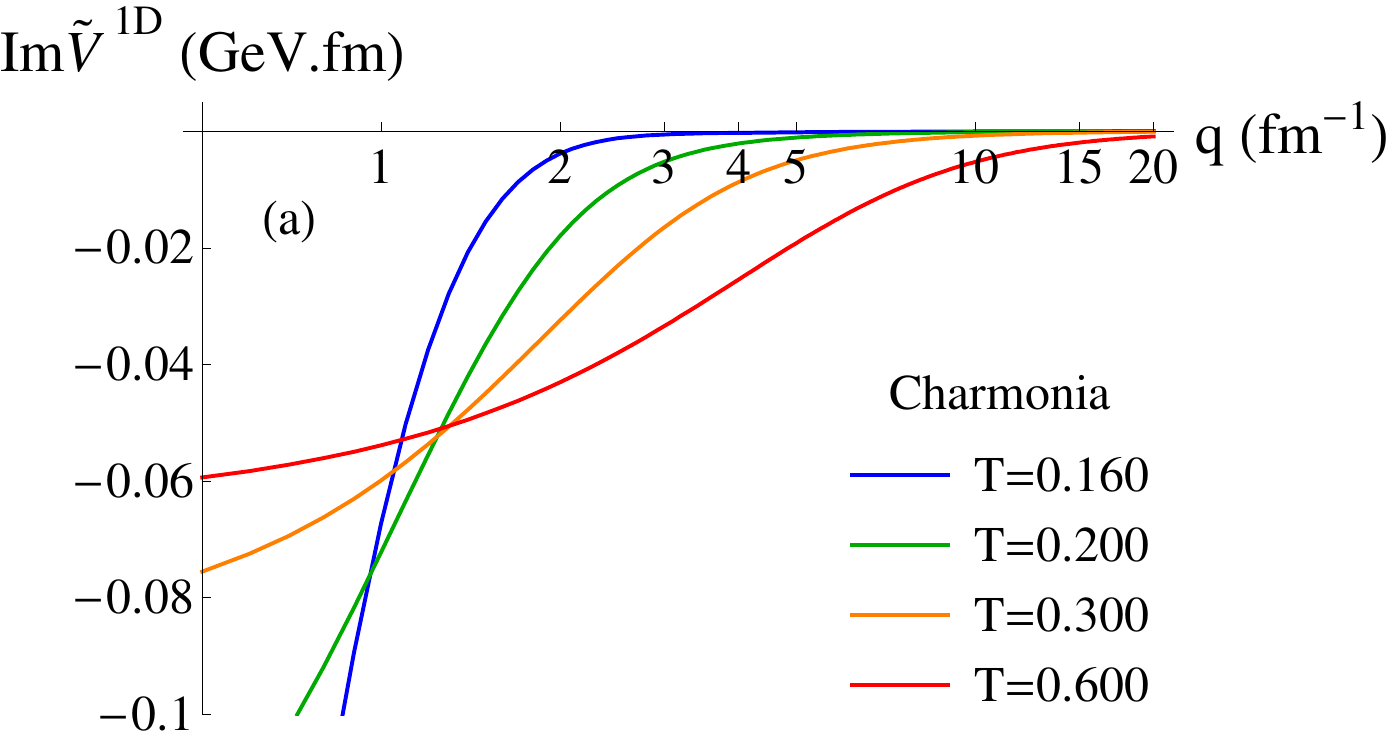}
    \hspace{4mm}
     \includegraphics[width=0.48\textwidth]{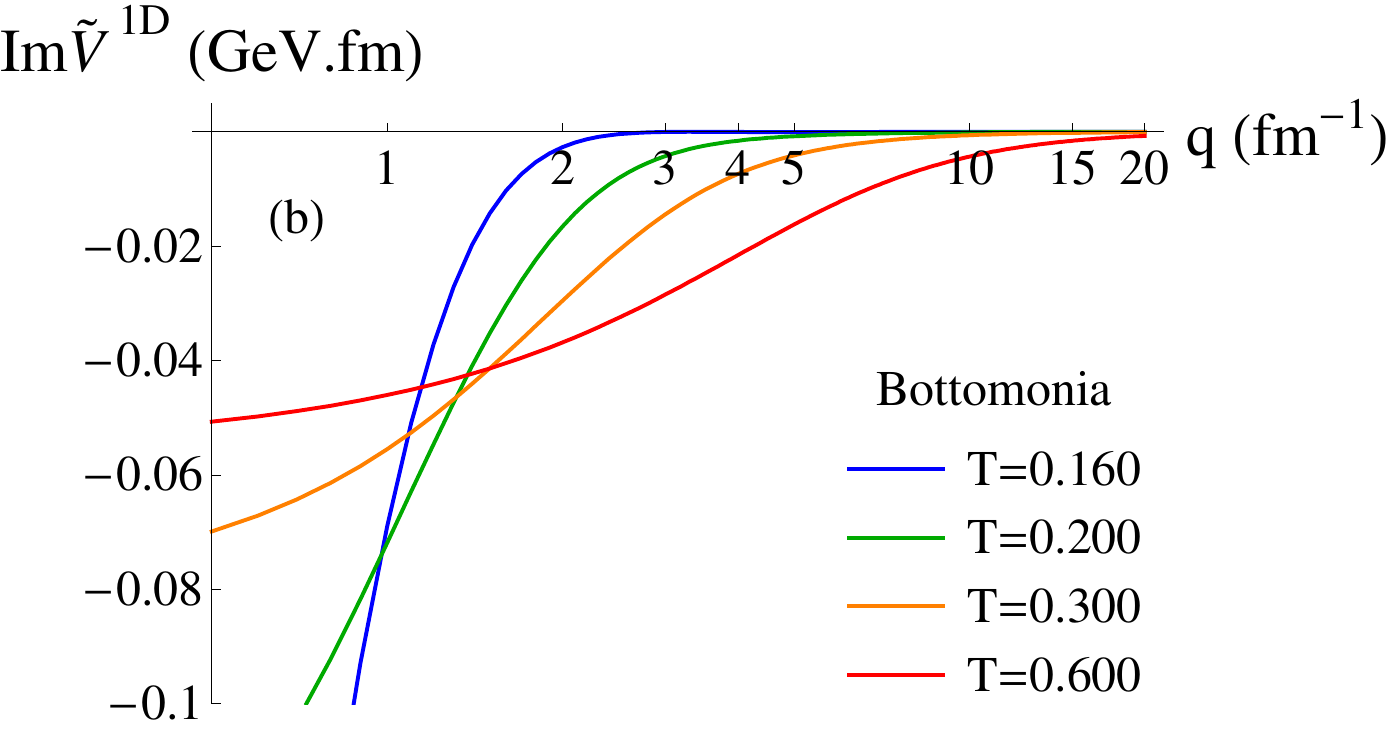}
    \caption{Spectral decomposition of the imaginary part of the potential ${\rm Im}V^{\rm 1D}=\alpha\,{\rm Im}V_c^{\rm 1D}+\beta\,{\rm Im}V_s^{\rm 1D}$} at different temperatures (in GeV) for the (a) charmonia and (b) bottomonia.
    \label{fig:1DspectraldecompoVtot}
\end{figure}

The spectral decomposition ${\rm Im}{\tilde V}^{\rm 1D}(q)=\alpha\, {\rm Im}{\tilde V}_c^{\rm 1D} + \beta\, {\rm Im}{\tilde V}_s^{\rm 1D}$ of this one-dimensional potential can be calculated analytically. For the Coulombic part, the Fourier transform is
\begin{eqnarray}\nonumber
{\rm Im}{\tilde V}_c^{\rm 1D} =\frac{\alpha_s T}{2\pi} \int_{-\infty}^\infty dx \,\phi\left(m_D |x|\right) \cos(q x).
\end{eqnarray}
Integrating by parts, we get:
\begin{eqnarray}\nonumber
{\rm Im}{\tilde V}_c^{\rm 1D} =\frac{\alpha_s T}{2}\left[\frac{q}{q^2+m_D^2} + \frac{1}{m_D}\left(\arctan\left(\frac{q}{m_D} \right)-\frac{\pi}{2}\right)\right],
\end{eqnarray}
where $-\frac{\pi}{2}$ is added to the primitive to get rid of its asymptotic value. Similarly, the spectral decomposition of the string part is given by
\begin{eqnarray}\nonumber
{\rm Im}{\tilde V}_s^{\rm 1D} =\frac{\sigma T }{2\pi m_D^2} \int_{-\infty}^\infty dx \,\chi\left(m_D |x|\right) \cos(q x),
\end{eqnarray}
and the integration by parts yields
\begin{eqnarray}\nonumber
{\rm Im}{\tilde V}_s^{\rm 1D} =-\frac{\sigma T }{m_D^3}\left[\frac{2}{\sqrt{{\Delta_D}^2+\frac{q^2}{m_D^2} }\left(1+\frac{q^2}{m_D^2}\right)^2}+\frac{{\Delta_D}^2 \left(1 - 3 \frac{q^2}{m_D^2} \right) -4 \frac{q^4}{m_D^4} }{\left(1 + \frac{q^2}{m_D^2} \right)^3 \left({\Delta_D}^2 + \frac{q^2}{m_D^2} \right)^{3/2}}\right].
\end{eqnarray}

As shown in Fig.\ \ref{fig:1DspectraldecompoVtot}, the spectral decomposition ${\rm Im}{\tilde V}^{\rm 1D}$ is negative for all q and at any temperatures considered here. We can thus conclude that this one-dimensional potential satisfies the positivity requirement of the Lindblad equations.

\section{Comparison with other one-dimensional potential models} \label{Comparing1DModels}

\begin{figure}[htb!]
    \centering
    \includegraphics[width=0.48\textwidth]{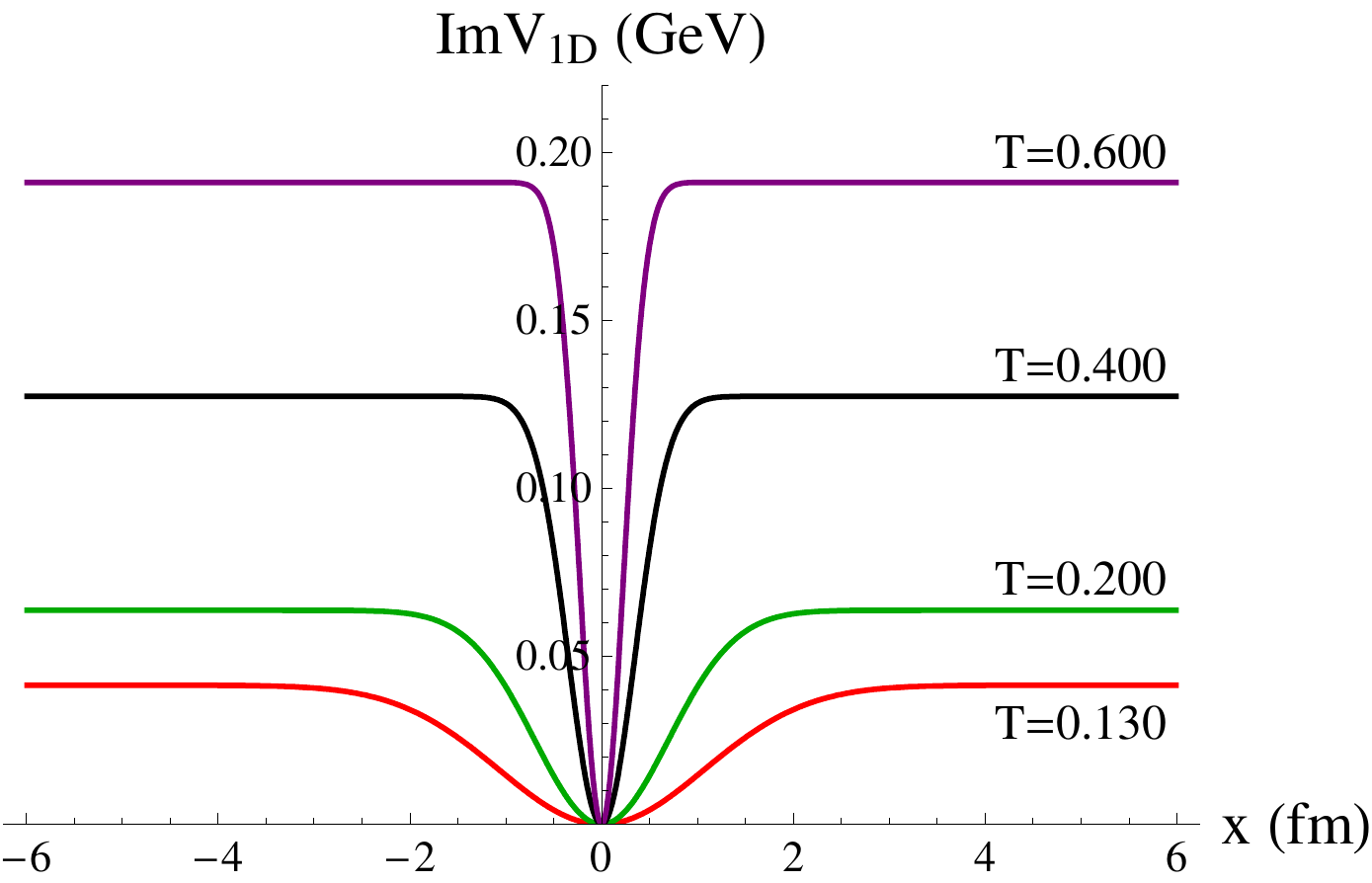}
    \caption{Imaginary part of the one-dimensional potential used in \cite{Miura:2019ssi,Alund:2020ctu}. It can be compared to the potential proposed in Sec.\ \ref{3Dto1Dim} shown in Fig.\ \ref{fig:1DImPot} : their behaviour are qualitatively different for $T< 200$ MeV. }
    \label{fig:ImPot1DAkamatsu}
\end{figure}

The one-dimensional potential described in Sec.\ \ref{3Dto1Dreal} and \ref{3Dto1Dim} (from now on called ``model A") can be compared to the reduction of the HTL potential proposed in Sec.\ \ref{pQCDHTL1D} (from now on called ``model C"), and to other potentials found in the literature \cite{DeBoni:2017ocl,Kajimoto:2017rel,Miura:2019ssi,Alund:2020ctu}. Among the latter we choose the potential used in \cite{Miura:2019ssi,Alund:2020ctu} (``model B"), given by:
\begin{eqnarray}\label{eq:1DReVakama}
{\rm Re}V^{\rm 1D}(x,T)=-\frac{\alpha}{\sqrt{x^2+x_c^2}}e^{-m_D|x|},
\end{eqnarray}
for the real part, and
\begin{eqnarray}\label{eq:1DImVakama}
{\rm Im}V^{\rm 1D}(x,T)=\gamma-\gamma \exp(-x^2/l_{\rm corr}^2).
\end{eqnarray}
for the imaginary part. The parameters are taken to be $x_c=1/m$, $l_{\rm corr}=1/T$, $m_D=2T$ and $\gamma = T/\pi$. The imaginary part given by Eq.\ \ref{eq:1DImVakama} is shown in Fig.\ \ref{fig:ImPot1DAkamatsu} for different temperatures. The three models A, B and C are summed up in Tab.\ \ref{tab:3models}.

\begin{table}[htb!]
\renewcommand{\arraystretch}{1.5}
\begin{center}
    \begin{tabular}{| P{3cm}|| P{4.5cm}|P{3.5cm}|P{4.2cm}|}
    \hline
1D potential & Model A & Model B & Model C  \\
    \hline
    \hline
Described in & Sec.\ \ref{3Dto1Dreal} and \ref{3Dto1Dim}  & Sec.\ \ref{Comparing1DModels} \cite{Miura:2019ssi,Alund:2020ctu} & Sec.\ \ref{pQCDHTL1D} (HTL 1D) \\   
   \hline
 Real part & $1/2\,K|x|\,e^{-\mu_1|x|} + \mu_2$  saturated from some $|x|$ & $-\frac{\alpha}{\sqrt{x^2+x_c^2}}e^{-m_D|x|}$ & $\frac{\sigma}{m_{D}}\bigl(1 - e^{-m_{D}|x|}\bigr)$ \\ 
    \hline
    Imaginary part & $\alpha\,\mathrm{Im}V_C\left(|x|\right) + \beta\,\mathrm{Im}V_S\left(|x|\right)$ (see Eq.~\ref{eq:Coulombic3D} and \ref{eq:String3D})  & $\gamma-\gamma \exp(-x^2/l_{\rm corr}^2)$ & $- 2\sigma m_{D}^{2} T \int_{\,0}^{\,\infty} dq \frac{1 - \cos{q|x|}}{q(q^{2} + m_{D}^{2})^{2}}$ \\ 
    \hline
    \end{tabular}
\caption {\label{tab:3models} 
\small The 3 models of one-dimensional potential that are compared in this section.}
\end{center}
\end {table}

\begin{figure}[htb!]
    \centering
    \includegraphics[width=0.49\textwidth]{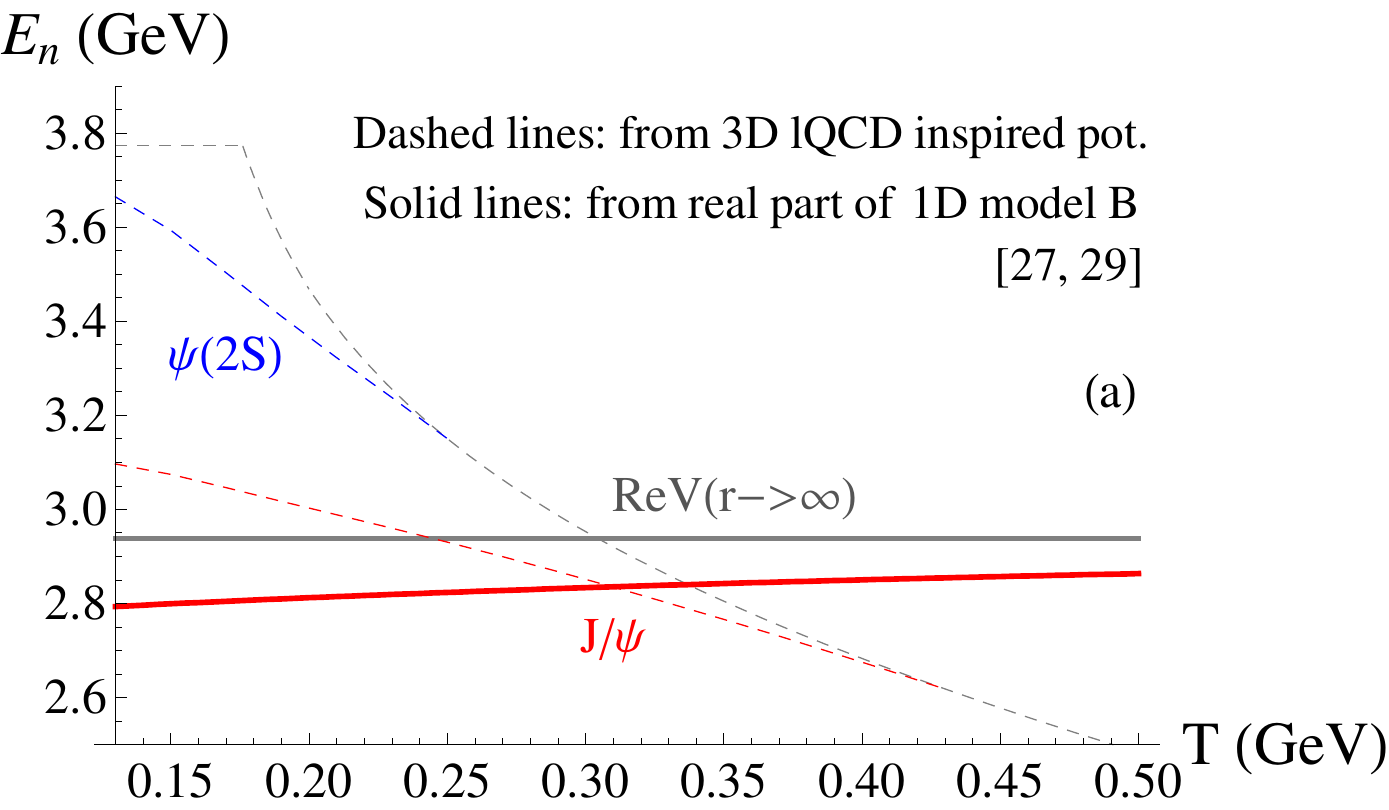}
    \includegraphics[width=0.49\textwidth]{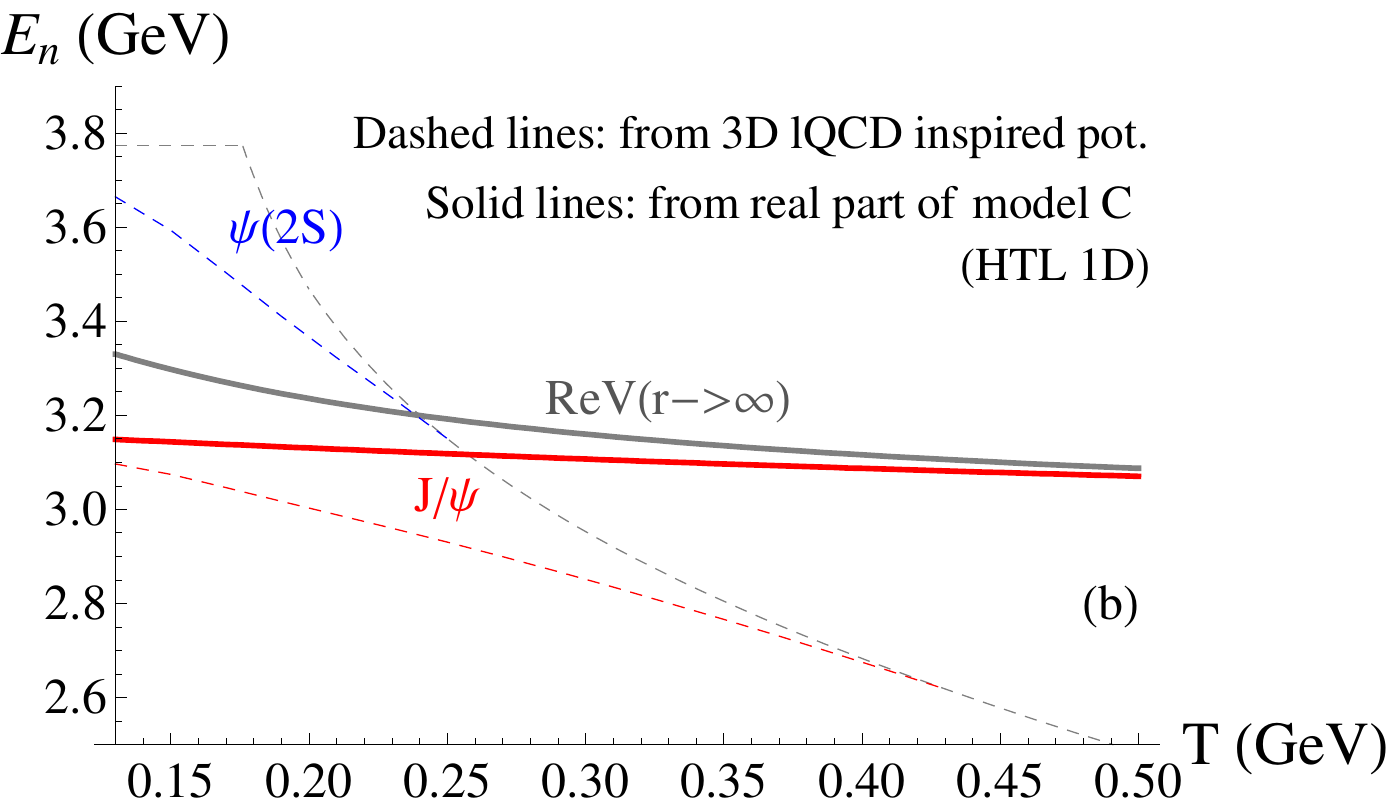}
    \caption{(a) Charmonium mass spectrum of the S states obtained with the real part of the one-dimensional model B \cite{Miura:2019ssi,Alund:2020ctu} (solid lines) using the expectation values of the Hamiltonian compared to the results obtained with the three-dimensional potential  (dashed lines). (b) Same comparison with the one-dimensional model C (1D HTL).}
    \label{fig:EnSpectraModelBC}
\end{figure}

As shown in Fig.\ \ref{fig:EnSpectraModelBC}, the energy spectra $E_n$ - and the binding energies given by ${\rm Re}V(r\rightarrow \infty)-E_n$ - obtained from the real parts of the models B and C are qualitatively and quantitatively very different from what is expected from the three-dimensional potential inspired by lattice QCD (described in Sec.\ \ref{Rothkopf3D}). Within the range of temperature studied here, these two models only give one bound S state. At the opposite, the model A was parameterized to fit at best the energy spectrum given by the three-dimensional potential (see Fig.\ \ref{fig:MassSpectra1D}). In terms of phenomenology, the real parts of the models B and C are therefore irrelevant.\\ 

\begin{figure}[htb!]
    \centering
    \includegraphics[width=0.49\textwidth]{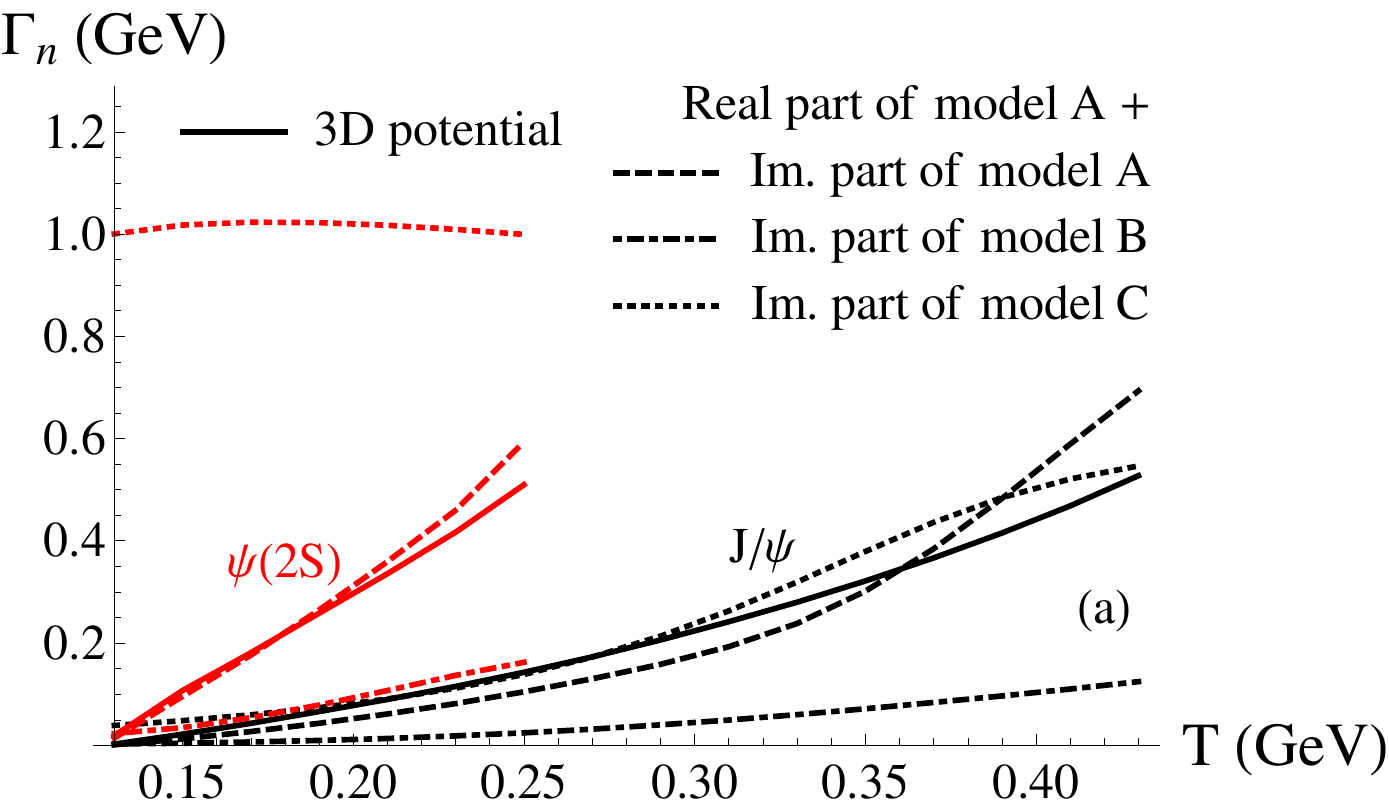}
    \includegraphics[width=0.49\textwidth]{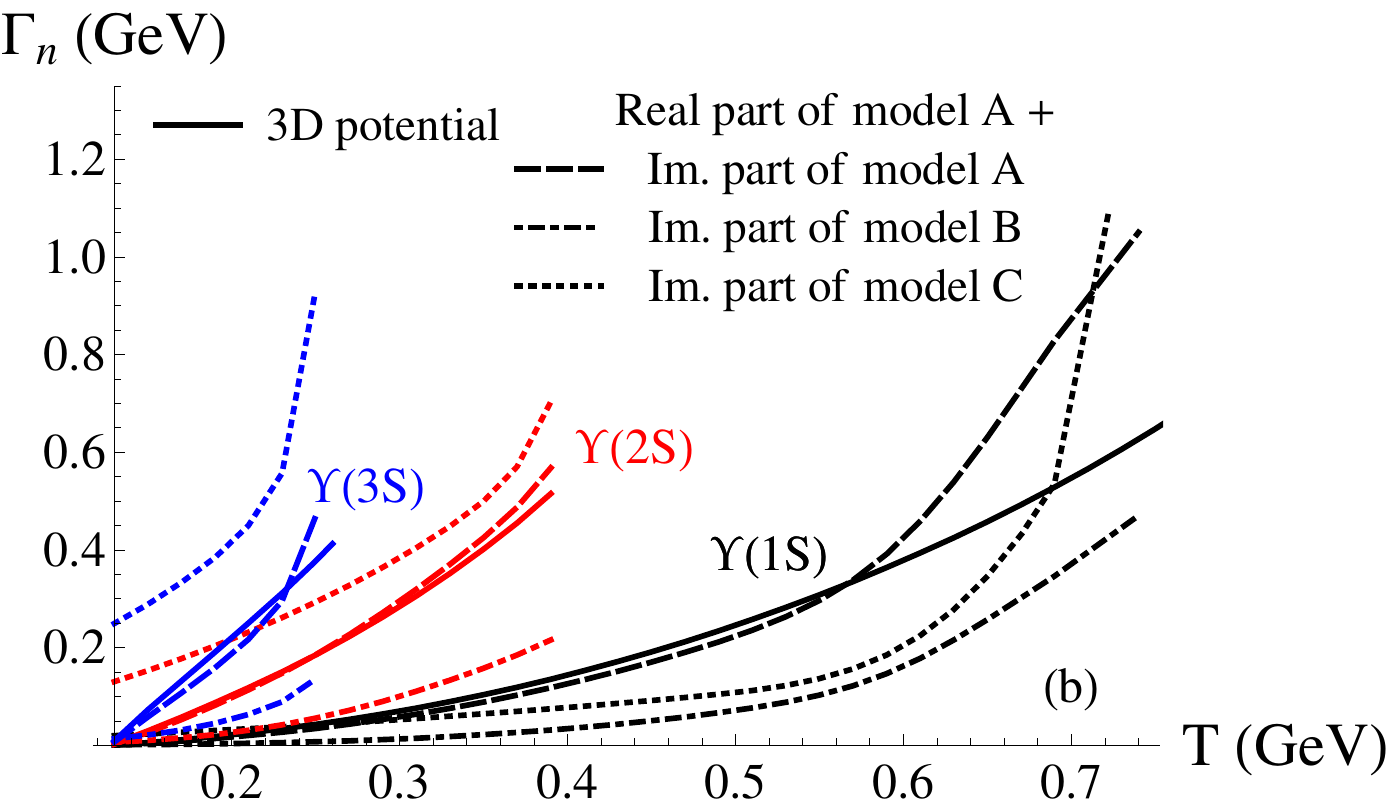}
    \caption{(a) Charmonium decay widths obtained with the imaginary parts of the three models and the real part of the model A compared to the results obtained with the three-dimensional potential (solid lines). (b) Same for bottomonia.}
    \label{fig:DecayModelABC}
\end{figure}

\begin{figure}[htb!]
    \centering
 \includegraphics[width=0.49\textwidth]{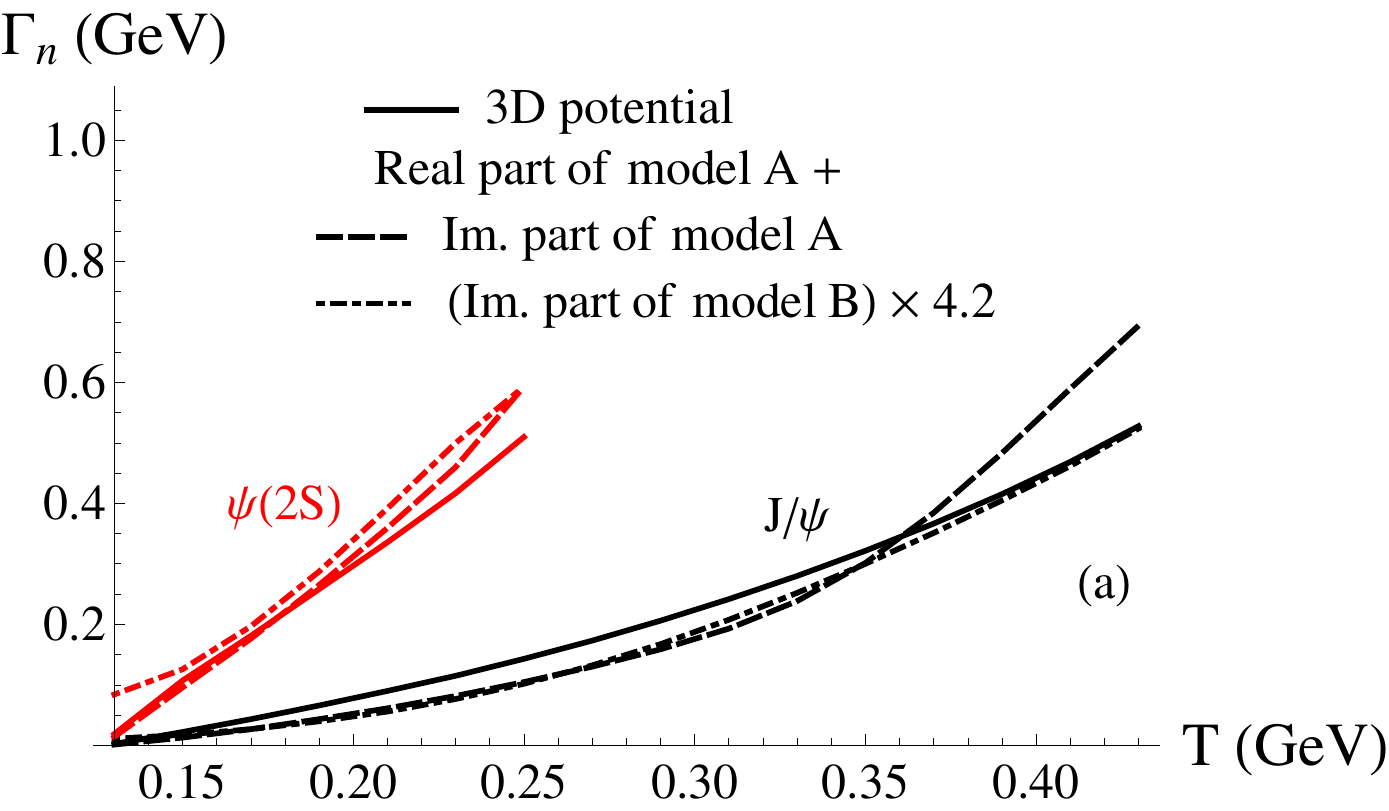}
    \includegraphics[width=0.49\textwidth]{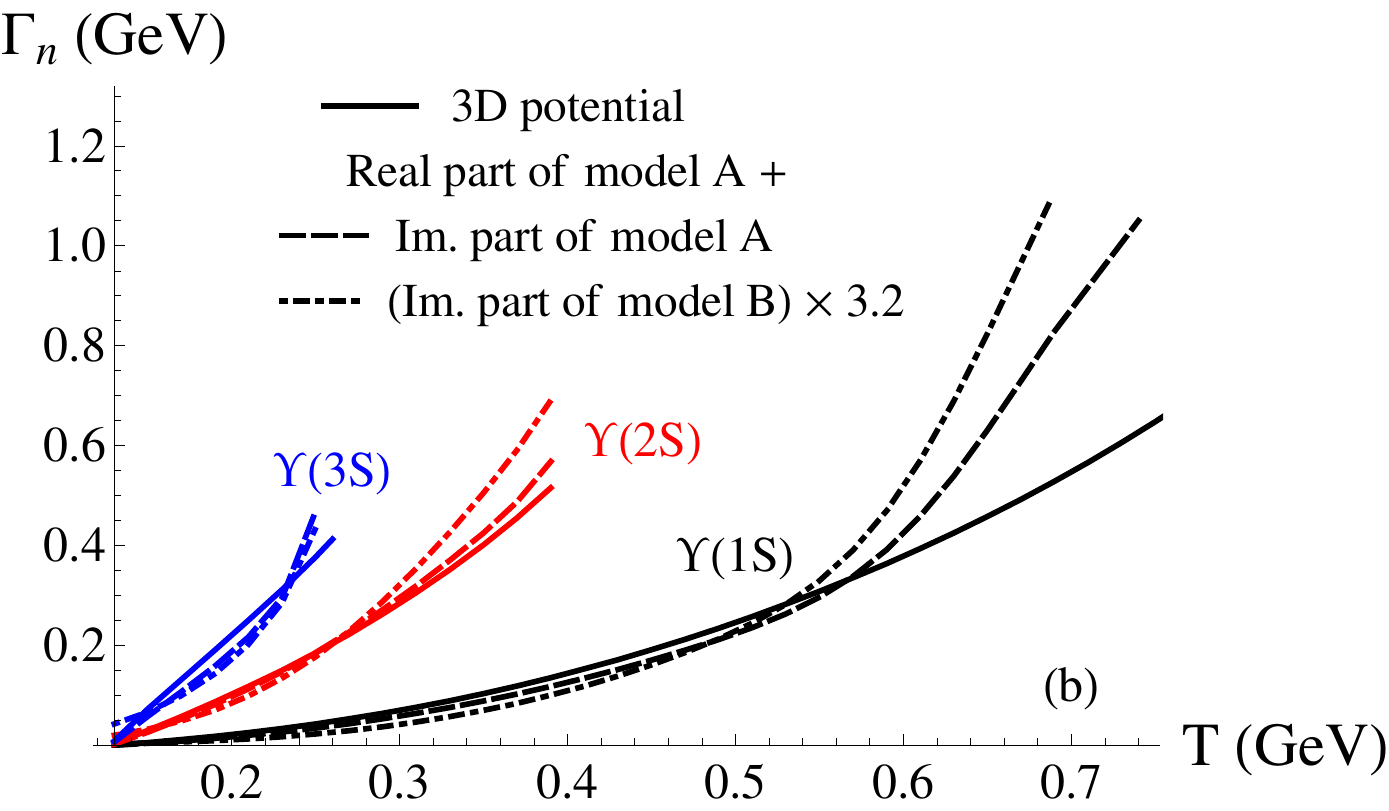}
    \caption{(a) Same as Fig.\ \ref{fig:DecayModelABC} (a), with an extra global factor for the imaginary part of the model B, i.e. by multiplying the $\gamma$ parameter by $4.2$. (b) Same for bottomonia, with a global factor of $3.2$.}
    \label{fig:DecayModelAB}
\end{figure}

Because the eigenstates obtained with the three models are very different, comparing the decay widths given by each full model - i.e.~using the eigenstates given by the real part to compute the expectation value of the imaginary part - is meaningless. 
To compare the imaginary parts of the three models on a common basis, we will only use the eigenstates given by the model A (see Sec.\ \ref{3Dto1Dreal}). In Fig.\ \ref{fig:DecayModelABC}, the decay widths obtained with the imaginary parts of the three models and the real part of the model A are compared to the ones of the three-dimensional potential. The decay widths given by the model C (1D HTL) overestimate or underestimate quite strongly the three-dimensional results and show a different qualitative behaviour (exception made of the $J/\psi$-like state). The decay widths obtained with the model B globally underestimate the three-dimensional results but seems qualitatively relevant. As shown in Fig.\ \ref{fig:DecayModelAB}, if one re-scales the imaginary part of the model B, i.e.~multiplies the $\gamma$ parameter by some factor, one can get a much better quantitative match. Although the correspondence with the three-dimensional results is slightly less good with the adjusted model B than with the model A (the $J/\psi$-like state excepted), the adjusted model B can be seen as a simple alternative to model A (if one considers the decay widths as the only criterion of choice).

\section*{Conclusion} 

In this paper, we have analysed the behavior and properties of two three-dimensional complex potentials for in-medium quarkonium physics, derived from the hard thermal loop perturbation theory or inspired by lattice QCD data \cite{Burnier:2015tda,Lafferty:2019jpr}. Based on this analysis, we proposed a one-dimensional complex potential parameterized to reproduce at best two key properties -- the temperature-dependent masses of the eigenstates and their decay widths -- of the three-dimensional lattice QCD inspired potential. The real part of this one-dimensional potential is given by Eqs.\ \ref{V1D} and \ref{V1Dmu} with the set of parameters summed up in Tab.\ \ref{tab:1DRePotParameters}. The vacuum spectra, the temperature-dependent masses and the dissociation temperatures of the states correspond in a good approximation to the ones given by the three-dimensional potential. The features of the one- and three-dimensional eigenstates are roughly similar at any temperature relevant for phenomenology, despite their difference of nature (Gaussian-like for a linear-like potential and exponential-like for the radial part of a Coulombic-like potential). The imaginary part of this one-dimensional potential, given by Eq.\ \ref{eq:1DImV} and the parameters in Tab.\ \ref{tab:1DImPotParameters}, is a simple ``symmetrization" of the three-dimensional potential radial component. The thermal decay widths obtained with the one- and three-dimensional potentials are similiar up to $T=400$ MeV ($T=600$ MeV) for the charmonia (bottomonia). Note that within this model, the one-dimensional potential for charmonia and bottomonia are different. As a sanity check, we calculated the spectral decompositions of both the one- and three-dimensional potentials and ensured their compatibility with the positivity of the Lindblad equation. The proposed potential can thus be used in the resolution of one-dimensional Lindblad equations, or related stochastic Schr\"odinger equations, in order to reproduce in a good approximation the quarkonium phenomenology in heavy ion collisions. We also compared the proposed potential to a possible reduction of the HTL potential in one-dimensional and to a potential found in the literature \cite{Miura:2019ssi,Alund:2020ctu}. The behaviour of the real parts of these two potentials is incompatible with the energy spectra given by the three-dimensional lattice QCD inspired potential. Nevertheless, combining the real part of the proposed potential and the imaginary part of the model found in the literature (rescaled by some factor) reasonably matches the thermal widths of the three-dimensional potential. This combination can thus be seen as a simple alternative to the main proposal in this work. We finally emphasize that one-dimensional dynamics remains inherently limited and does not aim to reproduce the whole physics at stake. Nevertheless, it can still be used as a good proxy for phenomenology while being much less demanding from a computational viewpoint.

\section*{Acknowledgments} 

We wish to thank Thierry Gousset and Alexander Rothkopf for discussions and help. The authors thank the Région Pays de la Loire and Subatech for support. R.K.~is under contract No.~2015-08473. S.D.~is supported by the Centre national de la recherche scientifique (CNRS) and Région Pays de la Loire.

\bibliography{BIG}
\end{document}